\documentclass[pra,twocolumn,amsmath,amssymb,superscriptaddress]{revtex4-1}
\usepackage{epsfig,amsmath}
\usepackage{subfigure}
\usepackage{graphicx}
\usepackage{dcolumn}
\usepackage{stmaryrd}
\usepackage{mathrsfs}
\usepackage{pifont}
\usepackage{amsthm}
\usepackage{amssymb}
\usepackage{bm}
\usepackage{latexsym}
\usepackage[colorlinks=true,linkcolor=blue,citecolor=blue]{hyperref}
\usepackage{color}
\usepackage{epstopdf}

\usepackage{booktabs}
\usepackage{threeparttable}
\usepackage{multirow}

\begin{document}

\title{Preparing Greenberger-Horne-Zeilinger state on ground levels of neutral atoms}

\author{Zhu-yao Jin}
\affiliation{School of Physics, Zhejiang University, Hangzhou 310027, Zhejiang, China}

\author{Jun Jing}
\email{Contact author: jingjun@zju.edu.cn}
\affiliation{School of Physics, Zhejiang University, Hangzhou 310027, Zhejiang, China}

\date{\today}

\begin{abstract}
We propose a scalable protocol in the Rydberg atomic system to generate the Greenberger-Horne-Zeilinger (GHZ) states, where the qubits are encoded on the hyperfine ground levels. Our system is featured with off-resonant driving fields rather than strong Rydberg interaction sufficient for blockade. Closely it can follow the desired nonadiabatic passage during the time evolution and avoid the unwanted transition by imposing the passage-dependent and fast-varying global phase, that serves as an error correction to the universal quantum control. In our protocol, an $N$-qubit GHZ state is prepared in $N-1$ steps, which is found to be robust against both environmental noise and systematic deviation. Our protocol therefore provides an avenue toward large-scale entanglement, which is essential for quantum information processing and computation based on neutral atoms.
\end{abstract}

\maketitle

\section{Introduction}

Efficient control~\cite{Kral2007Colloquium,Li2020Aboost} over large-scale quantum system plays a critical role in exhibiting quantum advantages. Creating large-scale entanglement~\cite{Einstein1935Can,Bohr1935Can,Horodecki2009Quantum} as one of the main tasks of quantum control, is a necessary step toward quantum information processing~\cite{Ekert1996Quantum,Edward2001Aquantum,Sergey2018Quantum}, quantum teleportation~\cite{Bennett1993Teleporting}, and one-way quantum computation~\cite{Raussendorf2001Oneway}. In practice, large-scale entanglement is required to be immune to both decoherence from environmental noises and imperfect operations from systematic errors~\cite{Gidney2019HowTF}.

In the pursuit of quantum computation~\cite{Ladd2010Quantum}, numerous platforms have been developed over past decades, such as trapped-ion systems~\cite{Benhelm2008Towards,Monz2011QubitEntanglement,Harty2014Highfiedelity,Ballance2016Highfidelity,
Gaebler2016Highfidelity}, superconducting qubit systems~\cite{Chow2012Universal,Barends2014Superconducting,Chao2019Generation,Wright2019Benchmarking,Zhong2021Deterministic}, and neutral atoms~\cite{Saffman2010Quantum,Xia2015Randomized,Wang2016Single,Sheng2018Highfidelity,Weiss2017Quantum,
Levine2018Highfidelity,Madjarove2020Highfidelity,Omran2019Generation,Barredo2016Anatombyatom,Ebadi2021Quantum,
Barredo2018Barredo,Barredo2020Threedimensional}. The Bell states can be prepared with a fidelity $\mathcal{F}>0.999$ in trapped-ion systems~\cite{Ballance2016Highfidelity,Gaebler2016Highfidelity} and with $\mathcal{F}>0.99$ in superconducting circuits~\cite{Barends2014Superconducting}. However, large-scale entanglement in these two platforms is constrained by the growing crosstalk and unwanted interaction between qubits as their number increases~\cite{Li2020Aboost}. For example, the fidelity of the Greenberger-Horne-Zeilinger (GHZ) state of six qubits is now upper-bounded by $\mathcal{F}=0.892$ in trapped-ion systems~\cite{Monz2011QubitEntanglement} and by $\mathcal{F}=0.722$ in superconducting systems~\cite{Zhong2021Deterministic}. Almost all of these results have to resort to error correction.

In contrast, neutral atoms have recently attracted attention for exhibiting both scalability and controllability. A lot of atoms can be loaded into one-, two-, or three-dimensional arrays~\cite{Xia2015Randomized,Barredo2016Anatombyatom,Ebadi2021Quantum,Barredo2018Barredo,
Barredo2020Threedimensional}. They can be individually addressed by laser fields and little crosstalk is found among the non-nearest-neighboring atoms~\cite{Saffman2010Quantum,Weiss2017Quantum}. Bottom-up methods have been significantly improved for the neutral atoms, including accurate initialization, manipulation, and readout of the internal and motional states~\cite{Weiss2017Quantum,Levine2018Highfidelity,Madjarove2020Highfidelity}. The coherence time of single-qubit gates encoded on the hyperfine ground states of Rydberg atoms is over seconds in both three-dimensional~\cite{Wang2016Single} and two-dimensional~\cite{Sheng2018Highfidelity} arrays, which is three orders in magnitude longer than that of gates on the Rydberg states. The single-excitation Bell state of two Rydberg atoms, which is encoded on a ground level and a Rydberg level, has been demonstrated after error correction~\cite{Levine2018Highfidelity,Madjarove2020Highfidelity}. However, the conventional protocols rely heavily on the strong, tunable, and long-range Rydberg interaction~\cite{Browaeys2016Experimental}, and the qubits encoded on the Rydberg states are highly susceptible to errors and decoherence.

Rydberg interactions~\cite{Browaeys2016Experimental} among neutral atoms can be classified to the van der Waals (vdW) interaction as $V_{\rm vdw}|r_{a(b)}r_{a(b)}\rangle\langle r_{a(b)}r_{a(b)}|$ and the dipole-dipole (DD) interaction as $V_{dd}|r_ar_b\rangle\langle r_br_a|+{\rm H.c.}$, where $|r_a\rangle$ and $|r_b\rangle$ represent different Rydberg states. The coupling strengths scale as $V_{\rm vdw}\sim1/d^6$ and $V_{dd}\sim1/d^3$ with the atomic distance $d$. When the atoms occupy the same Rydberg state~\cite{Walker2008Consequences,Saffman2010Quantum,Shao2017Ground,Shao2017Dissipation,Zhao2017Rydberg,Li2024Simulation,Shao2024Rydberg}, the vdW interaction dominates and the DD interaction is negligible. Using resonant driving field, the strong vdW interaction induces a large energy shift of the double Rydberg state $|rr\rangle$. It can be sufficient to observe the Rydberg blockade~\cite{Walker2008Consequences,Saffman2010Quantum,Dong2019Adiabatic} that the excitation of one atom to the Rydberg state inhibits the excitation of nearby atoms. Conventionally, $V_{\rm vdW}\sim100\Omega$ ($\Omega$ is the driving field intensity) is the foundation of two-qubit gates~\cite{Jaksch2000Fast}, Bell-state generation~\cite{Levine2019parallel,Graham2019Rydberg}, and multiparticle entanglement via adiabatic passage~\cite{Moller2008Quantumgates}. It can be relieved to about $10\Omega$ when compensated by the detunings of driving fields~\cite{Saffman2010Quantum,Su2020Rydberg,Su2021Dipole,Zhao2017Robust}. The neutral atoms can be coupled by a strong DD interaction $V_{dd}\sim10^{2}\Omega$ and a weak vdW interaction, when they are excited to different Rydberg states, yielding the asymmetric Rydberg blockade~\cite{Brion2007Conditional,Saffman2009Efficient,Implementation2010Wu,Rao2014Deterministice,Young2021Asymmetric}, which was used to generate the GHZ state of $N$ particles~\cite{Saffman2009Efficient,Implementation2010Wu,Rao2014Deterministice,Young2021Asymmetric}.

Both the Rydberg blockade and its asymmetric counterpart, however, require the atoms to be excited to the high-lying Rydberg state and held within a close distance. In this case, the distance fluctuation could induce significant fluctuations in coupling strength~\cite{Robicheaux2021Photon}. In addition, the asymmetric Rydberg blockade can be broken down in the presence of a third atom within the blockade radius~\cite{Pohl2009Breaking,Urban2009Observation}. The generation of large-scale entanglement of high fidelity and stability is then challenging in neutral atoms with the conventional protocols.

In this paper, we use off-resonant driving fields to generate the GHZ state of multiple Rydberg atoms within the subspace of the hyperfine ground states, which only requires a Rydberg interaction of about $5\Omega$. Using the Magnus expansion and the strong large-detuning driving, we obtain the same effective Hamiltonian as that by the James' method~\cite{James2007Effective} for quantum systems with a large-detuning control Hamiltonian. Then we apply our universal control theory~\cite{Jin2025Universal,Jin2025Entangling,Jin2025ErrCorr} to design nonadiabatic and transitionless passages of multiple Rydberg qubits. With a simple method of error correction, the double-excitation Bell state can be stabilized with a high fidelity, even in the presence of significant global or local systematic deviations. As an extension, the general GHZ state of $N$ qubits can be generated with $N-1$ steps. Each step is characterized by the driving pulses applied to the selected pair of qubits.

The rest of this paper is structured as follows. In Sec.~\ref{general}, we derive the effective Hamiltonian with strong and large-detuning control components, and briefly recall the general framework of our universal quantum control in the absence and in the presence of systematic errors. In Sec.~\ref{Two}, we prepare a double-excitation Bell state of two Rydberg atoms in the ideal situation. Transitionless passages immune to both environment noises and systematic errors are presented in Sec.~\ref{Num}. Our protocol is extended in Sec.~\ref{EntangleN} to generate a multiparticle GHZ state. The whole work is concluded in Sec.~\ref{conclusion}. Appendixes~\ref{Efftheory} and \ref{UnivErr} provide the detailed derivations of the effective Hamiltonian and the error correction condition for the universal quantum control theory, respectively.

\section{Effective Hamiltonian and universal quantum control}\label{general}

Our study starts from a $K$-dimensional closed quantum system, where $K$ is arbitrary and finite. In the rotating frame with respect to the free Hamiltonian, the system dynamics can be described by the time-dependent Schr\"odinger equation as ($\hbar\equiv1$)
\begin{equation}\label{Sch}
i\frac{d|\phi_m(t)\rangle}{dt}=H_I(t)|\phi_m(t)\rangle, \quad 1\leq m\leq K,
\end{equation}
where $H_I(t)$ is the time-dependent control Hamiltonian and $|\phi_m(t)\rangle$'s are the pure-state solutions. According to the Magnus expansion~\cite{Blanes2009Magnus}, the evolution operator can be expressed as
\begin{equation}\label{Umagnus}
U_I(t)=\exp\left[\Lambda(t)\right]=\exp\left[\sum_{l=1}^{\infty}\Lambda_l(t)\right],
\end{equation}
where the first- and second-order components are
\begin{equation}\label{MagnusTerm}
\begin{aligned}
\Lambda_1(t)&=-i\int_0^tdt_1H_I(t_1),\\
\Lambda_2(t)&=\frac{(-i)^2}{2}\int_0^tdt_1\int_0^{t_1}dt_2\left[H_I(t_1), H_I(t_2)\right],
\end{aligned}
\end{equation}
respectively. In many realistic scenario, the control Hamiltonian $H_I(t)$ can be written as
\begin{equation}\label{HI}
H_I(t)=\sum_j\Omega_j(t)e^{i\Delta_j t}A_j+{\rm H.c.},
\end{equation}
where $\Omega_j(t)$, $\Delta_j$, and $A_j$ are coupling strength, detuning, and coupling operator, respectively. With sufficiently large detunings, i.e., $\Delta_j\gg\Omega_j(t)$, the first-order term and a part of the second-order term in the expansion of $U_I(t)$ in Eq.~(\ref{Umagnus}) can be eliminated (see appendix~\ref{Efftheory} for details). Consequently, the system dynamics described by Eq.~(\ref{HI}) is equivalent to that governed by the effective Hamiltonian
\begin{equation}\label{Heff}
H_{\rm eff}(t)=-\frac{i}{2}\left[H_I(t), \int_0^tH_I(t_1)dt_1\right].
\end{equation}
The James' method~\cite{James2007Effective}, originally applied to the systems of time-independent Hamiltonian, is formally extended to those of time-dependent Hamiltonian by our theory, as justified with an illustrative example in Sec.~\ref{Entangle}. Typically the degree of freedoms of the effective Hamiltonian can be much reduced by large detunings since the transition components of high orders have been omitted. In other words, the rank of the matrix for $H_{\rm eff}(t)$ is smaller than that for $H_I(t)$. With no loss of generality, $H_{\rm eff}(t)$ can live in a $K'$-dimensional, $K'\le K$, subspace spanned by a proper set of basis states.

The theory about the effective Hamiltonian~(\ref{Heff}) can be integrated with the universal control framework~\cite{Jin2025Universal,Jin2025Entangling}. For the system dynamics, we now have
\begin{equation}\label{SchNon}
i\frac{d|\psi_n(t)\rangle}{dt}=H_{\rm eff}(t)|\psi_n(t)\rangle,
\end{equation}
where $|\psi_n(t)\rangle$'s are the pure-state solutions in the $K'$-dimensional subspace with $1\leq n\leq K'$. As a generalization of the d'Alembert principle about virtual displacement, the dynamics described by Eq.~(\ref{SchNon}) can be alternatively treated in a rotated picture spanned by the ancillary basis states $|\mu_k(t)\rangle$'s, $1\leq k\leq K'$. Rotated by $V(t)\equiv\sum_{k=1}^{K'}|\mu_k(t)\rangle\langle\mu_k(0)|$, Eq.~(\ref{SchNon}) is transformed to be
\begin{equation}\label{Schrot}
i\frac{d|\psi_n(t)\rangle_{\rm rot}}{dt}=H_{\rm rot}(t)|\psi_n(t)\rangle_{\rm rot}
\end{equation}
with the rotated pure-state solutions $|\psi_n(t)\rangle_{\rm rot}=V^\dagger(t)|\psi_n(t)\rangle$ and the rotated system Hamiltonian under a fixed base,
\begin{equation}\label{Hamrot}
\begin{aligned}
&H_{\rm rot}(t)=V^\dagger(t)H_{\rm eff}(t)V(t)-iV^\dagger(t)\frac{d}{dt}V(t)\\
&=-\sum_{k=1}^{K'}\sum_{n=1}^{K'}\left[\mathcal{G}_{kn}(t)-\mathcal{D}_{kn}(t)\right]|\mu_k(0)\rangle\langle\mu_n(0)|.
\end{aligned}
\end{equation}
Here $\mathcal{G}_{kn}(t)\equiv i\langle\mu_k(t)|\dot{\mu}_n(t)\rangle$ and $\mathcal{D}_{kn}(t)\equiv\langle\mu_k(t)|H_{\rm eff}(t)|\mu_n(t)\rangle$ represent the geometrical and dynamical parts of the matrix elements, respectively.

If $H_{\rm rot}(t)$ is diagonalized, i.e., $\mathcal{G}_{kn}(t)-\mathcal{D}_{kn}(t)=0$ for $k\ne n$, then Eq.~(\ref{Schrot}) can be exactly solved~\cite{Jin2025Universal,Jin2025Entangling,Jin2025ErrCorr}. In particular, Eq.~(\ref{Hamrot}) can be simplified as
\begin{equation}\label{Hdigfull}
H_{\rm rot}(t)=-\sum_{k=1}^{K'}\left[\mathcal{G}_{kk}(t)-\mathcal{D}_{kk}(t)\right]|\mu_k(0)\rangle\langle\mu_k(0)|.
\end{equation}
Consequently, the time-evolution operator $U_{\rm rot}(t)$ can be directly obtained as
\begin{equation}\label{Urot1}
U_{\rm rot}(t)=\sum_{k=1}^{K'}e^{if_k(t)}|\mu_k(0)\rangle\langle\mu_k(0)|,
\end{equation}
where the $k$th global phase is defined as $f_k(t)\equiv\int_0^t[\mathcal{G}_{kk}(t_1)-\mathcal{D}_{kk}(t_1)]dt_1$. Rotating back to the original picture, the time-evolution operator can be written as
\begin{equation}\label{U0}
U_0(t)=V(t)U_{\rm rot}(t)=\sum_{k=1}^{K'}e^{if_k(t)}|\mu_k(t)\rangle\langle\mu_k(0)|.
\end{equation}
This equation implies that if the system starts from any passage $|\mu_k(0)\rangle$, it will closely follow the instantaneous state $|\mu_k(t)\rangle$ and accumulate a global phase $f_k(t)$. During this time evolution, there exists no transition among different passages.

A previous work of ours proved that the diagonalization of $H_{\rm rot}(t)$ in Eq.~(\ref{Hdigfull}) can be practically implemented by the von Neumann equation~\cite{Jin2025Universal,Jin2025Entangling,Jin2025ErrCorr}
\begin{equation}\label{von}
\frac{d}{dt}\Pi_k(t)=-i\left[H_{\rm eff}(t), \Pi_k(t)\right]
\end{equation}
with the effective system Hamiltonian $H_{\rm eff}(t)$ and the projection operator $\Pi_k(t)\equiv|\mu_k(t)\rangle\langle\mu_k(t)|$. The static dark states $|\mu_k(t)\rangle\rightarrow|\mu_k\rangle$ can be regarded as trivial solutions to Eq.~(\ref{von}). Armed with the time dependence, they can be activated to be useful passages, e.g., see the recipe of $|\mu_k(t)\rangle$'s for the general two-band systems~\cite{Jin2025Entangling}.

The universal control framework can be extended to cope with the inevitable systematic errors arising from the control setting. Under this nonideal situation, the system evolves according to
\begin{equation}\label{HamNon}
i\frac{d|\psi_m(t)\rangle}{dt}=H(t)|\psi_m(t)\rangle, \quad H(t)=H_{\rm eff}(t)+\epsilon H_e(t),
\end{equation}
where $H_e(t)$ is the error Hamiltonian in the effective subspace and $\epsilon$ is a perturbative coefficient measuring the error magnitude. Then the rotated Hamiltonian~(\ref{Hdigfull}) becomes
\begin{equation}\label{HamNonRot}
\begin{aligned}
&H_{\rm rot}(t)=-\sum_{n=1}^{K'}\dot{f}_k(t)|\mu_k(0)\rangle\langle\mu_k(0)|\\
&+\epsilon\sum_{k=1}^{K'}\sum_{n=1}^{K'}\mathcal{D}_{kn}^{(e)}(t)|\mu_k(0)\rangle\langle\mu_n(0)|,
\end{aligned}
\end{equation}
where $\mathcal{D}_{kn}^{(e)}(t)\equiv\langle\mu_k(t)|H_e(t)|\mu_n(t)\rangle$. Indicated by the off-diagonal elements of $H_{\rm rot}(t)$, the systematic errors yield unwanted transitions among the passages $|\mu_k(t)\rangle$. The adverse effects induced by systematic errors can be estimated and neutralized~\cite{Jin2025ErrCorr} (see Appendix~\ref{UnivErr} for details). Specifically, when the passage-dependent global phase is a fast-varying function of time in comparison to the transition rate of the passages,
\begin{equation}\label{OptGeneral}
|\dot{f}_k(t)-\dot{f_n}(t)|\gg \frac{d}{dt}\left[\langle\mu_k(t)|H_e(t)|\mu_n(t)\rangle\right],
\end{equation}
the undesirable transitions among the passages can be significantly suppressed.

\section{Maximally entangling two qubits via universal nonadiabatic passages}\label{Entangle}

\subsection{Ideal situation}\label{Two}

\begin{figure}[htbp]
\centering
\includegraphics[width=0.8\linewidth]{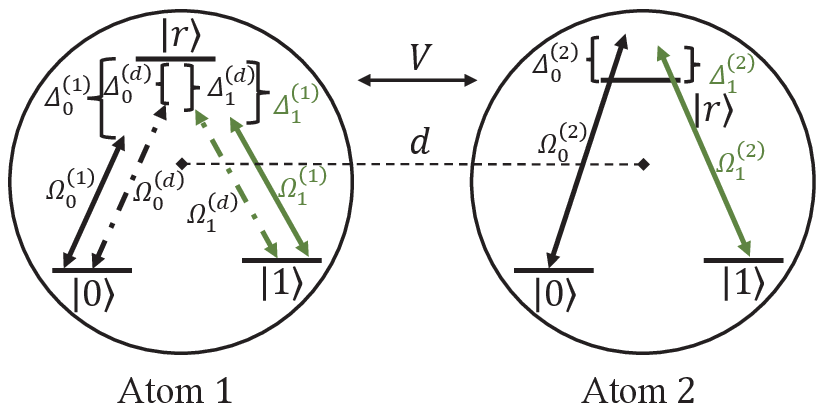}
\caption{Sketch of a pair of coupled Rydberg atoms under largely detuned driving fields. $V$ represents the Rydberg interaction between them. Each Rydberg atom consists of two ground states $|0\rangle$ and $|1\rangle$ and one Rydberg state $|r\rangle$. $d$ is atomic distance. }\label{model}
\end{figure}

Consider two coupled Rydberg atoms under driving fields in Fig.~\ref{model}. The atom consists of two stable ground states $|0\rangle$ and $|1\rangle$ and one Rydberg state $|r\rangle$. The first and second atoms are driven by the red-detuned and blue-detuned laser fields, respectively, with time-dependent Rabi frequencies $\Omega_{0,1}^{(j)}(t)$ and time-independent detuning $\Delta_{0,1}^{(j)}$, $j=1,2$. In addition, the transitions $|r\rangle\leftrightarrow|n\rangle$, $n=0,1$, of the first atom have two extra driving fields with Rabi frequency $\Omega_n^{(d)}(t)$ and detuning $\Delta_n^{(d)}$. The full Hamiltonian in the rotating frame with respect to the free Hamiltonian of the two atoms can be written as
\begin{equation}\label{Ham}
H(t)=\left[H_1(t)+H_d(t)\right]\otimes \mathcal{I}_2+\mathcal{I}_1\otimes H_2(t)+V|rr\rangle\langle rr|,
\end{equation}
with
\begin{equation}\label{H1H2}
\begin{aligned}
&H_j(t)=\sum_{n=0}^1\Omega_n^{(j)}(t)e^{i\Delta_n^{(j)}t}|r\rangle_j\langle n|+{\rm H.c.}, \quad j=1,2\\
&H_d(t)=\sum_{n=0}^1\Omega_n^{(d)}(t)e^{i\Delta_n^{(d)}t}|r\rangle_1\langle n|+{\rm H.c.},
\end{aligned}
\end{equation}
where the Hamiltonian $H_1(t)$ and $H_d(t)$ describe the driving fields on the first atom and $H_2(t)$ acts on the second one. Here $\mathcal{I}$ denotes the identity operator. Conventional Rydberg-blockade protocols~\cite{Muller2011Optimizing,Goerz2014Robustness,Muller2014Implementation,Moller2008Quantumgates,
Rao2014Deterministice,Levine2019parallel,Graham2019Rydberg} use resonant driving fields and take advantage of the substantial energy shift about the double Rydberg state $|rr\rangle$~\cite{Walker2008Consequences} induced by a strong vdW interaction $V$. However, a stronger atomic interaction is associated with a larger relative distance fluctuation $\delta d/d$ and a more significant relative interaction fluctuation $\delta V/V$~\cite{Robicheaux2021Photon}, given a fixed distance fluctuation $\delta d$ for the atoms in close proximity. In contrast, we use the detunings of driving fields to compensate a much weaker Rydberg interaction and then our protocol does not require an atomic coupling strength much larger than the Rabi frequency of driving fields.

\begin{figure}[htbp]
\centering
\includegraphics[width=0.8\linewidth]{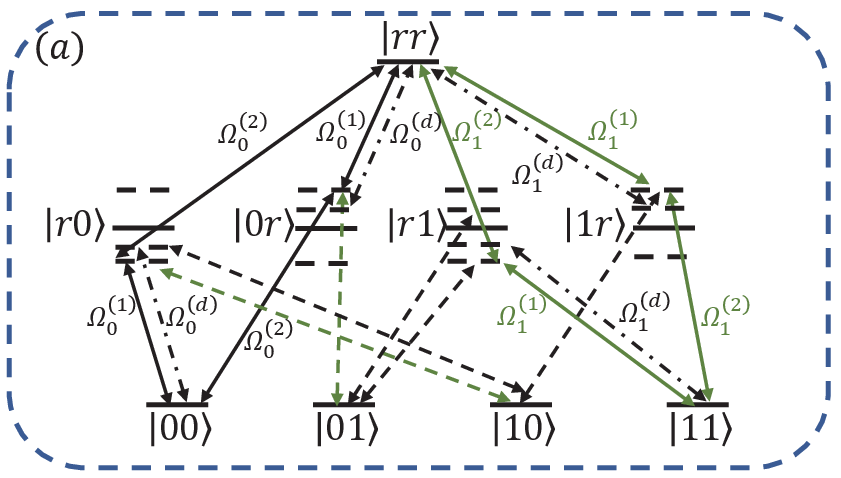}
\includegraphics[width=0.8\linewidth]{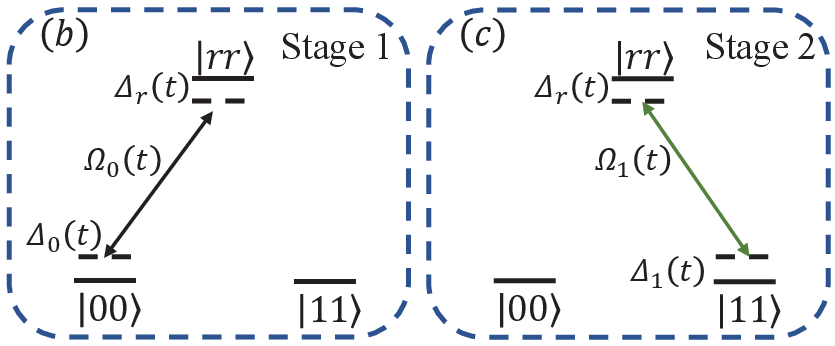}
\caption{(a) Transition diagram for the two Rydberg atomic system driven by largely-detuned driving fields. The black and green solid lines describe the effective transitions $|rr\rangle\leftrightarrow|00\rangle$ (Stage 1) with coupling strength $\Omega_0(t)$ and $|rr\rangle\leftrightarrow|11\rangle$ (Stage 2) with $\Omega_1(t)$, respectively. The round trips plotted with dotted-dashed lines contribute to the effective detunings $\Delta_0(t)$, $\Delta_r(t)$, and $\Delta_1(t)$ for the states $|00\rangle$, $|rr\rangle$, and $|11\rangle$. The transitions indicated by the black and green dashed lines can be strongly suppressed under large detunings, i.e., $\Delta_n^{(j)}\gg\Omega_n^{(j)}$, $n=0,1$. (b) and (c) Effective transition diagram in the three-dimensional subspace for Stages 1 and 2, respectively.}\label{transition}
\end{figure}

In the second-rotated frame with respect to $\mathcal{U}(t)=\exp(-iVt|rr\rangle\langle rr|)$, the Hamiltonian in Eq.~(\ref{Ham}) can be transformed as
\begin{equation}\label{Hrot}
\begin{aligned}
&H_I(t)=H_R(t)+H_D(t)\\
\end{aligned}
\end{equation}
with
\begin{equation}\label{Hrot12}
\begin{aligned}
&H_R(t)=\sum_{n=0}^1\Omega_n^{(1)}(t)e^{i\Delta_n^{(1)}t}(|r0\rangle\langle n0|+|r1\rangle\langle n1|\\ &+e^{iVt}|rr\rangle\langle nr|)+\sum_{n=0}^1\Omega_n^{(2)}(t)e^{-i\Delta_n^{(2)}t}(|0r\rangle\langle0n|\\ &+|1r\rangle\langle1n|+e^{iVt}|rr\rangle\langle rn|)+{\rm H.c.},
\end{aligned}
\end{equation}
and
\begin{equation}\label{HrotD}
\begin{aligned}
&H_D(t)=\sum_{n=0}^{1}\Omega_n^{(d)}(t)e^{i\Delta_n^{(d)}t}(|r0\rangle\langle n0|+|r1\rangle\langle n1|\\
&+e^{iVt}|rr\rangle\langle nr|)+{\rm H.c.},
\end{aligned}
\end{equation}
where $H_D(t)$ is used to distinguish the extra driving fields on the first atom. The transition diagram in the whole Hilbert space is demonstrated in Fig.~\ref{transition}(a).

The following protocol for generating a double-excitation Bell state with the ground levels of the Rydberg system assumes a large detuning for all the fields, i.e., $\Delta_n^{(j)}\gg\Omega_n^{(j)}(t)$ and $\Delta_n^{(d)}\gg\Omega_n^{(d)}(t)$. It consists of two stages of the same period $T$, by which all other parameters including coupling strengths, Rabi frequencies, and detunings can be accordingly set. Our protocol of state generation is then scale free at least in the ideal situation.

\emph{Stage 1.} On this stage, the two-atom system is forced to evolve from the ground state $|00\rangle$ to the superposed state $(|00\rangle+|rr\rangle)/\sqrt{2}$. By switching on (off) the driving fields on the transition $|r\rangle\leftrightarrow|0\rangle$ ($|r\rangle\leftrightarrow|1\rangle$) of both atoms, the control Hamiltonian in Eq.~(\ref{Hrot}) is reduced to
\begin{equation}\label{HamrotS1}
\begin{aligned}
&H_I(t)=\Omega_0^{(1)}(t)e^{i\Delta_0^{(1)}t}(|r0\rangle\langle00|+|r1\rangle\langle01|\\ &+e^{iVt}|rr\rangle\langle0r|)
+\Omega_0^{(2)}(t)e^{-i\Delta_0^{(2)}t}(|0r\rangle\langle00|+|1r\rangle\langle10|\\ &+e^{iVt}|rr\rangle\langle r0|)
+\Omega_0^{(d)}(t)e^{i\Delta_0^{(d)}t}(|r0\rangle\langle00|+|r1\rangle\langle01|\\ &+e^{iVt}|rr\rangle\langle0r|)+{\rm H.c.}.
\end{aligned}
\end{equation}
With the setting that $\Delta_0^{(1)}+V=\Delta_0^{(2)}$ and $\Omega_0^{(1)}(t)=\Omega_0^{(2)}(t)$, the effective Hamiltonian can be obtained as
\begin{equation}\label{HeffS1}
\begin{aligned}
H_{\rm eff}(t)&=\Delta_r(t)|rr\rangle\langle rr|+\Delta_0(t)|00\rangle\langle00|\\
&+\Omega_0(t)|rr\rangle\langle 00|+{\rm H.c.}
\end{aligned}
\end{equation}
by Eq.~(\ref{Heff}). Here the effective detunings $\Delta_r(t)$ and $\Delta_0(t)$ and Rabi-frequency $\Omega_0(t)$ are
\begin{equation}\label{effDetuS1}
\begin{aligned}
&\Delta_r(t)=\frac{[\Omega_0^{(d)}(t)]^2}{\Delta_0^{(d)}+V}, \quad\Delta_0(t)=-\frac{[\Omega_0^{(d)}(t)]^2}{\Delta_0^{(d)}}\approx-\Delta_r(t),\\
&\Omega_0(t)=[\Omega_0^{(1)}(t)]^2\left[\frac{1}{\Delta_0^{(2)}}-\frac{1}{\Delta_0^{(1)}}\right],
\end{aligned}
\end{equation}
where the approximation holds under the condition of $\Delta_0^{(d)}\gg V$. Then the transition diagram in Fig.~\ref{transition}(a) is reduced to that in Fig.~\ref{transition}(b).

Equivalently we are left with a two-level system in the subspace spanned by $|00\rangle$ and $|rr\rangle$, in which the energy splitting $2\Delta_r(t)$ and the transition rate $\Omega_0(t)$ can be manipulated by the driving intensities $\Omega_0^{(d)}(t)$ and $\Omega_0^{(1)}(t)$, respectively. Applying the universal control theory in Sec.~\ref{general}, the system dynamics can be described by the ancillary basis states~\cite{Jin2025Universal,Jin2025Entangling}. They can be written as
\begin{equation}\label{Anci}
\begin{aligned}
|\mu_1(t)\rangle&=\cos\theta(t)e^{i\frac{\alpha(t)}{2}}|00\rangle-\sin\theta(t)e^{-i\frac{\alpha(t)}{2}}|rr\rangle,\\
|\mu_2(t)\rangle&=\sin\theta(t)e^{i\frac{\alpha(t)}{2}}|00\rangle+\cos\theta(t)e^{-i\frac{\alpha(t)}{2}}|rr\rangle,
\end{aligned}
\end{equation}
where the parameters $\theta(t)$ and $\alpha(t)$ are used to control the population and the relative phase between the states $|00\rangle$ and $|rr\rangle$, respectively.

Substituting the ancillary basis states in Eq.~(\ref{Anci}) into the von Neumann equation (\ref{von}) with the effective Hamiltonian in Eq.~(\ref{HeffS1}), we have
\begin{equation}\label{Condition}
\begin{aligned}
&2\Delta_r(t)=\dot{\alpha}(t)+2\Omega_0(t)\cot2\theta(t)\cos\alpha(t),\\
&\Omega_0(t)=-\frac{\dot{\theta}(t)}{\sin\alpha(t)}.
\end{aligned}
\end{equation}
Due to Eq.~(\ref{U0}), the time-evolution operator reads,
\begin{equation}\label{U}
U(t,0)=e^{if(t)}|\mu_1(t)\rangle\langle\mu_1(0)|+e^{-if(t)}|\mu_2(t)\rangle\langle\mu_2(0)|,
\end{equation}
where the global phase satisfies
\begin{equation}\label{phase}
\dot{f}(t)=\Omega_0(t)\frac{\cos\alpha(t)}{\sin2\theta(t)}=-\dot{\theta}(t)\frac{\cot\alpha(t)}{\sin2\theta(t)}.
\end{equation}
Regarding $\theta(t)$, $\alpha(t)$, and $f(t)$ as independent variables, the conditions in Eq.~(\ref{Condition}) are expressed as
\begin{subequations}\label{ConditionInv}
\begin{align}
&2\Delta_r(t)=\dot{\alpha}(t)+2\dot{f}(t)\cos2\theta(t), \label{Deltar} \\
&\Omega_0(t)=-\sqrt{\dot{\theta}(t)^2+\dot{f}(t)^2\sin^22\theta(t)}, \label{Om0}\\
&\dot{\alpha}(t)=-\frac{\ddot{\theta}\dot{f}\sin2\theta-\ddot{f}\dot{\theta}\sin2\theta
-2\ddot{f}\dot{\theta}^2\cos2\theta}{\dot{f}^2\sin^22\theta+\dot{\theta}^2}.\label{dotalpha}
\end{align}
\end{subequations}
With Eqs.~(\ref{effDetuS1}) and (\ref{ConditionInv}), the Rabi-frequencies $\Omega_0^{(1)}(t)$ and $\Omega_0^{(d)}(t)$ are practically formulated as
\begin{equation}\label{ConditionOri}
\begin{aligned}
\Omega_0^{(1)}(t)&=\sqrt{\Omega_0(t)\left[\Delta_0^{(1)}\Delta_0^{(2)}\right]/\left[\Delta_0^{(2)}-\Delta_0^{(1)}\right]},\\
\Omega_0^{(d)}(t)&=\sqrt{\Delta_r(t)\Delta_0^{(d)}}.
\end{aligned}
\end{equation}
The task of Stage 1 can be accomplished by selecting either passage $|\mu_1(t)\rangle$ or $|\mu_2(t)\rangle$ in Eq.~(\ref{Anci}), through appropriate setting of $\theta(t)$, $\alpha(t)$, and $f(t)$. For example, when $\theta(0)=0$, $\alpha(0)=0$, $\theta(T)=\pi/4$, and $\alpha(T)=\pi$, the system can evolve along the passage $|\mu_1(t)\rangle$ and the state $|00\rangle$ becomes $(|00\rangle+|rr\rangle)/\sqrt{2}$ when $t=T$.

\emph{Stage 2.} On this stage, we target to push the system from the superposed state $(|00\rangle+|rr\rangle)/\sqrt{2}$ to the double-excitation Bell state encoded with the ground states $|00\rangle$ and $|11\rangle$. Contrary to Stage 1, now we switch on (off) the driving fields on the transition $|r\rangle\leftrightarrow|1\rangle$ ($|r\rangle\leftrightarrow|0\rangle$) of both atoms. The rotated Hamiltonian in Eq.~(\ref{Hrot}) is then reduced to
\begin{equation}\label{HamrotS2}
\begin{aligned}
&H_I(t)=\Omega_1^{(1)}(t)e^{i\Delta_1^{(1)}t}(|r0\rangle\langle10|+|r1\rangle\langle11|\\ &+e^{iVt}|rr\rangle\langle1r|)
+\Omega_1^{(2)}(t)e^{-i\Delta_1^{(2)}t}(|0r\rangle\langle01|+|1r\rangle\langle11|\\ &+e^{iVt}|rr\rangle\langle r1|)
+\Omega_1^{(d)}(t)e^{i\Delta_1^{(d)}t}(|r0\rangle\langle10|+|r1\rangle\langle11|\\ &+e^{iVt}|rr\rangle\langle1r|)+{\rm H.c.}.
\end{aligned}
\end{equation}
Similar to Stage 1, the detunings and Rabi frequencies are set as $\Delta_1^{(1)}+V=\Delta_1^{(2)}$ and $\Omega_1^{(1)}(t)=\Omega_1^{(2)}(t)$, then the effective Hamiltonian becomes
\begin{equation}\label{HeffS2}
\begin{aligned}
H_{\rm eff}(t)&=\Delta_r(t)|rr\rangle\langle rr|+\Delta_1(t)|11\rangle\langle11|\\
&+\Omega_1(t)|rr\rangle\langle 11|+{\rm H.c.},
\end{aligned}
\end{equation}
where
\begin{equation}\label{effDetuS2}
\begin{aligned}
&\Delta_r(t)=\frac{[\Omega_1^{(d)}(t)]^2}{\Delta_1^{(d)}+V}, \quad\Delta_1(t)=-\frac{[\Omega_1^{(d)}(t)]^2}{\Delta_1^{(d)}}\approx-\Delta_r(t),\\
&\Omega_1(t)=[\Omega_1^{(1)}(t)]^2\left[\frac{1}{\Delta_1^{(2)}}-\frac{1}{\Delta_1^{(1)}}\right].
\end{aligned}
\end{equation}
Consequently, the transition diagram in Fig.~\ref{transition}(a) is reduced to that in Fig.~\ref{transition}(c), in which the ground state $|00\rangle$ is decoupled from the system dynamics on this stage. Similar to Eq.~(\ref{Anci}), the dynamics can be described by the ancillary basis states superposed by $|11\rangle$ and $|rr\rangle$:
\begin{equation}\label{Anci1}
\begin{aligned}
|\mu_1(t)\rangle&=\cos\theta(t)e^{i\frac{\alpha(t)}{2}}|11\rangle-\sin\theta(t)e^{-i\frac{\alpha(t)}{2}}|rr\rangle,\\
|\mu_2(t)\rangle&=\sin\theta(t)e^{i\frac{\alpha(t)}{2}}|11\rangle+\cos\theta(t)e^{-i\frac{\alpha(t)}{2}}|rr\rangle.
\end{aligned}
\end{equation}

Replacing $\Omega_0(t)$ in Eqs.~(\ref{Condition}), (\ref{phase}), and (\ref{ConditionInv}) with $\Omega_1(t)$, one can again obtain the same time-evolution operator in Eq.~(\ref{U}), if the Rabi frequencies $\Omega_1^{(1)}(t)$ and $\Omega_1^{(d)}(t)$ satisfy the same conditions in Eq.~(\ref{ConditionOri}) for $\Omega_0^{(1)}(t)$ and $\Omega_0^{(d)}(t)$, respectively. Under the boundary conditions that $\theta(T+0^+)=\pi/2$, $\theta(2T)=0$, and $\alpha(T+0^+)=\alpha(2T)=\pi$, the system can evolve from $(|00\rangle+|rr\rangle)/\sqrt{2}$ to $(|00\rangle+\exp[if(2T)]|11\rangle)/\sqrt{2}$ along $|\mu_1(t)\rangle$.

With no loss of generality, the two-stage $\theta(t)$ can be set as
\begin{equation}\label{theta}
\theta(t)=\left\{
\begin{aligned}
&\frac{\pi t}{4T},\quad t\in\left[0, T\right]\\
&\frac{\pi t}{2T}.\quad t\in\left[T, 2T\right]
\end{aligned}
\right.
\end{equation}
in the following numerical simulation. And after $t=2T$, the driving fields are tuned off, i.e., $\theta(t)=0$, to verify the stability of the entangled states created by our protocol, in which the qubits are encoded on the ground states.

\subsection{Nonideal situation and numerical calculation}\label{Num}

In this section, we first examine the robustness of our protocol against the external noises, that arises primarily from the leakage to the Rydberg state and could be avoided through postselection. Next we consider the systematic errors, including the laser phase noise and the interatomic distance fluctuation, that might affect real-world implementations of our protocol. Then we apply the correction mechanism in Eq.~(\ref{OptGeneral}) to suppress these adverse effects.

We use the Lindblad master equation to take the environmental noise into account~\cite{Carmichael1999statistical}, which reads,
\begin{equation}\label{master_2qubit}
\begin{aligned}
&\frac{\partial \rho}{\partial t}=-i[H(t), \rho]+\sum_{j=1}^2\Big[\frac{\kappa_0}{2}\mathcal{L}(|0\rangle_j\langle r|)\\
&+\frac{\kappa_1}{2}\mathcal{L}(|1\rangle_j\langle r|)+\frac{\kappa_z}{2}\mathcal{L}(\sigma_j^z)\Big].
\end{aligned}
\end{equation}
Here $\rho$ is the density matrix for the two coupled Rydberg atoms, $H(t)$ is the full Hamiltonian in Eq.~(\ref{Ham}), and $\mathcal{L}(o)$ is the Lindblad superoperator defined as $\mathcal{L}(o)\equiv2o\rho o^\dagger-o^\dagger o\rho-\rho o^\dagger o$~\cite{Scully1997quantum}, where $o=|0\rangle_j\langle r|,|1\rangle_j\langle r|,\sigma_j^z$. Particularly, $|0\rangle_j\langle r|$ and $|1\rangle_j\langle r|$ represent the spontaneous emission of the $j$th atom from its Rydberg state $|r\rangle$ to the ground states $|0\rangle$ and $|1\rangle$ with decay rates $\kappa_0$ and $\kappa_1$, respectively. The dephasing of the $j$th atom is described by $\sigma_j^z=|0\rangle_j\langle0|+|1\rangle_j\langle1|-|r\rangle_j\langle r|$ with a rate $\kappa_z$. For simplicity, we assume $\kappa_0=\kappa_1=\kappa$ and $\kappa_z/\kappa=0.1$~\cite{Sun2021Onestep}. The decoherence rate $\kappa$ for the Rydberg states of the alkali-metal atoms~\cite{Beterov2009Quasiclassical,Isenhower2010Demonstration,Adams2020Rydberg} with a low angular momentum is relevant to the principle and azimuthal quantum numbers~\cite{Adams2020Rydberg}, which is typically in the range of $0.75\sim2.48$ kHz~\cite{Beterov2009Quasiclassical,Isenhower2010Demonstration} around the room temperature.

\begin{figure}[htbp]
\centering
\includegraphics[width=0.8\linewidth]{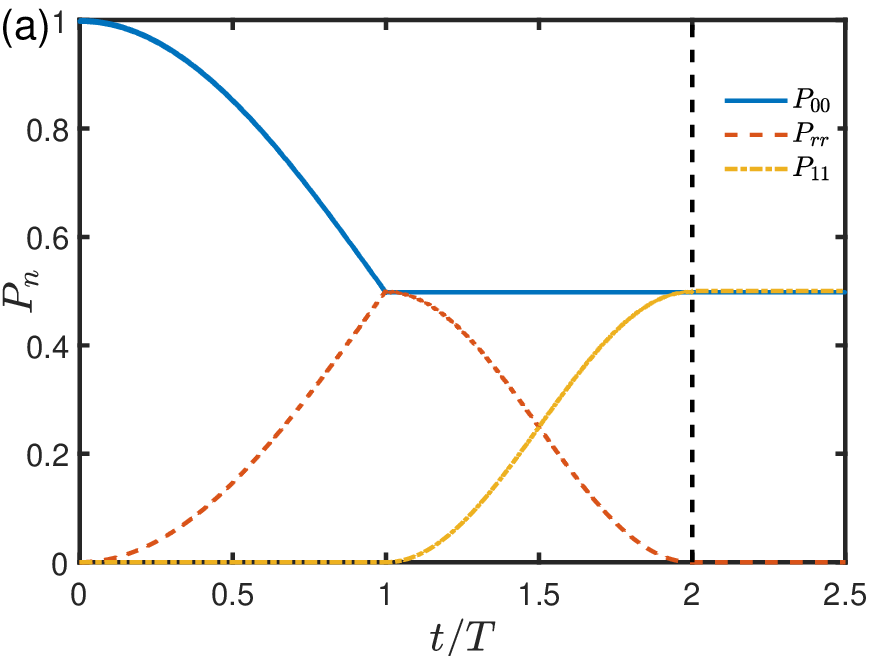}
\includegraphics[width=0.8\linewidth]{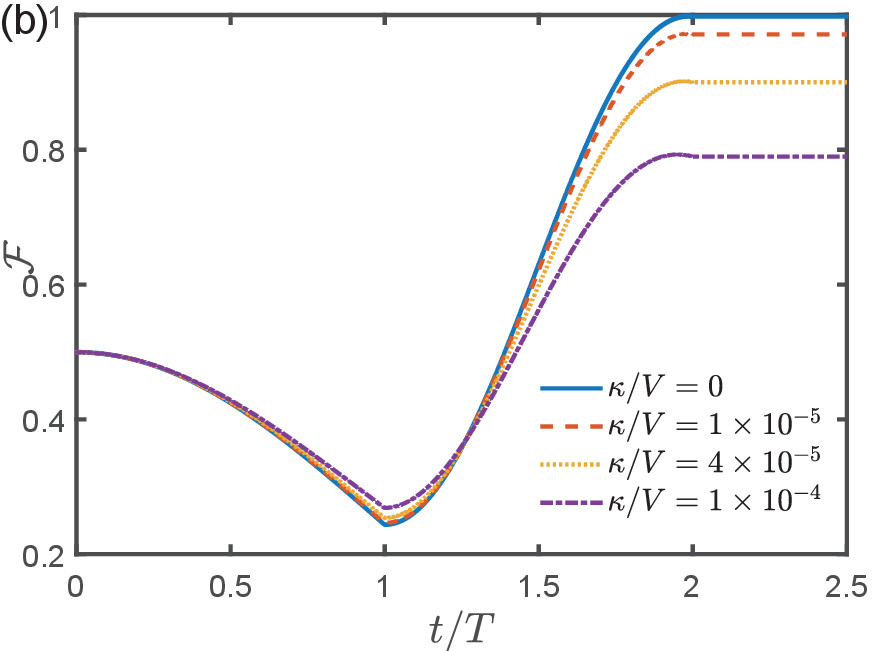}
\caption{(a) Population dynamics $P_n$, $n=00, rr, 11$, under the decoherence-free situation and (b) Fidelity dynamics $\mathcal{F}$ for the target state $(|00\rangle+|11\rangle)/\sqrt{2}$ for both closed and open systems. The period for each stage is set as $T=50$ns. The detunings are set as $\Delta_n^{(1)}=5V$, $\Delta_n^{(2)}=\Delta_n^{(1)}+V$, and $\Delta_n^{(d)}=\Delta_n^{(1)}/2$, with $n=0,1$ and $V/2\pi=20$ MHz. The Rabi frequencies are maintained as $\Omega_0^{(1)}(t)=\Omega_0^{(2)}(t)$ (Stage 1) and $\Omega_1^{(1)}(t)=\Omega_1^{(2)}(t)$ (Stage 2), where $\Omega_0^{(1)}(t)$ and $\Omega_1^{(1)}(t)$ are obtained by Eq.~(\ref{ConditionOri}). The effective detuning $\Delta_r(t)$ in Eq.~(\ref{Deltar}), the Rabi-frequencies $\Omega_0(t)$ (Stage 1) and $\Omega_1(t)$ (Stage 2) in Eq.~(\ref{Om0}), and the phase $\alpha(t)$ in Eq.~(\ref{dotalpha}) are under the conditions of $\theta(t)$ in Eq.~(\ref{theta}) and $f(t)=0$. During $t\in[2T, 2.5T]$, all the driving fields are turned off to verify the stability of the prepared entangled state under the master equation~(\ref{master_2qubit}).}\label{Bell}
\end{figure}

The performance of our protocol can be evaluated by the state population $P_n=|\langle n|\rho(t)|n\rangle|$, $n=00, rr, 11$, and the fidelity for the target state $|\psi_{\rm tar}\rangle=(|00\rangle+|11\rangle)/\sqrt{2}$. The fidelity is defined as $\mathcal{F}(t)={\rm Tr}[\rho_s^2(t)]$, where $\rho_s(t)=\Pi\rho(t)\Pi$ with $\Pi\equiv|00\rangle\langle00|+|11\rangle\langle11|$. Figure~\ref{Bell}(a) demonstrates the dynamics of system populations $P_n$ for the closed system. During the first stage, i.e., $t\in[0,T]$, the initial state $|00\rangle$ can gradually evolve to the superposed state which equally populates the states $|00\rangle$ and $|rr\rangle$. Then, during the second stage $t\in[T,2T]$, the population on the Rydberg state $|rr\rangle$ can be completely transferred to the ground state $|11\rangle$ and the population on $|00\rangle$ is untouched. In Fig.~\ref{Bell}(b), the dynamics of fidelity $\mathcal{F}$ about the Bell state $|\psi_{\rm tar}\rangle=(|00\rangle+|11\rangle)/\sqrt{2}$ is demonstrated under various decay rates $\kappa$. In the ideal situation, i.e., $\kappa=0$, the fidelity is about $\mathcal{F}(2T)=0.997$ on account of the leakage out of the ground-level subspace.

With no error correction, our protocol presents a robust performance when the system is exposed to the environmental noises including dissipation and dephasing. Under an experimentally practical decay rate, i.e., $\kappa/V=1\times10^{-5}$, the final fidelity is still as high as $\mathcal{F}(2T)=0.970$. Even when $\kappa/V=4\times10^{-5}$, which is about four times as high as the practical one, $\mathcal{F}(2T)=0.901$. In addition, after the entangled state is generated ($t\geq2T$), the fidelity could be stable as long as we turn off the driving fields. In the conventional protocols based on Rydberg blockade~\cite{Jaksch2000Fast,Levine2019parallel,Graham2019Rydberg,Moller2008Quantumgates} and asymmetric blockade~\cite{Brion2007Conditional,Saffman2009Efficient,Implementation2010Wu,Rao2014Deterministice,Young2021Asymmetric}, the Rydberg interaction $V$ is required to be two orders higher than the Rabi frequencies. In our protocol, the ratio of Rydberg interaction to Rabi frequency can be as low as $V/\Omega_0^{(1)}(t)\sim5$, as determined by the parametric setting in Fig.~\ref{Bell}.

In the presence of the systematical errors, the original Hamiltonian in Eq.~(\ref{Ham}) will be perturbed as
\begin{equation}\label{HamErr}
H(t)\rightarrow H(t)+H_e(t),
\end{equation}
where $H_e(t)$ is the error Hamiltonian. Without loss of generality, it can be categorized to the global error
\begin{equation}\label{CommunErr}
H_e(t)=\epsilon\left[H_1(t)+H_d(t)\right]\otimes \mathcal{I}_2+\epsilon\mathcal{I}_1\otimes H_2(t),
\end{equation}
and the local error
\begin{equation}\label{NoncomErr}
H_e(t)=\epsilon\left[H_1(t)+H_d(t)\right]\otimes\mathcal{I}_2.
\end{equation}
They describe the global fluctuation of the driving fields on both atoms and the local fluctuation of the driving fields on the first atom, respectively. By Eq.~(\ref{effDetuS1}), it is found that the global and local errors induce different consequence to the effective Hamiltonian~(\ref{HeffS1}) for Stage 1. They read,
\begin{equation}\label{HeffFluGlob}
\begin{aligned}
H_{\rm eff}'(t)&=(1+\epsilon)^2\Big[\Delta_r(t)|rr\rangle\langle rr|+\Delta_0(t)|00\rangle\langle00|\\
&+\Omega_0(t)|rr\rangle\langle00|+{\rm H.c.}\Big],
\end{aligned}
\end{equation}
and
\begin{equation}\label{HeffFluLocal}
\begin{aligned}
H_{\rm eff}'(t)&=(1+\epsilon)^2\Big[\Delta_r(t)|rr\rangle\langle rr|+\Delta_0(t)|00\rangle\langle00|\Big]\\
&+(1+\epsilon)\Big[\Omega_0(t)|rr\rangle\langle00|+{\rm H.c.}\Big],
\end{aligned}
\end{equation}
respectively. Similarly, the effective Hamiltonian~(\ref{HeffS2}) for Stage 2 become Eqs.~(\ref{HeffFluGlob}) and (\ref{HeffFluLocal}) under global and local errors, respectively, with $|00\rangle\rightarrow|11\rangle$, $\Omega_0\rightarrow\Omega_1$, and $\Delta_0\rightarrow\Delta_1$.

The instability of the driving field can induce the fluctuations of the phase $\alpha(t)$ in Eq.~(\ref{Condition}). For example, in the presence of the phase fluctuation, the driving field on the transition $|r\rangle\leftrightarrow|1\rangle$ of the first atom will deviate as $\Omega_1^{(1)}(t)\rightarrow\Omega_1^{(1)}(t)e^{i\delta\varphi}$. By Eqs.~(\ref{HeffS2}) and (\ref{effDetuS2}), one can directly find that its effect is similar to the fluctuation in the Rabi frequencies $\Omega_1^{(1)}(t)$ and the deviation amplitude is proportional to $\delta\varphi$, which can be described by Eq.~(\ref{NoncomErr}) or Eq.~(\ref{HeffFluLocal}). In experiments, the stability of laser phase can be ensured by the technologies such as the modern frequency comb~\cite{Cundiff2003Colloquium}.

\begin{figure}[htbp]
\centering
\includegraphics[width=0.9\linewidth]{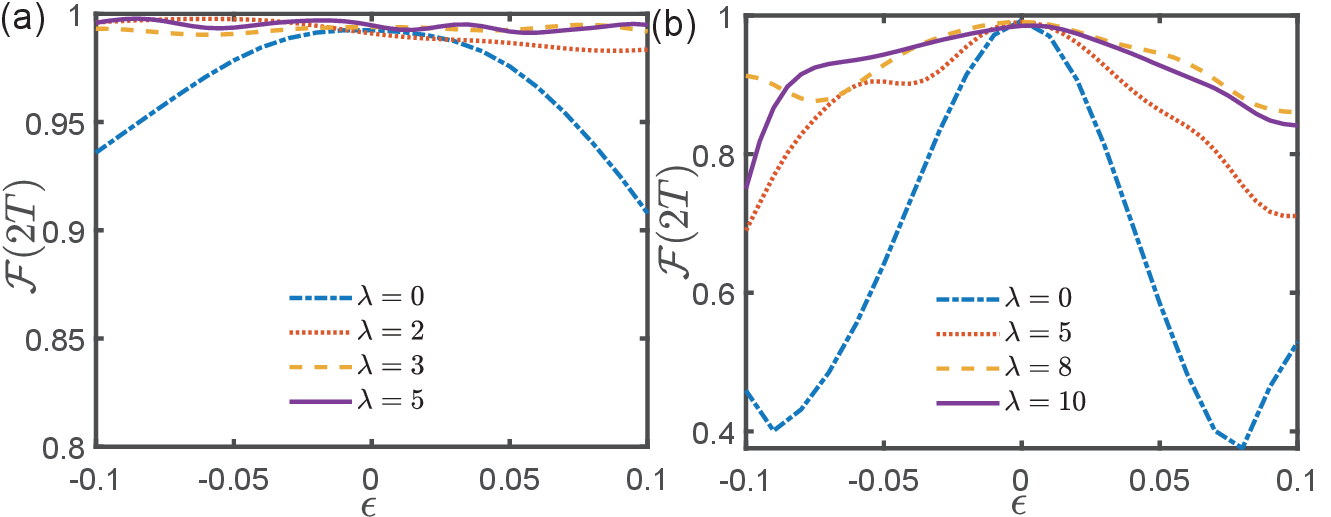}
\caption{Fidelity $\mathcal{F}(2T)$ about the target Bell state $(|00\rangle+\exp[if(2T)]|11\rangle)/\sqrt{2}$ versus the error magnitude $\epsilon$ with (a) the global error and (b) the local error for both stages under various control magnitudes $\lambda$ in Eq.~(\ref{phasef}). $\kappa=0$ and $\kappa_z=0$. The other parameters are set the same as Fig.~\ref{Bell}.}\label{ComErr}
\end{figure}

In our model, a simple solution for the correction mechanism~(\ref{OptGeneral}) is to set the global phase as
\begin{equation}\label{phasef}
\dot{f}(t)=\lambda\dot{\theta}(t),
\end{equation}
where $\dot{\theta}(t)$ can be used to estimate the evolution and transition rates of the universal passages in Eqs.~(\ref{Anci}) and (\ref{Anci1}). The adverse effects induced by perturbative errors can be suppressed as long as $\lambda$ is sufficiently large, that is shown in Fig.~\ref{ComErr} about the final fidelity for the target state $(|00\rangle+\exp[if(2T)]|11\rangle)/\sqrt{2}$ with a nonvanishing local phase. Under the global error, Fig.~\ref{ComErr}(a) demonstrates that for $\lambda=0$ (with no correction), the fidelity decreases to $\mathcal{F}(2T)=0.936$ when $\epsilon=-0.1$, $\mathcal{F}(2T)=0.979$ when $\epsilon=-0.05$, $\mathcal{F}(2T)=0.976$ when $\epsilon=0.05$, and $\mathcal{F}(2T)=0.908$ when $\epsilon=0.1$. With correction, the passage exhibits a less susceptibility to the global error. It becomes $\mathcal{F}(2T)>0.983$ when $\lambda=2$, $\mathcal{F}(2T)>0.991$ when $\lambda=3$, and $\mathcal{F}(2T)>0.993$ when $\lambda=5$, over the whole range of $\epsilon\in[-0.1,0.1]$. Figure~\ref{ComErr}(b) demonstrates the results under the local errors, which are more harmful than the global error. For $\lambda=0$ (with no correction), the fidelity will be as low as $\mathcal{F}(2T)=0.641$ when $\epsilon=-0.05$ and $\mathcal{F}(2T)=0.595$ when $\epsilon=0.05$. A larger $\lambda$ than that for the global error is required to hold the nonadiabatic passage. Over the range of $\epsilon\in [-0.05,0.05]$, the fidelity is enhanced to $\mathcal{F}>0.863$ when $\lambda=5$ and $\mathcal{F}>0.928$ when $\lambda=8$.

\begin{figure}[htbp]
\centering
\includegraphics[width=0.9\linewidth]{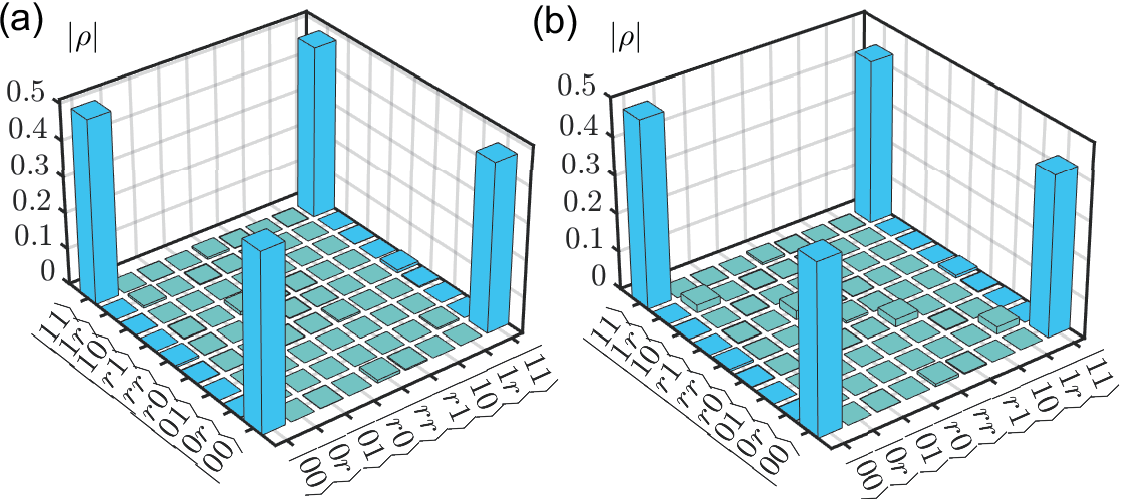}
\caption{Absolute value of density matrix of the atomic system $|\rho(t=2T)|$ under the global error in Eq.~(\ref{CommunErr}) with $\epsilon=-0.1$. The control magnitude in Eq.~(\ref{phasef}) is set as $\lambda=5$. The dephasing rate is set as $\kappa_z/\kappa=0.1$, with the decay rate $\kappa/V=1\times10^{-5}$ in (a) and $\kappa/V=4\times10^{-5}$ in (b). The other parameters are set the same as Fig.~\ref{Bell}.}\label{Bellbar}
\end{figure}

Figure~\ref{Bellbar} shows the final density matrix $|\rho(t=2T)|$ for the open atomic system in the presence of both environmental noise and global error in Hamiltonian. $|\rho|$ means taking the absolute value for each element of the $9\times9$ density matrix of the two Rydberg atoms. In Fig.~\ref{Bellbar}(a) with $\kappa/V=1\times10^{-5}$, the numerical calculation demonstrates that $|\langle00|\rho(2T)|11\rangle|=0.482$, $|\langle00|\rho(2T)|00\rangle|=0.507$, and $|\langle11|\rho(2T)|11\rangle|=0.465$. The fidelity is found to be $\mathcal{F}(2T)=0.968$ on account of the leakage to the Rydberg state $|r\rangle$. In Fig.~\ref{Bellbar}(b) with a much larger decay rate, the matrix elements are found to be $|\langle00|\rho(2T)|11\rangle|=0.439$, $|\langle00|\rho(2T)|00\rangle|=0.489$, and $|\langle11|\rho(2T)|11\rangle|=0.426$. The fidelity is about $\mathcal{F}(2T)=0.895$. Thus our protocol with error correction shows remarkable performance in comparison to the ideal-parameter case in Fig.~\ref{Bell}, even under a deviation of about $10\%$ in parameter. These results also show the insensitivity of our protocol to the environmental noise.

In addition to the global error in Eq.~(\ref{CommunErr}) and local error in Eq.~(\ref{NoncomErr}), the impact of the atomic interaction fluctuation on our protocol can be estimated and corrected. In this case, the perturbed Hamiltonian becomes
\begin{equation}\label{HamErrDeltaV}
H(t)\rightarrow H(t)+H_e,\quad H_e=\delta V|rr\rangle\langle rr|,
\end{equation}
where the interaction fluctuation $\delta V$ is induced by the atomic distance fluctuation $\delta d$.

\begin{figure}[htbp]
\centering
\includegraphics[width=0.9\linewidth]{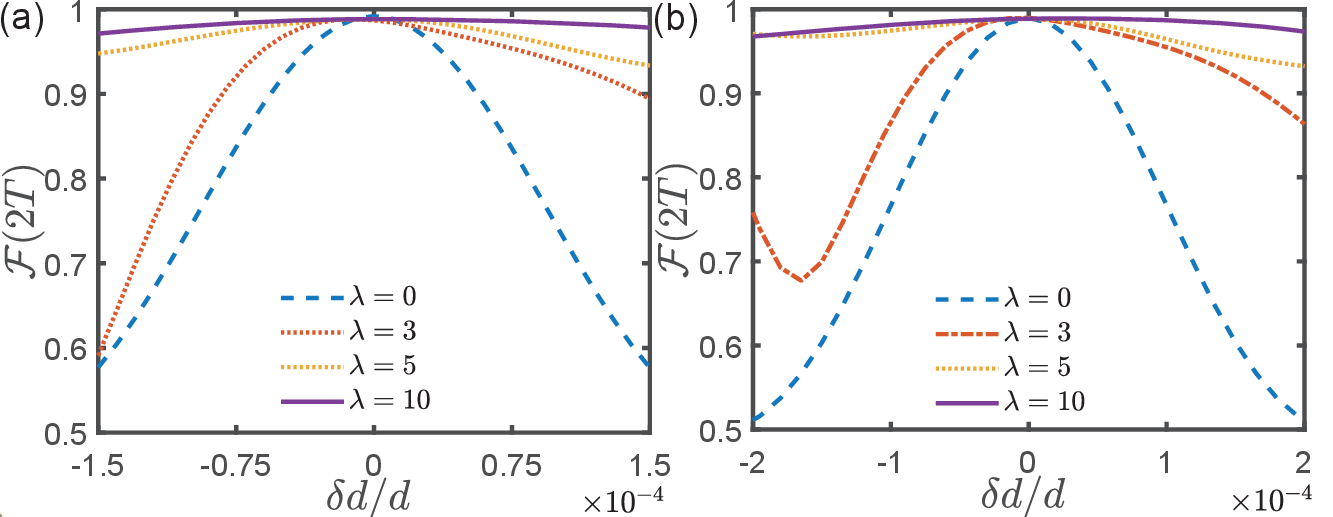}
\caption{Fidelity $\mathcal{F}(2T)$ about the target Bell state $(|00\rangle+\exp[if(2T)]|11\rangle)/\sqrt{2}$ vs the relative distance fluctuation $\delta d/d$ for both stages under various control magnitudes $\lambda$ in Eq.~(\ref{phasef}). In (a) $T=50$ ns and $V/2\pi=20$ MHz, and in (b) $T=1\mu$s and $V/2\pi=200$ MHz. The other parameters are set the same as Fig.~\ref{ComErr}.}\label{ComErrFlu}
\end{figure}

With no loss of generality, we consider a pair of Rubidium 87 atoms with a principle quantum number $n=79$ and the interaction coefficient $C_6/2\pi=5.36\times10^3$ GHz $\mu m^6$~\cite{Vaillant2012Longrange,Zheng2017Entangling}. The vdW interaction $V/2\pi=C_6/d^6$ between them is about $V/2\pi=200$ MHz when the atoms are separated with $d\approx 5.470\mu$m and about $V/2\pi=20$ MHz when $d\approx8.030\mu$m. The distance fluctuation $\delta d$ arises typically from the thermal motion~\cite{Levine2018Highfidelity,Madjarove2020Highfidelity,Nikolaus2021Raman,Evered2023Highfidelity} and the photon recoil caused by the laser-induced absorption and emission~\cite{Robicheaux2021Photon}. Particularly, for the atoms trapped in the optical tweezers~\cite{Nikolaus2021Raman}, their distance fluctuation $\delta d$ induced by the thermal motion is proportional to the square-root of temperature and inversely proportional to the trapping frequency. As an example, for the temperature $10\mu$K and the trapping frequency $400$kHz, the distance fluctuation induced by the thermal motion is about $\delta d\sim10^{-3}\mu$m~\cite{Evered2023Highfidelity}. When the trap pulses are turned off during the coherent control of Rydberg atoms, the distance fluctuation induced by photon recoil~\cite{Robicheaux2021Photon} depends on the moment kick $\hbar K=2\pi/\lambda$ with $\lambda$ the wavelength of driving fields, the duration of driving fields on the Rydberg atoms $\tau$, and the mass of atom $M$. Using $\lambda=822.9$ nm, $\tau=1\mu$s, and $M=132.91$ atomic mass units~\cite{Robicheaux2021Photon}, it is found that $\delta d\sim10^{-3}\mu$m. Thus we can fix the distance fluctuation as $\delta d=10^{-3}\mu$m, by which the atoms coupled by the Rydberg interaction $V/2\pi=20$ MHz and $V/2\pi=200$ MHz are accompanied with the fluctuation $\delta V/2\pi=0.015$ MHz and $\delta V/2\pi=0.220$ MHz, respectively.

To the first order in the deviation magnitude, it is found that the detuning fluctuation induced by Doppler shifts in $\Delta_0^{(d)}$ or $\Delta_1^{(d)}$ is proportional to the atomic interaction fluctuation. It means that the detuning fluctuation can also be described by Eq.~(\ref{HamErrDeltaV}). In experiments, the Doppler shifts can be suppressed through locking the lasers to the optical references cavities with a high finesse~\cite{Bohlouli2006Optical,Johnson2008Rabi} or to the Rydberg atoms themselves~\cite{Abel2009Laser,Weichman2024Doppler}.

The fidelity $\mathcal{F}(2T)$ for the preparation of a Bell state as a function of the relative distance fluctuation $\delta d/d$ is shown in Fig.~\ref{ComErrFlu}, where $T=50$ ns and $V/2\pi=20$ MHz in Fig.~\ref{ComErrFlu}(a) and $T=1\mu$s and $V/2\pi=200$ MHz in Fig.~\ref{ComErrFlu}(b). Our protocol and the correction condition turn out to be nearly scale free with respect to the atomic interaction. For example, in Fig.~\ref{ComErrFlu}(a) with $\lambda=0$ (parallel transport condition), the fidelity is as low as $\mathcal{F}(2T)=0.577$ for $\delta d/d=\pm 1.5\times10^{-4}$. Using our correction protocol in Eq.~(\ref{phasef}), the fidelity can be lower bounded by $\mathcal{F}\geq0.933$ when $\lambda=5$ and $\mathcal{F}\geq0.971$ when $\lambda=10$, in the whole range of $\delta d/d\in[-1.5\times10^{-4}, 1.5\times10^{-4}]$. And in Fig.~\ref{ComErrFlu}(b) the fidelity can be lower bounded by $\mathcal{F}\geq0.933$ when $\lambda=5$ and $\mathcal{F}\geq0.968$ when $\lambda=10$, respectively.

\section{Maximally entangling multiple qubits via robust universal passages}\label{EntangleN}

In this section, our protocol is applied to prepare a stable GHZ state of $N$ Rydberg atoms that are coupled through the nearest-neighboring Rydberg interaction, i.e., $H_{\rm int}=V\sum_{k=1}^{N-1}|rr\rangle_{k,k+1}\langle rr|$. The protocol is completed within $N-1$ steps, the $k$th step of which, $1\le k\le N-1$, is characterized by only driving the $k$th and $(k+1)$th Rydberg atoms with largely detuned laser fields. Initially, the whole system is assumed to be at the ground state $|0\rangle^{\otimes N}$.

\emph{Step 1.} The atom pair under driving fields has exactly the same configuration space as that in Fig.~\ref{model}. Following the same process as the two-atom case in Sec.~\ref{Two}, this step is divided into two stages of equal duration $T$. In Stage 1, the atomic system evolves from the ground state $|0\rangle^{\otimes N}$ to the superposed state $(|00\rangle+|rr\rangle)\otimes|0\rangle^{\otimes(N-2)}/\sqrt{2}$. In Stage 2, it is prepared to be the state equally superposed by the ground states $|0\rangle^{\otimes N}$ and $|11\rangle\otimes|0\rangle^{\otimes(N-2)}$. The passages for these two stages [see Eqs.~(\ref{Anci}) and (\ref{Anci1})] are slightly modified to be
\begin{equation}\label{AnciStep1P1}
\begin{aligned}
|\mu_1(t)\rangle&=\Big[\cos\theta(t)e^{i\frac{\alpha(t)}{2}}|00\rangle\\
&-\sin\theta(t)e^{-i\frac{\alpha(t)}{2}}|rr\rangle\Big]\otimes|0\rangle^{\otimes (N-2)},\\
|\mu_2(t)\rangle&=\Big[\sin\theta(t)e^{i\frac{\alpha(t)}{2}}|00\rangle\\
&+\cos\theta(t)e^{-i\frac{\alpha(t)}{2}}|rr\rangle\Big]\otimes|0\rangle^{\otimes (N-2)},
\end{aligned}
\end{equation}
and
\begin{equation}\label{AnciStep1P2}
\begin{aligned}
|\mu_1(t)\rangle&=\Big[\cos\theta(t)e^{i\frac{\alpha(t)}{2}}|rr\rangle\\
&-\sin\theta(t)e^{-i\frac{\alpha(t)}{2}}|11\rangle\Big]\otimes|0\rangle^{\otimes(N-2)},\\
|\mu_2(t)\rangle&=\Big[\sin\theta(t)e^{i\frac{\alpha(t)}{2}}|rr\rangle\\
&+\cos\theta(t)e^{-i\frac{\alpha(t)}{2}}|11\rangle\Big]\otimes|0\rangle^{\otimes(N-2)},
\end{aligned}
\end{equation}
respectively.

\begin{figure}[htbp]
\centering
\includegraphics[width=0.9\linewidth]{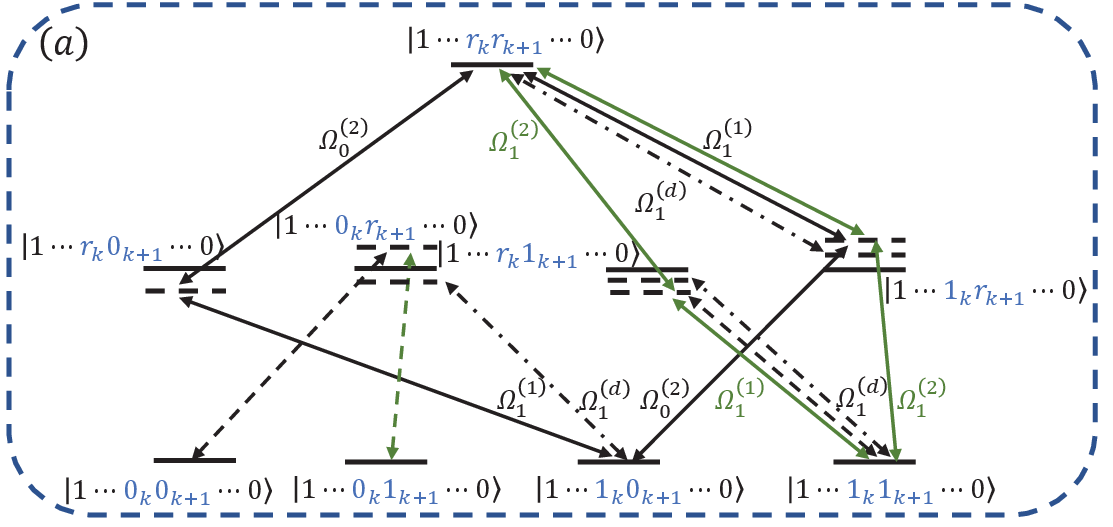}
\includegraphics[width=0.9\linewidth]{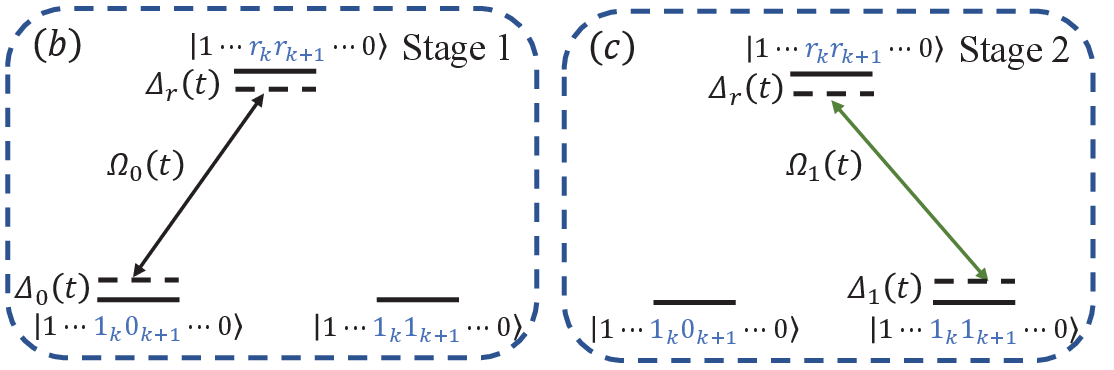}
\caption{(a) Transition diagram for the multiple Rydberg atomic system driven by largely-detuned driving fields in the $k$th step. The black and green solid lines describe the effective transitions $|1\ldots r_kr_{k+1}\ldots0\rangle\leftrightarrow|1\ldots1_k0_{k+1}\ldots0\rangle$ (Stage 1) with coupling strength $\Omega_0(t)$ and $|1\ldots r_kr_{k+1}\ldots0\rangle\leftrightarrow|1\ldots1_k1_{k+1}\ldots0\rangle$ (Stage 2) with coupling strength $\Omega_1(t)$, respectively. The round trips plotted with dotted-dashed lines contribute to the effective detunings $\Delta_0(t)$, $\Delta_r(t)$, and $\Delta_1(t)$ for the states $|1\ldots1_k0_{k+1}\ldots0\rangle$, $|1\ldots r_kr_{k+1}\ldots0\rangle$, and $|1\ldots1_k1_{k+1}\ldots0\rangle$. The transitions indicated by the black and green dashed lines can be strongly suppressed under large detunings, i.e., $\Delta_n^{(j)}\gg\Omega_n^{(j)}$, $n=0,1$. (b) and (c) Effective transition diagrams in the 3-dimensional subspace for Stages 1 and 2, respectively.}\label{transitionMulti}
\end{figure}

\emph{Step $k$}, $2\le k\le N-1$. Regarding the first and the second atoms in Fig.~\ref{model} as the $k$th and the $k+1$th atoms in this step, respectively, their configuration under driving is slightly different from that in Fig.~\ref{model} by switching off the driving fields for the transition $|r\rangle\leftrightarrow|0\rangle$ in the $k$th atom, i.e., $\Omega_0^{(1)}(t)=0$ and $\Omega_0^{(d)}(t)=0$. Consequently, the rotated Hamiltonian $H_I(t)$~(\ref{Hrot}) is modified to be
\begin{equation}\label{Hrotmod}
\begin{aligned}
&H_I(t)=\Omega_1^{(1)}(t)e^{i\Delta_1^{(1)}t}(|r0\rangle_{k,k+1}\langle 10|+|r1\rangle_{k,k+1}\langle 11|\\
&+e^{iVt}|rr\rangle_{k,k+1}\langle 1r|)+\sum_{n=0}^1\Omega_n^{(2)}(t)e^{-i\Delta_n^{(2)}t}(|0r\rangle_{k,k+1}\langle0n|\\ &+|1r\rangle_{k,k+1}\langle1n|+e^{iVt}|rr\rangle_{k,k+1}\langle rn|)+\Omega_1^{(d)}(t)e^{i\Delta_1^{(d)}t}\times\\
&(|r0\rangle_{k,k+1}\langle 10|+|r1\rangle_{k,k+1}\langle 11|+e^{iVt}|rr\rangle_{k,k+1}\langle 1r|)+{\rm H.c.}
\end{aligned}
\end{equation}
Therefore, the transition diagram in Fig.~\ref{transition}(a) becomes Fig.~\ref{transitionMulti}(a) within the subspace spanned by the states where the atoms $k'<k$ are in the state $|1\rangle$, and the atoms $k'>k+1$ are in the state $|0\rangle$.

Note that one half of the population occupies the ground state $|0\rangle^{\otimes N}$ throughout all the steps $k>1$, as it is decoupled from the system dynamics. Each step also consists of two stages with an equal duration $T$. In Stage 1, the population on $|1\ldots1_k0_{k+1}\ldots0\rangle$ is transferred to $|1\ldots r_kr_{k+1}\ldots0\rangle$, where the subscript $k$ indexes the $k$th atom. Then in Stage 2, the population on $|1\ldots r_kr_{k+1}\ldots0\rangle$ is transferred to $|1\ldots 1_k1_{k+1}\ldots0\rangle$. Thus a general $N$-qubits GHZ state, i.e., the state equally superposed by the ground states $|0\rangle^{\otimes N}$ and $|1\rangle^{\otimes N}$, can be prepared at the end of the $(N-1)$th step.

In particular, in Stage 1, i.e., $t\in[2(k-1)T,(2k-1)T]$, the driving field on the transition $|r\rangle\leftrightarrow|1\rangle$ of the $(k+1)$th atom is switched off, and the other driving fields satisfy the conditions of $\Delta_1^{(1)}+V=\Delta_0^{(2)}$ and $\Omega_1^{(1)}(t)=\Omega_0^{(2)}(t)$. Using Eq.~(\ref{Heff}) and the modified $H_I(t)$ in Eq.~(\ref{Hrotmod}), the effective Hamiltonian can be obtained as
\begin{equation}\label{HeffNk}
\begin{aligned}
H_{\rm eff}(t)&=\Delta_r(t)|rr\rangle_{k,k+1}\langle rr|+\Delta_0(t)|10\rangle_{k,k+1}\langle 10|\\
&+\Omega_0(t)|rr\rangle_{k,k+1}\langle 10|+{\rm H.c.},
\end{aligned}
\end{equation}
where $\Delta_r(t)$, $\Delta_0(t)$, and $\Omega_0(t)$ follow similar formations as in Eq.~(\ref{effDetuS1}) under the substitutions $\Omega_0^{(d)}(t)\rightarrow\Omega_1^{(d)}(t)$, $\Delta_0^{(d)}(t)\rightarrow\Delta_1^{(d)}(t)$, $\Omega_0^{(1)}(t)\rightarrow\Omega_1^{(1)}(t)$, and $\Delta_0^{(1)}(t)\rightarrow\Delta_1^{(1)}(t)$. Then the transition diagram for the subspace in Fig.~\ref{transitionMulti}(a) can be reduced to that in Fig.~\ref{transitionMulti}(b). Following Eqs.~(\ref{Anci}), (\ref{Anci1}), (\ref{AnciStep1P1}), and (\ref{AnciStep1P2}), the ancillary basis states for Fig.~\ref{transitionMulti}(b) can be chosen as
\begin{equation}\label{AnciStepkP1}
\begin{aligned}
&|\mu_1(t)\rangle=|1\rangle^{\otimes(k-1)}\otimes\Big[\cos\theta(t)e^{i\frac{\alpha(t)}{2}}|10\rangle\\
&-\sin\theta(t)e^{-i\frac{\alpha(t)}{2}}|rr\rangle\Big]\otimes|0\rangle^{\otimes(N-k-1)},\\
&|\mu_2(t)\rangle=|1\rangle^{\otimes(k-1)}\otimes\Big[\sin\theta(t)e^{i\frac{\alpha(t)}{2}}|10\rangle\\
&+\cos\theta(t)e^{-i\frac{\alpha(t)}{2}}|rr\rangle\Big]\otimes|0\rangle^{\otimes(N-k-1)},
\end{aligned}
\end{equation}
When $\Delta_r(t)$ and $\Omega_0(t)$ as in Eq.~(\ref{HeffNk}) satisfy the conditions in Eqs.~(\ref{Condition}) or (\ref{ConditionInv}) with $\theta[2(k-1)T+0^+]=0$ and $\theta[(2k-1)T]=\pi/2$, the population on the state $|1\ldots1_k0_{k+1}\ldots0\rangle$ can be completely transferred to the state $|1\ldots r_kr_{k+1}\ldots0\rangle$ along the passage $|\mu_1(t)\rangle$ when $t=(2k-1)T$.

On Stage 2, i.e., $t\in[(2k-1)T,2kT]$, we switch off the driving field on the transition $|r\rangle\leftrightarrow|0\rangle$ of the $(k+1)$th atom and tune the other driving fields to satisfy the conditions of $\Delta_1^{(1)}+V=\Delta_1^{(2)}$ and $\Omega_1^{(1)}(t)=\Omega_1^{(2)}(t)$. Then we achieve the same effective Hamiltonian as in Eq.~(\ref{HeffS2}) only with a different pair of atoms:
\begin{equation}\label{HeffNkP2}
\begin{aligned}
H_{\rm eff}(t)&=\Delta_r(t)|rr\rangle_{k,k+1}\langle rr|+\Delta_1(t)|11\rangle_{k,k+1}\langle11|\\
&+\Omega_1(t)|rr\rangle_{k,k+1}\langle 11|+{\rm H.c.},
\end{aligned}
\end{equation}
where $\Delta_r(t)$, $\Delta_1(t)$, and $\Omega_1(t)$ satisfy the same conditions in Eq.~(\ref{effDetuS2}). Then the transition diagram in Fig.~\ref{transitionMulti}(a) reduces to that in Fig.~\ref{transitionMulti}(c). Similar to Eqs.~(\ref{Anci}), (\ref{Anci1}), (\ref{AnciStep1P1}), (\ref{AnciStep1P2}), and (\ref{AnciStepkP1}), the ancillary basis states for Fig.~\ref{transitionMulti}(c) can be chosen as
\begin{equation}\label{AnciStepkP2}
\begin{aligned}
&|\mu_1(t)\rangle=|1\rangle^{\otimes(k-1)}\otimes\Big[\cos\theta(t)e^{i\frac{\alpha(t)}{2}}|rr\rangle\\
&-\sin\theta(t)e^{-i\frac{\alpha(t)}{2}}|11\rangle\Big]\otimes|0\rangle^{\otimes(N-k-1)},\\
&|\mu_2(t)\rangle=|1\rangle^{\otimes(k-1)}\otimes\Big[\sin\theta(t)e^{i\frac{\alpha(t)}{2}}|rr\rangle\\
&+\cos\theta(t)e^{-i\frac{\alpha(t)}{2}}|11\rangle\Big]\otimes|0\rangle^{\otimes(N-k-1)}.
\end{aligned}
\end{equation}
Under the conditions in Eqs.~(\ref{Condition}) or (\ref{ConditionInv}) with the substitution of $\Omega_1(t)$ for $\Omega_0(t)$, the population on $|1\ldots r_kr_{k+1}\ldots0\rangle$ can be transferred to $|1\ldots 1_k1_{k+1}\ldots0\rangle$ along the passage $|\mu_2(t)\rangle$ when $t=2kT$, for the boundary conditions $\theta[(2k-1)T]=\pi/2$ and $\theta(2kT)=\pi$. Thus one can set $\theta(t)=\pi[t-2(k-1)T]/(2T)$ for $k>1$.

\begin{table}[htbp]
\centering
\caption{Fidelity $\mathcal{F}$ about the target $N$-qubit GHZ state encoded with the ground states for the open systems with no correction, in the presence of the systematic error in Eq.~(\ref{HeffFlu}) with $\epsilon=-0.1$ and $\delta V=\gamma$ ($\gamma\equiv2\pi\times1$ MHz is an energy unit). $\kappa_z/\kappa=0.1$. The Rabi frequency is in the order of $\gamma$.}\label{table}
\begin{tabular}{ccccccc}
\hline \hline
$\kappa/\gamma$ & $N=2$ & $N=3$ & $N=4$ & $N=5$ & $N=6$ & $N=7$ \\
\hline
$1\times10^{-3}$ &0.970& $0.929$ & $0.890$ & $0.854$ & $0.820$ & $0.786$ \\
$5\times10^{-3}$ &0.969& $0.927$ & $0.887$ & $0.850$ & $0.816$ & $0.782$ \\
$1\times10^{-2}$ &0.968& $0.924$ & $0.884$ & $0.846$ & $0.811$ & $0.778$ \\
\hline \hline
\end{tabular}
\end{table}

In Table~\ref{table}, we present the fidelity $\mathcal{F}$ of the $N$-qubit GHZ state for the open systems in the presence of systematic errors. With no correction, the target state reads $[|0\rangle^{\otimes N}+(-1)^{N-1}|1\rangle^{\otimes N}]/\sqrt{2}$. The dynamics of the multiple atomic system is described by the master equation,
\begin{equation}\label{mastermulti}
\begin{aligned}
\frac{\partial \rho}{\partial t}&=-i[H_{\rm eff}(t)+H_e(t), \rho]+\sum_{k=1}^N\Big\{\frac{\kappa}{2}\Big[\mathcal{L}(|0\rangle_k\langle r|)\\
&+\mathcal{L}(|1\rangle_k\langle r|)\Big]+\frac{\kappa_z}{2}\mathcal{L}(\sigma_k^z)\Big\},
\end{aligned}
\end{equation}
where $H_{\rm eff}(t)$ can be set as Eqs.~(\ref{HamrotS1}), (\ref{HamrotS2}), (\ref{HeffNk}), and (\ref{HeffNkP2}) during the relevant stages of evolution and
\begin{equation}\label{HeffFlu}
H_e(t)=\epsilon H_{\rm eff}(t)+\sum_{k=1}^{N-1}\delta V|rr\rangle_{k,k+1}\langle rr|
\end{equation}
describing the global systematic errors and the interaction fluctuation in the same time. The fidelity is obtained in the retained encoded subspace spanned by the ground states $|0\rangle$ and $|1\rangle$. Again the results justify that our protocol is not sensitive to the environmental noise. Under the practical decay-rate $\kappa/\gamma=1\times10^{-3}$, the fidelity is found to be about $\mathcal{F}=0.929$ when $N=3$, $\mathcal{F}=0.890$ when $N=4$, $\mathcal{F}=0.854$ when $N=5$, $\mathcal{F}=0.820$ when $N=6$, and $\mathcal{F}=0.786$ when $N=7$. When the decay rate is as large as $\kappa/\gamma=1\times10^{-2}$ (10 times as the practical one), the fidelity is about $\mathcal{F}=0.924$ for $N=3$, $\mathcal{F}=0.884$ for $N=4$, $\mathcal{F}=0.846$ for $N=5$, $\mathcal{F}=0.811$ for $N=6$, and $\mathcal{F}=0.778$ when $N=7$.

We also check the performance of our error-correction mechanism~(\ref{OptGeneral}) in the preparation of the $N$-qubit GHZ state, e.g., $[|0\rangle^{\otimes 6}+\exp[i\sum_{k=2}^{6}f(2kT)]|1\rangle^{\otimes 6}]/\sqrt{2}$. It is found that the target-state fidelity can be enhanced to $\mathcal{F}\ge0.990$ as long as the control magnitude $\lambda$ in Eq.~(\ref{phasef}) is larger than $5$.

\section{Conclusion and discussion}\label{conclusion}

In summary, we designed a stable and scalable protocol under the universal quantum control framework to prepare a high-fidelity $N$-qubit GHZ state encoded on the hyperfine ground states of the Rydberg atoms. The protocol is performed along the nonadiabatic passages using off-resonant driving fields, which avoids the conventional requirement about the strong Rydberg interaction in the generation of Bell state~\cite{Levine2018Highfidelity,Madjarove2020Highfidelity}. The passages are universal since the ancillary basis states can be mapped to any desired target states. Our protocol shows that robustness against the environmental noise for the target state does not involve the sensitive Rydberg state, which can be further improved by increasing the atomic interaction and Rabi frequency of driving fields. In addition, a passage-dependent and rapidly varying global phase can be used to counteract the systematic errors. Our protocol therefore provides a powerful tool of state generation in the neutral atomic systems for quantum information processing and computation.

Our protocol depends on the time modulation over the Rabi frequency, the driving frequency, and the phase of driving fields. In practice, they can be precisely manipulated by the acousto-optic modulators (AOM)~\cite{Evered2023Highfidelity,Oliveira2025Demonstration} or electro-optic modulators (EOM)~\cite{Zeiher2016Manybody}, which are driven by the arbitrary waveform generator (AWG)~\cite{Evered2023Highfidelity,Oliveira2025Demonstration}. The response time of the combination of an AOM and an AWG is in the order of $10$ ns, while that of an EOM combined with an AWG can be as low as $1$ ns. These setups allow an accurate timing control of pulse sequences used in our control over Rydberg atoms.

Our protocol is superior to the existing schemes in the number of steps for preparing the maximally entangled states on the ground levels of neutral atoms~\cite{Levine2019parallel,Evered2023Highfidelity}. For instance, to prepare a double-excitation Bell state, our protocol employs merely one step consisting of two stages (see different colors for driving in Fig.~\ref{model}), each of which uses two driving fields on the first atom and another one on the second atom. The previous schemes~\cite{Levine2019parallel,Evered2023Highfidelity} require four steps, with each step consisting of two or three stages. Each stage performs a control-$Z$ gate, double individual $X$ gates, or double individual $Z$ gates. From the scaling perspective, only $N-1$ steps or $2(N-1)$ stages are required to prepare a general $N$-qubit GHZ state by our protocol. In contrast, over ten multi stage steps have to be used to prepare a three-qubit GHZ state by the existing schemes~\cite{Levine2019parallel,Evered2023Highfidelity}. In between these steps, the Rydberg atoms have to be cooled down to suppress the atomic distance fluctuations induced by the coherent control~\cite{Robicheaux2021Photon,Levine2019parallel,Evered2023Highfidelity}. More steps in state preparation can give rise to a larger probability of exponentially growing systematic errors.

\section*{Acknowledgments}

We acknowledge grant support from the ``Pioneer'' and ``Leading Goose'' R\&D Program of Zhejiang (Grant No. 2025C01028).

\appendix

\section{Effective Hamiltonian theory}\label{Efftheory}

This appendix provides the derivation details about the effective Hamiltonian theory in Eq.~(\ref{Heff}). We start with the evolution operator $U_I(T)$ in Eq.~(\ref{Umagnus}) determined by the Hamiltonian $H_I(t)$. For the sake of deduction, we set $H_I(t)\rightarrow\epsilon H_I(t)$ with $\epsilon$ a constant number. Expanding the evolution operator $U_I(t)$ in Eq.~(\ref{Umagnus}) with respect to $\epsilon$, we have~\cite{Blanes2009Magnus}
\begin{equation}\label{Udet}
U_I(t)=1+\left[\frac{\partial U_I(t)}{\partial\epsilon}\right]_{\epsilon=0}\epsilon
+\left[\frac{\partial^2U_I(t)}{\partial\epsilon^2}\right]_{\epsilon=0}\frac{\epsilon^2}{2}+\cdots
\end{equation}
with
\begin{equation}
\frac{\partial U_I(t)}{\partial\epsilon}=U_I(t)\sum_{k=0}^{\infty}\frac{1}{(k+1)!}
ad_{\Lambda(t)}^k\left[\frac{\partial\Lambda(t)}{\partial\epsilon}\right],
\end{equation}
where $ad_B^0(A)\equiv A$ and $ad_B^{k+1}(A)=[B, ad_B^{k}(A)]$. When $\epsilon=1$, Eq.~(\ref{Udet}) becomes
\begin{equation}\label{Usecond}
\begin{aligned}
&U_I(t)=1-i\int_0^tdt_1H_I(t_1)-\frac{1}{2}\Big\{\left[\int_0^tdt_1H_I(t_1)\right]^2\\
&+\int_0^tdt_1\left[H_I(t_1),\int_0^{t_1}dt_2H_I(t_2)\right]\Big\}+\cdots.
\end{aligned}
\end{equation}

With no loss of generality, the Hamiltonian $H_I(t)$ is assumed to be in the form of Eq.~(\ref{HI}). With sufficiently large detunings, i.e., $\Delta_j\gg\Omega_j(t)$, the first-order component in Eq.~(\ref{Usecond}) can be estimated as
\begin{equation}\label{estifirst}
\begin{aligned}
\left|\int_0^tdt_1H_I(t_1)\right|&\approx\sum_j\left|\frac{\Omega_j(t)e^{i\Delta_j t}-\Omega_j(0)}{i\Delta_j}A_j+{\rm H.c.} \right|,\\ &\le\sum_j\left|\frac{\Omega_j(t)-\Omega_j(0)}{\Delta_j}A_j+{\rm H.c.} \right|,
\end{aligned}
\end{equation}
and the two second-order components are
\begin{equation}\label{estisecond}
\begin{aligned}
&\left[\int_0^tdt_1H_I(t_1)\right]^2\approx-\sum_{j,k}\frac{1}{\Delta_j\Delta_k}
\Big\{\left[\Omega_j(t)e^{i\Delta_jt}-\Omega_j(0)\right]\\
&\times\left[\Omega_k(t)e^{i\Delta_kt}-\Omega_k(0)\right]A_jA_k+\left[\Omega_j(t)e^{-i\Delta_jt}-\Omega_j(0)\right]\\
&\times\left[\Omega_k(t)e^{i\Delta_k(t)}-\Omega_k(0)\right]A_j^\dagger A_k+{\rm H.c.}\Big\},\\
\end{aligned}
\end{equation}
and
\begin{equation}\label{estimated}
\begin{aligned}
&\int_0^tdt_1\left[H_I(t_1),\int_0^{t_1}dt_2H_I(t_2)\right]\\
&\approx\sum_{j,k}\int_0^t\left[\frac{\Omega_j(t_1)\Omega_k(t_1)}{\Delta_j}A_jA_k^\dagger+{\rm H.c.}\right]dt_1\\
&\le\sum_{j,k}\int_0^t\left[\frac{\Omega^2_j(t_1)}{\Delta_j}A_jA_k^\dagger+{\rm H.c.}\right]dt_1.
\end{aligned}
\end{equation}
The approximation and estimation in Eq.~(\ref{estimated}) hold under the resonant condition $\Delta_j=\Delta_k$ and the condition $\Omega_j(t_1)\simeq\Omega_k(t_1)$. In comparison to Eq.~(\ref{estimated}), all the terms in Eq.~(\ref{estisecond}) can be neglected because they scale as $\sim1/\Delta_j^2$ with large detunings. The terms in Eq.~(\ref{estifirst}) can be conditionally omitted under a strong control Hamiltonian or a large duration of control, i.e., $|\int_0^t\Omega^2_j(t_1)dt_1|\gg|\Omega_j(t)-\Omega_j(0)|$, which shares the same idea as the adiabatic theorem~\cite{Vitanove2017Stimulated}. Alternatively they can be omitted by a periodical control such that $\Omega_j(t)=\Omega_j(0)$. The ignorance of both Eqs.~(\ref{estifirst}) and (\ref{estisecond}) is consistent with the idea of nonperturbative dynamical decoupling~\cite{Jing2013Nonperturbative,Jing2015Nonperturbative} that both a large time-integral over detuning and the strong control Hamiltonian can be used to suppress the unwanted transitions to the second order. Consequently, Eq.~(\ref{Usecond}) can be reduced to
\begin{equation}\label{Ureduce}
U_{\rm eff}(t)\approx1-\frac{1}{2}\left\{\int_0^tdt_1\left[H_I(t_1), \int_0^{t_1}dt_2H_I(t_2)\right]\right\}+\cdots.
\end{equation}
The system dynamics described by Eq.~(\ref{Ureduce}) is equivalent to that governed by the effective Hamiltonian
\begin{equation}\label{HeffApp}
H_{\rm eff}(t)=-\frac{i}{2}\left[H_I(t), \int_0^tH_I(t_1)dt_1\right],
\end{equation}
which is exactly Eq.~(\ref{Heff}) in the main text.

\section{Universal passages with error correction}\label{UnivErr}

This appendix contributes to the detailed derivation of the error correction condition in Eq.~(\ref{OptGeneral}). We begin with the system Hamiltonian in the first rotated picture, i.e., $H_{\rm rot}(t)$ in Eq.~(\ref{HamNonRot}). To find an error-suppression method, the system dynamics can be considered in the second-rotated picture with respect to $U_{\rm rot}(t)$ in Eq.~(\ref{Urot1}). In particular, we have
\begin{equation}\label{HamNonSecRot}
\begin{aligned}
\tilde{H}_e(t)&=U_{\rm rot}^\dagger(t)H_{\rm rot}(t)U_{\rm rot}(t)-iU_{\rm rot}^\dagger(t)\frac{d}{dt}U_{\rm rot}(t)\\
&=\epsilon\sum_{k=1}^{K'}\sum_{n=1}^{K'}\tilde{\mathcal{D}}_{kn}^{(e)}(t)|\mu_k(0)\rangle\langle\mu_n(0)|
\end{aligned}
\end{equation}
where
\begin{equation}\label{ErrElement}
\tilde{\mathcal{D}}_{kn}^{(e)}(t)\equiv\langle\mu_k(t)|H_e(t)|\mu_n(t)\rangle e^{-i[f_k(t)-f_n(t)]}.
\end{equation}
Using the Magnus expansion in Eqs.~(\ref{Umagnus}), (\ref{MagnusTerm}), and (\ref{Udet}), the time-evolution operator for $\tilde{H}_e(t)$~(\ref{HamNonSecRot}) is found to be similar to Eq.~(\ref{Usecond}):
\begin{equation}\label{Uerr}
\begin{aligned}
&U_e(t)=1-i\int_0^tdt_1\tilde{H}_e(t_1)-\frac{1}{2}\Big\{\left[\int_0^tdt_1\tilde{H}_e(t_1)\right]^2\\
&+\int_0^tdt_1\left[\tilde{H}_e(t_1), \int_0^{t_1}dt_2\tilde{H}_e(t_2)\right]\Big\}+\cdots.
\end{aligned}
\end{equation}
Subsequently, one can find that the leading order of the infidelity between the instantaneous state governed by $H(t)$ and the target state is up to the second order of the error magnitude: $1-|\langle\mu_k(0)|U_e(t)|\mu_k(0)\rangle|^2=\epsilon^2\sum_{n=1,n\ne k}^{K'}|\int_0^t\tilde{\mathcal{D}}_{kn}^{(e)}(t_1)dt_1|^2+\mathcal{O}(\epsilon^3)$. According to the correction mechanism~\cite{Jin2025ErrCorr}, the adverse effects that arise from the systematic errors can be corrected, when the passage-dependent global phase varies rapidly with time in comparison to the transition rate of the passages. In particular, we have
\begin{equation}\label{OptGeneralApp}
|\dot{f}_k(t)-\dot{f_n}(t)|\gg \frac{d}{dt}\left[\langle\mu_k(t)|H_e(t)|\mu_n(t)\rangle\right],
\end{equation}
which is exactly Eq.~(\ref{OptGeneral}) in the main text.

\bibliographystyle{apsrevlong}
\bibliography{ref}

\begin{thebibliography}{85}%
\makeatletter
\providecommand \@ifxundefined [1]{%
 \@ifx{#1\undefined}
}%
\providecommand \@ifnum [1]{%
 \ifnum #1\expandafter \@firstoftwo
 \else \expandafter \@secondoftwo
 \fi
}%
\providecommand \@ifx [1]{%
 \ifx #1\expandafter \@firstoftwo
 \else \expandafter \@secondoftwo
 \fi
}%
\providecommand \natexlab [1]{#1}%
\providecommand \enquote  [1]{``#1''}%
\providecommand \bibnamefont  [1]{#1}%
\providecommand \bibfnamefont [1]{#1}%
\providecommand \citenamefont [1]{#1}%
\providecommand \href@noop [0]{\@secondoftwo}%
\providecommand \href [0]{\begingroup \@sanitize@url \@href}%
\providecommand \@href[1]{\@@startlink{#1}\@@href}%
\providecommand \@@href[1]{\endgroup#1\@@endlink}%
\providecommand \@sanitize@url [0]{\catcode `\\12\catcode `\$12\catcode
  `\&12\catcode `\#12\catcode `\^12\catcode `\_12\catcode `\%12\relax}%
\providecommand \@@startlink[1]{}%
\providecommand \@@endlink[0]{}%
\providecommand \url  [0]{\begingroup\@sanitize@url \@url }%
\providecommand \@url [1]{\endgroup\@href {#1}{\urlprefix }}%
\providecommand \urlprefix  [0]{URL }%
\providecommand \Eprint [0]{\href }%
\providecommand \doibase [0]{http://dx.doi.org/}%
\providecommand \selectlanguage [0]{\@gobble}%
\providecommand \bibinfo  [0]{\@secondoftwo}%
\providecommand \bibfield  [0]{\@secondoftwo}%
\providecommand \translation [1]{[#1]}%
\providecommand \BibitemOpen [0]{}%
\providecommand \bibitemStop [0]{}%
\providecommand \bibitemNoStop [0]{.\EOS\space}%
\providecommand \EOS [0]{\spacefactor3000\relax}%
\providecommand \BibitemShut  [1]{\csname bibitem#1\endcsname}%
\let\auto@bib@innerbib\@empty
\bibitem [{\citenamefont {Kr\'al}\ \emph {et~al.}(2007)\citenamefont {Kr\'al},
  \citenamefont {Thanopulos},\ and\ \citenamefont
  {Shapiro}}]{Kral2007Colloquium}%
  \BibitemOpen
  \bibfield  {author} {\bibinfo {author} {\bibfnamefont {P.}~\bibnamefont
  {Kr\'al}}, \bibinfo {author} {\bibfnamefont {I.}~\bibnamefont {Thanopulos}},
  \ and\ \bibinfo {author} {\bibfnamefont {M.}~\bibnamefont {Shapiro}},\
  }\bibfield  {title} {\emph {\bibinfo {title} {Colloquium: Coherently
  controlled adiabatic passage},\ }}\href {\doibase 10.1103/RevModPhys.79.53}
  {\bibfield  {journal} {\bibinfo  {journal} {Rev. Mod. Phys.}\ }\textbf
  {\bibinfo {volume} {79}},\ \bibinfo {pages} {53} (\bibinfo {year}
  {2007})}\BibitemShut {NoStop}%
\bibitem [{\citenamefont {Li}(2020)}]{Li2020Aboost}%
  \BibitemOpen
  \bibfield  {author} {\bibinfo {author} {\bibfnamefont {W.}~\bibnamefont
  {Li}},\ }\bibfield  {title} {\emph {\bibinfo {title} {A boost to {R}ydberg
  quantum computing},\ }}\href {\doibase 10.1038/s41567-020-0907-8} {\bibfield
  {journal} {\bibinfo  {journal} {Nat. Phys.}\ }\textbf {\bibinfo {volume}
  {16}},\ \bibinfo {pages} {820} (\bibinfo {year} {2020})}\BibitemShut
  {NoStop}%
\bibitem [{\citenamefont {Einstein}\ \emph {et~al.}(1935)\citenamefont
  {Einstein}, \citenamefont {Podolsky},\ and\ \citenamefont
  {Rosen}}]{Einstein1935Can}%
  \BibitemOpen
  \bibfield  {author} {\bibinfo {author} {\bibfnamefont {A.}~\bibnamefont
  {Einstein}}, \bibinfo {author} {\bibfnamefont {B.}~\bibnamefont {Podolsky}},
  \ and\ \bibinfo {author} {\bibfnamefont {N.}~\bibnamefont {Rosen}},\
  }\bibfield  {title} {\emph {\bibinfo {title} {Can quantum-mechanical
  description of physical reality be considered complete?}\ }}\href {\doibase
  10.1103/PhysRev.47.777} {\bibfield  {journal} {\bibinfo  {journal} {Phys.
  Rev.}\ }\textbf {\bibinfo {volume} {47}},\ \bibinfo {pages} {777} (\bibinfo
  {year} {1935})}\BibitemShut {NoStop}%
\bibitem [{\citenamefont {Bohr}(1935)}]{Bohr1935Can}%
  \BibitemOpen
  \bibfield  {author} {\bibinfo {author} {\bibfnamefont {N.}~\bibnamefont
  {Bohr}},\ }\bibfield  {title} {\emph {\bibinfo {title} {Can
  quantum-mechanical description of physical reality be considered complete?}\
  }}\href {\doibase 10.1103/PhysRev.48.696} {\bibfield  {journal} {\bibinfo
  {journal} {Phys. Rev.}\ }\textbf {\bibinfo {volume} {48}},\ \bibinfo {pages}
  {696} (\bibinfo {year} {1935})}\BibitemShut {NoStop}%
\bibitem [{\citenamefont {Horodecki}\ \emph {et~al.}(2009)\citenamefont
  {Horodecki}, \citenamefont {Horodecki}, \citenamefont {Horodecki},\ and\
  \citenamefont {Horodecki}}]{Horodecki2009Quantum}%
  \BibitemOpen
  \bibfield  {author} {\bibinfo {author} {\bibfnamefont {R.}~\bibnamefont
  {Horodecki}}, \bibinfo {author} {\bibfnamefont {P.}~\bibnamefont
  {Horodecki}}, \bibinfo {author} {\bibfnamefont {M.}~\bibnamefont
  {Horodecki}}, \ and\ \bibinfo {author} {\bibfnamefont {K.}~\bibnamefont
  {Horodecki}},\ }\bibfield  {title} {\emph {\bibinfo {title} {Quantum
  entanglement},\ }}\href {\doibase 10.1103/RevModPhys.81.865} {\bibfield
  {journal} {\bibinfo  {journal} {Rev. Mod. Phys.}\ }\textbf {\bibinfo {volume}
  {81}},\ \bibinfo {pages} {865} (\bibinfo {year} {2009})}\BibitemShut
  {NoStop}%
\bibitem [{\citenamefont {Ekert}\ and\ \citenamefont
  {Jozsa}(1996)}]{Ekert1996Quantum}%
  \BibitemOpen
  \bibfield  {author} {\bibinfo {author} {\bibfnamefont {A.}~\bibnamefont
  {Ekert}}\ and\ \bibinfo {author} {\bibfnamefont {R.}~\bibnamefont {Jozsa}},\
  }\bibfield  {title} {\emph {\bibinfo {title} {Quantum computation and
  {S}hor's factoring algorithm},\ }}\href {\doibase 10.1103/RevModPhys.68.733}
  {\bibfield  {journal} {\bibinfo  {journal} {Rev. Mod. Phys.}\ }\textbf
  {\bibinfo {volume} {68}},\ \bibinfo {pages} {733} (\bibinfo {year}
  {1996})}\BibitemShut {NoStop}%
\bibitem [{\citenamefont {Farhi}\ \emph {et~al.}(2001)\citenamefont {Farhi},
  \citenamefont {Goldstone}, \citenamefont {Gutmann}, \citenamefont {Lapan},
  \citenamefont {Lundgren},\ and\ \citenamefont {Preda}}]{Edward2001Aquantum}%
  \BibitemOpen
  \bibfield  {author} {\bibinfo {author} {\bibfnamefont {E.}~\bibnamefont
  {Farhi}}, \bibinfo {author} {\bibfnamefont {J.}~\bibnamefont {Goldstone}},
  \bibinfo {author} {\bibfnamefont {S.}~\bibnamefont {Gutmann}}, \bibinfo
  {author} {\bibfnamefont {J.}~\bibnamefont {Lapan}}, \bibinfo {author}
  {\bibfnamefont {A.}~\bibnamefont {Lundgren}}, \ and\ \bibinfo {author}
  {\bibfnamefont {D.}~\bibnamefont {Preda}},\ }\bibfield  {title} {\emph
  {\bibinfo {title} {A quantum adiabatic evolution algorithm applied to random
  instances of an {NP}-complete problem},\ }}\href {\doibase
  10.1126/science.1057726} {\bibfield  {journal} {\bibinfo  {journal}
  {Science}\ }\textbf {\bibinfo {volume} {292}},\ \bibinfo {pages} {472}
  (\bibinfo {year} {2001})}\BibitemShut {NoStop}%
\bibitem [{\citenamefont {Bravyi}\ \emph {et~al.}(2018)\citenamefont {Bravyi},
  \citenamefont {Gosset},\ and\ \citenamefont {König}}]{Sergey2018Quantum}%
  \BibitemOpen
  \bibfield  {author} {\bibinfo {author} {\bibfnamefont {S.}~\bibnamefont
  {Bravyi}}, \bibinfo {author} {\bibfnamefont {D.}~\bibnamefont {Gosset}}, \
  and\ \bibinfo {author} {\bibfnamefont {R.}~\bibnamefont {König}},\
  }\bibfield  {title} {\emph {\bibinfo {title} {Quantum advantage with shallow
  circuits},\ }}\href {\doibase 10.1126/science.aar3106} {\bibfield  {journal}
  {\bibinfo  {journal} {Science}\ }\textbf {\bibinfo {volume} {362}},\ \bibinfo
  {pages} {308} (\bibinfo {year} {2018})}\BibitemShut {NoStop}%
\bibitem [{\citenamefont {Bennett}\ \emph {et~al.}(1993)\citenamefont
  {Bennett}, \citenamefont {Brassard}, \citenamefont {Cr\'epeau}, \citenamefont
  {Jozsa}, \citenamefont {Peres},\ and\ \citenamefont
  {Wootters}}]{Bennett1993Teleporting}%
  \BibitemOpen
  \bibfield  {author} {\bibinfo {author} {\bibfnamefont {C.~H.}\ \bibnamefont
  {Bennett}}, \bibinfo {author} {\bibfnamefont {G.}~\bibnamefont {Brassard}},
  \bibinfo {author} {\bibfnamefont {C.}~\bibnamefont {Cr\'epeau}}, \bibinfo
  {author} {\bibfnamefont {R.}~\bibnamefont {Jozsa}}, \bibinfo {author}
  {\bibfnamefont {A.}~\bibnamefont {Peres}}, \ and\ \bibinfo {author}
  {\bibfnamefont {W.~K.}\ \bibnamefont {Wootters}},\ }\bibfield  {title} {\emph
  {\bibinfo {title} {Teleporting an unknown quantum state via dual classical
  and {Einstein-Podolsky-Rosen} channels},\ }}\href {\doibase
  10.1103/PhysRevLett.70.1895} {\bibfield  {journal} {\bibinfo  {journal}
  {Phys. Rev. Lett.}\ }\textbf {\bibinfo {volume} {70}},\ \bibinfo {pages}
  {1895} (\bibinfo {year} {1993})}\BibitemShut {NoStop}%
\bibitem [{\citenamefont {Raussendorf}\ and\ \citenamefont
  {Briegel}(2001)}]{Raussendorf2001Oneway}%
  \BibitemOpen
  \bibfield  {author} {\bibinfo {author} {\bibfnamefont {R.}~\bibnamefont
  {Raussendorf}}\ and\ \bibinfo {author} {\bibfnamefont {H.~J.}\ \bibnamefont
  {Briegel}},\ }\bibfield  {title} {\emph {\bibinfo {title} {A one-way quantum
  computer},\ }}\href {\doibase 10.1103/PhysRevLett.86.5188} {\bibfield
  {journal} {\bibinfo  {journal} {Phys. Rev. Lett.}\ }\textbf {\bibinfo
  {volume} {86}},\ \bibinfo {pages} {5188} (\bibinfo {year}
  {2001})}\BibitemShut {NoStop}%
\bibitem [{\citenamefont {Gidney}\ and\ \citenamefont
  {Eker{\aa}}(2019)}]{Gidney2019HowTF}%
  \BibitemOpen
  \bibfield  {author} {\bibinfo {author} {\bibfnamefont {C.}~\bibnamefont
  {Gidney}}\ and\ \bibinfo {author} {\bibfnamefont {M.}~\bibnamefont
  {Eker{\aa}}},\ }\bibfield  {title} {\emph {\bibinfo {title} {How to factor
  2048 bit {RSA} integers in 8 hours using 20 million noisy qubits},\ }}\href
  {https://api.semanticscholar.org/CorpusID:162183806} {\bibfield  {journal}
  {\bibinfo  {journal} {Quantum}\ }\textbf {\bibinfo {volume} {5}},\ \bibinfo
  {pages} {433} (\bibinfo {year} {2019})}\BibitemShut {NoStop}%
\bibitem [{\citenamefont {Ladd}\ \emph {et~al.}(2010)\citenamefont {Ladd},
  \citenamefont {Jelezko}, \citenamefont {Laflamme}, \citenamefont {Nakamura},
  \citenamefont {Monroe},\ and\ \citenamefont {O’Brien}}]{Ladd2010Quantum}%
  \BibitemOpen
  \bibfield  {author} {\bibinfo {author} {\bibfnamefont {T.~D.}\ \bibnamefont
  {Ladd}}, \bibinfo {author} {\bibfnamefont {F.}~\bibnamefont {Jelezko}},
  \bibinfo {author} {\bibfnamefont {R.}~\bibnamefont {Laflamme}}, \bibinfo
  {author} {\bibfnamefont {Y.}~\bibnamefont {Nakamura}}, \bibinfo {author}
  {\bibfnamefont {C.}~\bibnamefont {Monroe}}, \ and\ \bibinfo {author}
  {\bibfnamefont {J.~L.}\ \bibnamefont {O’Brien}},\ }\bibfield  {title}
  {\emph {\bibinfo {title} {Quantum computers},\ }}\href {\doibase
  10.1038/nature08812} {\bibfield  {journal} {\bibinfo  {journal} {Nature}\
  }\textbf {\bibinfo {volume} {464}},\ \bibinfo {pages} {45} (\bibinfo {year}
  {2010})}\BibitemShut {NoStop}%
\bibitem [{\citenamefont {Benhelm}\ \emph {et~al.}(2008)\citenamefont
  {Benhelm}, \citenamefont {Kirchmair}, \citenamefont {Roos},\ and\
  \citenamefont {Blatt}}]{Benhelm2008Towards}%
  \BibitemOpen
  \bibfield  {author} {\bibinfo {author} {\bibfnamefont {J.}~\bibnamefont
  {Benhelm}}, \bibinfo {author} {\bibfnamefont {G.}~\bibnamefont {Kirchmair}},
  \bibinfo {author} {\bibfnamefont {C.~F.}\ \bibnamefont {Roos}}, \ and\
  \bibinfo {author} {\bibfnamefont {R.}~\bibnamefont {Blatt}},\ }\bibfield
  {title} {\emph {\bibinfo {title} {Towards fault-tolerant quantum computing
  with trapped ions},\ }}\href {\doibase 10.1038/nphys961} {\bibfield
  {journal} {\bibinfo  {journal} {Nat. Phys.}\ }\textbf {\bibinfo {volume}
  {4}},\ \bibinfo {pages} {463} (\bibinfo {year} {2008})}\BibitemShut {NoStop}%
\bibitem [{\citenamefont {Monz}\ \emph {et~al.}(2011)\citenamefont {Monz},
  \citenamefont {Schindler}, \citenamefont {Barreiro}, \citenamefont {Chwalla},
  \citenamefont {Nigg}, \citenamefont {Coish}, \citenamefont {Harlander},
  \citenamefont {H\"ansel}, \citenamefont {Hennrich},\ and\ \citenamefont
  {Blatt}}]{Monz2011QubitEntanglement}%
  \BibitemOpen
  \bibfield  {author} {\bibinfo {author} {\bibfnamefont {T.}~\bibnamefont
  {Monz}}, \bibinfo {author} {\bibfnamefont {P.}~\bibnamefont {Schindler}},
  \bibinfo {author} {\bibfnamefont {J.~T.}\ \bibnamefont {Barreiro}}, \bibinfo
  {author} {\bibfnamefont {M.}~\bibnamefont {Chwalla}}, \bibinfo {author}
  {\bibfnamefont {D.}~\bibnamefont {Nigg}}, \bibinfo {author} {\bibfnamefont
  {W.~A.}\ \bibnamefont {Coish}}, \bibinfo {author} {\bibfnamefont
  {M.}~\bibnamefont {Harlander}}, \bibinfo {author} {\bibfnamefont
  {W.}~\bibnamefont {H\"ansel}}, \bibinfo {author} {\bibfnamefont
  {M.}~\bibnamefont {Hennrich}}, \ and\ \bibinfo {author} {\bibfnamefont
  {R.}~\bibnamefont {Blatt}},\ }\bibfield  {title} {\emph {\bibinfo {title}
  {14-qubit entanglement: Creation and coherence},\ }}\href {\doibase
  10.1103/PhysRevLett.106.130506} {\bibfield  {journal} {\bibinfo  {journal}
  {Phys. Rev. Lett.}\ }\textbf {\bibinfo {volume} {106}},\ \bibinfo {pages}
  {130506} (\bibinfo {year} {2011})}\BibitemShut {NoStop}%
\bibitem [{\citenamefont {Harty}\ \emph {et~al.}(2014)\citenamefont {Harty},
  \citenamefont {Allcock}, \citenamefont {Ballance}, \citenamefont {Guidoni},
  \citenamefont {Janacek}, \citenamefont {Linke}, \citenamefont {Stacey},\ and\
  \citenamefont {Lucas}}]{Harty2014Highfiedelity}%
  \BibitemOpen
  \bibfield  {author} {\bibinfo {author} {\bibfnamefont {T.~P.}\ \bibnamefont
  {Harty}}, \bibinfo {author} {\bibfnamefont {D.~T.~C.}\ \bibnamefont
  {Allcock}}, \bibinfo {author} {\bibfnamefont {C.~J.}\ \bibnamefont
  {Ballance}}, \bibinfo {author} {\bibfnamefont {L.}~\bibnamefont {Guidoni}},
  \bibinfo {author} {\bibfnamefont {H.~A.}\ \bibnamefont {Janacek}}, \bibinfo
  {author} {\bibfnamefont {N.~M.}\ \bibnamefont {Linke}}, \bibinfo {author}
  {\bibfnamefont {D.~N.}\ \bibnamefont {Stacey}}, \ and\ \bibinfo {author}
  {\bibfnamefont {D.~M.}\ \bibnamefont {Lucas}},\ }\bibfield  {title} {\emph
  {\bibinfo {title} {High-fidelity preparation, gates, memory, and readout of a
  trapped-ion quantum bit},\ }}\href {\doibase 10.1103/PhysRevLett.113.220501}
  {\bibfield  {journal} {\bibinfo  {journal} {Phys. Rev. Lett.}\ }\textbf
  {\bibinfo {volume} {113}},\ \bibinfo {pages} {220501} (\bibinfo {year}
  {2014})}\BibitemShut {NoStop}%
\bibitem [{\citenamefont {Ballance}\ \emph {et~al.}(2016)\citenamefont
  {Ballance}, \citenamefont {Harty}, \citenamefont {Linke}, \citenamefont
  {Sepiol},\ and\ \citenamefont {Lucas}}]{Ballance2016Highfidelity}%
  \BibitemOpen
  \bibfield  {author} {\bibinfo {author} {\bibfnamefont {C.~J.}\ \bibnamefont
  {Ballance}}, \bibinfo {author} {\bibfnamefont {T.~P.}\ \bibnamefont {Harty}},
  \bibinfo {author} {\bibfnamefont {N.~M.}\ \bibnamefont {Linke}}, \bibinfo
  {author} {\bibfnamefont {M.~A.}\ \bibnamefont {Sepiol}}, \ and\ \bibinfo
  {author} {\bibfnamefont {D.~M.}\ \bibnamefont {Lucas}},\ }\bibfield  {title}
  {\emph {\bibinfo {title} {High-fidelity quantum logic gates using trapped-ion
  hyperfine qubits},\ }}\href {\doibase 10.1103/PhysRevLett.117.060504}
  {\bibfield  {journal} {\bibinfo  {journal} {Phys. Rev. Lett.}\ }\textbf
  {\bibinfo {volume} {117}},\ \bibinfo {pages} {060504} (\bibinfo {year}
  {2016})}\BibitemShut {NoStop}%
\bibitem [{\citenamefont {Gaebler}\ \emph {et~al.}(2016)\citenamefont
  {Gaebler}, \citenamefont {Tan}, \citenamefont {Lin}, \citenamefont {Wan},
  \citenamefont {Bowler}, \citenamefont {Keith}, \citenamefont {Glancy},
  \citenamefont {Coakley}, \citenamefont {Knill}, \citenamefont {Leibfried},\
  and\ \citenamefont {Wineland}}]{Gaebler2016Highfidelity}%
  \BibitemOpen
  \bibfield  {author} {\bibinfo {author} {\bibfnamefont {J.~P.}\ \bibnamefont
  {Gaebler}}, \bibinfo {author} {\bibfnamefont {T.~R.}\ \bibnamefont {Tan}},
  \bibinfo {author} {\bibfnamefont {Y.}~\bibnamefont {Lin}}, \bibinfo {author}
  {\bibfnamefont {Y.}~\bibnamefont {Wan}}, \bibinfo {author} {\bibfnamefont
  {R.}~\bibnamefont {Bowler}}, \bibinfo {author} {\bibfnamefont {A.~C.}\
  \bibnamefont {Keith}}, \bibinfo {author} {\bibfnamefont {S.}~\bibnamefont
  {Glancy}}, \bibinfo {author} {\bibfnamefont {K.}~\bibnamefont {Coakley}},
  \bibinfo {author} {\bibfnamefont {E.}~\bibnamefont {Knill}}, \bibinfo
  {author} {\bibfnamefont {D.}~\bibnamefont {Leibfried}}, \ and\ \bibinfo
  {author} {\bibfnamefont {D.~J.}\ \bibnamefont {Wineland}},\ }\bibfield
  {title} {\emph {\bibinfo {title} {High-fidelity universal gate set for
  ${^{9}\mathrm{Be}}^{+}$ ion qubits},\ }}\href {\doibase
  10.1103/PhysRevLett.117.060505} {\bibfield  {journal} {\bibinfo  {journal}
  {Phys. Rev. Lett.}\ }\textbf {\bibinfo {volume} {117}},\ \bibinfo {pages}
  {060505} (\bibinfo {year} {2016})}\BibitemShut {NoStop}%
\bibitem [{\citenamefont {Chow}\ \emph {et~al.}(2012)\citenamefont {Chow},
  \citenamefont {Gambetta}, \citenamefont {C\'orcoles}, \citenamefont {Merkel},
  \citenamefont {Smolin}, \citenamefont {Rigetti}, \citenamefont {Poletto},
  \citenamefont {Keefe}, \citenamefont {Rothwell}, \citenamefont {Rozen},
  \citenamefont {Ketchen},\ and\ \citenamefont {Steffen}}]{Chow2012Universal}%
  \BibitemOpen
  \bibfield  {author} {\bibinfo {author} {\bibfnamefont {J.~M.}\ \bibnamefont
  {Chow}}, \bibinfo {author} {\bibfnamefont {J.~M.}\ \bibnamefont {Gambetta}},
  \bibinfo {author} {\bibfnamefont {A.~D.}\ \bibnamefont {C\'orcoles}},
  \bibinfo {author} {\bibfnamefont {S.~T.}\ \bibnamefont {Merkel}}, \bibinfo
  {author} {\bibfnamefont {J.~A.}\ \bibnamefont {Smolin}}, \bibinfo {author}
  {\bibfnamefont {C.}~\bibnamefont {Rigetti}}, \bibinfo {author} {\bibfnamefont
  {S.}~\bibnamefont {Poletto}}, \bibinfo {author} {\bibfnamefont {G.~A.}\
  \bibnamefont {Keefe}}, \bibinfo {author} {\bibfnamefont {M.~B.}\ \bibnamefont
  {Rothwell}}, \bibinfo {author} {\bibfnamefont {J.~R.}\ \bibnamefont {Rozen}},
  \bibinfo {author} {\bibfnamefont {M.~B.}\ \bibnamefont {Ketchen}}, \ and\
  \bibinfo {author} {\bibfnamefont {M.}~\bibnamefont {Steffen}},\ }\bibfield
  {title} {\emph {\bibinfo {title} {Universal quantum gate set approaching
  fault-tolerant thresholds with superconducting qubits},\ }}\href {\doibase
  10.1103/PhysRevLett.109.060501} {\bibfield  {journal} {\bibinfo  {journal}
  {Phys. Rev. Lett.}\ }\textbf {\bibinfo {volume} {109}},\ \bibinfo {pages}
  {060501} (\bibinfo {year} {2012})}\BibitemShut {NoStop}%
\bibitem [{\citenamefont {Barends}\ \emph {et~al.}(2014)\citenamefont
  {Barends}, \citenamefont {Kelly}, \citenamefont {Megrant}, \citenamefont
  {Veitia}, \citenamefont {Sank}, \citenamefont {Jeffrey}, \citenamefont
  {White}, \citenamefont {Mutus}, \citenamefont {Fowler}, \citenamefont
  {Campbell}, \citenamefont {Chen}, \citenamefont {Chen}, \citenamefont
  {Chiaro}, \citenamefont {Dunsworth}, \citenamefont {Neill}, \citenamefont
  {O’Malley}, \citenamefont {Roushan}, \citenamefont {Vainsencher},
  \citenamefont {Wenner}, \citenamefont {Korotkov}, \citenamefont {Cleland},\
  and\ \citenamefont {Martinis}}]{Barends2014Superconducting}%
  \BibitemOpen
  \bibfield  {author} {\bibinfo {author} {\bibfnamefont {R.}~\bibnamefont
  {Barends}}, \bibinfo {author} {\bibfnamefont {J.}~\bibnamefont {Kelly}},
  \bibinfo {author} {\bibfnamefont {A.}~\bibnamefont {Megrant}}, \bibinfo
  {author} {\bibfnamefont {A.}~\bibnamefont {Veitia}}, \bibinfo {author}
  {\bibfnamefont {D.}~\bibnamefont {Sank}}, \bibinfo {author} {\bibfnamefont
  {E.}~\bibnamefont {Jeffrey}}, \bibinfo {author} {\bibfnamefont {T.~C.}\
  \bibnamefont {White}}, \bibinfo {author} {\bibfnamefont {J.}~\bibnamefont
  {Mutus}}, \bibinfo {author} {\bibfnamefont {A.~G.}\ \bibnamefont {Fowler}},
  \bibinfo {author} {\bibfnamefont {B.}~\bibnamefont {Campbell}}, \bibinfo
  {author} {\bibfnamefont {Y.}~\bibnamefont {Chen}}, \bibinfo {author}
  {\bibfnamefont {Z.}~\bibnamefont {Chen}}, \bibinfo {author} {\bibfnamefont
  {B.}~\bibnamefont {Chiaro}}, \bibinfo {author} {\bibfnamefont
  {A.}~\bibnamefont {Dunsworth}}, \bibinfo {author} {\bibfnamefont
  {C.}~\bibnamefont {Neill}}, \bibinfo {author} {\bibfnamefont
  {P.}~\bibnamefont {O’Malley}}, \bibinfo {author} {\bibfnamefont
  {P.}~\bibnamefont {Roushan}}, \bibinfo {author} {\bibfnamefont
  {A.}~\bibnamefont {Vainsencher}}, \bibinfo {author} {\bibfnamefont
  {J.}~\bibnamefont {Wenner}}, \bibinfo {author} {\bibfnamefont {A.~N.}\
  \bibnamefont {Korotkov}}, \bibinfo {author} {\bibfnamefont {A.~N.}\
  \bibnamefont {Cleland}}, \ and\ \bibinfo {author} {\bibfnamefont {J.~M.}\
  \bibnamefont {Martinis}},\ }\bibfield  {title} {\emph {\bibinfo {title}
  {Superconducting quantum circuits at the surface code threshold for fault
  tolerance},\ }}\href {\doibase 10.1038/nature13171} {\bibfield  {journal}
  {\bibinfo  {journal} {Nature}\ }\textbf {\bibinfo {volume} {508}},\ \bibinfo
  {pages} {500} (\bibinfo {year} {2014})}\BibitemShut {NoStop}%
\bibitem [{\citenamefont {Song}\ \emph {et~al.}(2019)\citenamefont {Song},
  \citenamefont {Xu}, \citenamefont {Li}, \citenamefont {Zhang}, \citenamefont
  {Zhang}, \citenamefont {Liu}, \citenamefont {Guo}, \citenamefont {Wang},
  \citenamefont {Ren}, \citenamefont {Hao}, \citenamefont {Feng}, \citenamefont
  {Fan}, \citenamefont {Zheng}, \citenamefont {Wang}, \citenamefont {Wang},\
  and\ \citenamefont {Zhu}}]{Chao2019Generation}%
  \BibitemOpen
  \bibfield  {author} {\bibinfo {author} {\bibfnamefont {C.}~\bibnamefont
  {Song}}, \bibinfo {author} {\bibfnamefont {K.}~\bibnamefont {Xu}}, \bibinfo
  {author} {\bibfnamefont {H.}~\bibnamefont {Li}}, \bibinfo {author}
  {\bibfnamefont {Y.-R.}\ \bibnamefont {Zhang}}, \bibinfo {author}
  {\bibfnamefont {X.}~\bibnamefont {Zhang}}, \bibinfo {author} {\bibfnamefont
  {W.}~\bibnamefont {Liu}}, \bibinfo {author} {\bibfnamefont {Q.}~\bibnamefont
  {Guo}}, \bibinfo {author} {\bibfnamefont {Z.}~\bibnamefont {Wang}}, \bibinfo
  {author} {\bibfnamefont {W.}~\bibnamefont {Ren}}, \bibinfo {author}
  {\bibfnamefont {J.}~\bibnamefont {Hao}}, \bibinfo {author} {\bibfnamefont
  {H.}~\bibnamefont {Feng}}, \bibinfo {author} {\bibfnamefont {H.}~\bibnamefont
  {Fan}}, \bibinfo {author} {\bibfnamefont {D.}~\bibnamefont {Zheng}}, \bibinfo
  {author} {\bibfnamefont {D.-W.}\ \bibnamefont {Wang}}, \bibinfo {author}
  {\bibfnamefont {H.}~\bibnamefont {Wang}}, \ and\ \bibinfo {author}
  {\bibfnamefont {S.-Y.}\ \bibnamefont {Zhu}},\ }\bibfield  {title} {\emph
  {\bibinfo {title} {Generation of multicomponent atomic {S}chr\"odinger cat
  states of up to 20 qubits},\ }}\href {\doibase 10.1126/science.aay0600}
  {\bibfield  {journal} {\bibinfo  {journal} {Science}\ }\textbf {\bibinfo
  {volume} {365}},\ \bibinfo {pages} {574} (\bibinfo {year}
  {2019})}\BibitemShut {NoStop}%
\bibitem [{\citenamefont {Wright}\ \emph {et~al.}(2019)\citenamefont {Wright},
  \citenamefont {Beck}, \citenamefont {Debnath}, \citenamefont {Amini},
  \citenamefont {Nam}, \citenamefont {Grzesiak}, \citenamefont {Chen},
  \citenamefont {Pisenti}, \citenamefont {Chmielewski}, \citenamefont
  {Collins}, \citenamefont {Hudek}, \citenamefont {Mizrahi}, \citenamefont
  {Wong-Campos}, \citenamefont {Allen}, \citenamefont {Apisdorf}, \citenamefont
  {Solomon}, \citenamefont {Williams}, \citenamefont {Ducore}, \citenamefont
  {Blinov}, \citenamefont {Kreikemeier}, \citenamefont {Chaplin}, \citenamefont
  {Keesan}, \citenamefont {Monroe},\ and\ \citenamefont
  {Kim}}]{Wright2019Benchmarking}%
  \BibitemOpen
  \bibfield  {author} {\bibinfo {author} {\bibfnamefont {K.}~\bibnamefont
  {Wright}}, \bibinfo {author} {\bibfnamefont {K.~M.}\ \bibnamefont {Beck}},
  \bibinfo {author} {\bibfnamefont {S.}~\bibnamefont {Debnath}}, \bibinfo
  {author} {\bibfnamefont {J.~M.}\ \bibnamefont {Amini}}, \bibinfo {author}
  {\bibfnamefont {Y.}~\bibnamefont {Nam}}, \bibinfo {author} {\bibfnamefont
  {N.}~\bibnamefont {Grzesiak}}, \bibinfo {author} {\bibfnamefont {J.-S.}\
  \bibnamefont {Chen}}, \bibinfo {author} {\bibfnamefont {N.~C.}\ \bibnamefont
  {Pisenti}}, \bibinfo {author} {\bibfnamefont {M.}~\bibnamefont
  {Chmielewski}}, \bibinfo {author} {\bibfnamefont {C.}~\bibnamefont
  {Collins}}, \bibinfo {author} {\bibfnamefont {K.~M.}\ \bibnamefont {Hudek}},
  \bibinfo {author} {\bibfnamefont {J.}~\bibnamefont {Mizrahi}}, \bibinfo
  {author} {\bibfnamefont {J.~D.}\ \bibnamefont {Wong-Campos}}, \bibinfo
  {author} {\bibfnamefont {S.}~\bibnamefont {Allen}}, \bibinfo {author}
  {\bibfnamefont {J.}~\bibnamefont {Apisdorf}}, \bibinfo {author}
  {\bibfnamefont {P.}~\bibnamefont {Solomon}}, \bibinfo {author} {\bibfnamefont
  {M.}~\bibnamefont {Williams}}, \bibinfo {author} {\bibfnamefont {A.~M.}\
  \bibnamefont {Ducore}}, \bibinfo {author} {\bibfnamefont {A.}~\bibnamefont
  {Blinov}}, \bibinfo {author} {\bibfnamefont {S.~M.}\ \bibnamefont
  {Kreikemeier}}, \bibinfo {author} {\bibfnamefont {V.}~\bibnamefont
  {Chaplin}}, \bibinfo {author} {\bibfnamefont {M.}~\bibnamefont {Keesan}},
  \bibinfo {author} {\bibfnamefont {C.}~\bibnamefont {Monroe}}, \ and\ \bibinfo
  {author} {\bibfnamefont {J.}~\bibnamefont {Kim}},\ }\bibfield  {title} {\emph
  {\bibinfo {title} {Benchmarking an 11-qubit quantum computer},\ }}\href
  {\doibase 10.1038/s41467-019-13534-2} {\bibfield  {journal} {\bibinfo
  {journal} {Nat. Commun.}\ }\textbf {\bibinfo {volume} {10}},\ \bibinfo
  {pages} {5464} (\bibinfo {year} {2019})}\BibitemShut {NoStop}%
\bibitem [{\citenamefont {Zhong}\ \emph {et~al.}(2021)\citenamefont {Zhong},
  \citenamefont {Chang}, \citenamefont {Bienfait}, \citenamefont {Dumur},
  \citenamefont {Chou}, \citenamefont {Conner}, \citenamefont {Grebel},
  \citenamefont {Povey}, \citenamefont {Yan}, \citenamefont {Schuster},\ and\
  \citenamefont {Cleland}}]{Zhong2021Deterministic}%
  \BibitemOpen
  \bibfield  {author} {\bibinfo {author} {\bibfnamefont {Y.}~\bibnamefont
  {Zhong}}, \bibinfo {author} {\bibfnamefont {H.-S.}\ \bibnamefont {Chang}},
  \bibinfo {author} {\bibfnamefont {A.}~\bibnamefont {Bienfait}}, \bibinfo
  {author} {\bibfnamefont {E.}~\bibnamefont {Dumur}}, \bibinfo {author}
  {\bibfnamefont {M.-H.}\ \bibnamefont {Chou}}, \bibinfo {author}
  {\bibfnamefont {C.~R.}\ \bibnamefont {Conner}}, \bibinfo {author}
  {\bibfnamefont {J.}~\bibnamefont {Grebel}}, \bibinfo {author} {\bibfnamefont
  {R.~G.}\ \bibnamefont {Povey}}, \bibinfo {author} {\bibfnamefont
  {H.}~\bibnamefont {Yan}}, \bibinfo {author} {\bibfnamefont {D.~I.}\
  \bibnamefont {Schuster}}, \ and\ \bibinfo {author} {\bibfnamefont {A.~N.}\
  \bibnamefont {Cleland}},\ }\bibfield  {title} {\emph {\bibinfo {title}
  {Deterministic multi-qubit entanglement in a quantum network},\ }}\href
  {\doibase 10.1038/s41586-021-03288-7} {\bibfield  {journal} {\bibinfo
  {journal} {Nature}\ }\textbf {\bibinfo {volume} {590}},\ \bibinfo {pages}
  {571} (\bibinfo {year} {2021})}\BibitemShut {NoStop}%
\bibitem [{\citenamefont {Saffman}\ \emph {et~al.}(2010)\citenamefont
  {Saffman}, \citenamefont {Walker},\ and\ \citenamefont
  {M\o{}lmer}}]{Saffman2010Quantum}%
  \BibitemOpen
  \bibfield  {author} {\bibinfo {author} {\bibfnamefont {M.}~\bibnamefont
  {Saffman}}, \bibinfo {author} {\bibfnamefont {T.~G.}\ \bibnamefont {Walker}},
  \ and\ \bibinfo {author} {\bibfnamefont {K.}~\bibnamefont {M\o{}lmer}},\
  }\bibfield  {title} {\emph {\bibinfo {title} {Quantum information with
  {R}ydberg atoms},\ }}\href {\doibase 10.1103/RevModPhys.82.2313} {\bibfield
  {journal} {\bibinfo  {journal} {Rev. Mod. Phys.}\ }\textbf {\bibinfo {volume}
  {82}},\ \bibinfo {pages} {2313} (\bibinfo {year} {2010})}\BibitemShut
  {NoStop}%
\bibitem [{\citenamefont {Xia}\ \emph {et~al.}(2015)\citenamefont {Xia},
  \citenamefont {Lichtman}, \citenamefont {Maller}, \citenamefont {Carr},
  \citenamefont {Piotrowicz}, \citenamefont {Isenhower},\ and\ \citenamefont
  {Saffman}}]{Xia2015Randomized}%
  \BibitemOpen
  \bibfield  {author} {\bibinfo {author} {\bibfnamefont {T.}~\bibnamefont
  {Xia}}, \bibinfo {author} {\bibfnamefont {M.}~\bibnamefont {Lichtman}},
  \bibinfo {author} {\bibfnamefont {K.}~\bibnamefont {Maller}}, \bibinfo
  {author} {\bibfnamefont {A.~W.}\ \bibnamefont {Carr}}, \bibinfo {author}
  {\bibfnamefont {M.~J.}\ \bibnamefont {Piotrowicz}}, \bibinfo {author}
  {\bibfnamefont {L.}~\bibnamefont {Isenhower}}, \ and\ \bibinfo {author}
  {\bibfnamefont {M.}~\bibnamefont {Saffman}},\ }\bibfield  {title} {\emph
  {\bibinfo {title} {Randomized benchmarking of single-qubit gates in a {2D}
  array of neutral-atom qubits},\ }}\href {\doibase
  10.1103/PhysRevLett.114.100503} {\bibfield  {journal} {\bibinfo  {journal}
  {Phys. Rev. Lett.}\ }\textbf {\bibinfo {volume} {114}},\ \bibinfo {pages}
  {100503} (\bibinfo {year} {2015})}\BibitemShut {NoStop}%
\bibitem [{\citenamefont {Wang}\ \emph {et~al.}(2016)\citenamefont {Wang},
  \citenamefont {Kumar}, \citenamefont {Wu},\ and\ \citenamefont
  {Weiss}}]{Wang2016Single}%
  \BibitemOpen
  \bibfield  {author} {\bibinfo {author} {\bibfnamefont {Y.}~\bibnamefont
  {Wang}}, \bibinfo {author} {\bibfnamefont {A.}~\bibnamefont {Kumar}},
  \bibinfo {author} {\bibfnamefont {T.-Y.}\ \bibnamefont {Wu}}, \ and\ \bibinfo
  {author} {\bibfnamefont {D.~S.}\ \bibnamefont {Weiss}},\ }\bibfield  {title}
  {\emph {\bibinfo {title} {Single-qubit gates based on targeted phase shifts
  in a {3D} neutral atom array},\ }}\href {\doibase 10.1126/science.aaf2581}
  {\bibfield  {journal} {\bibinfo  {journal} {Science}\ }\textbf {\bibinfo
  {volume} {352}},\ \bibinfo {pages} {1562} (\bibinfo {year}
  {2016})}\BibitemShut {NoStop}%
\bibitem [{\citenamefont {Sheng}\ \emph {et~al.}(2018)\citenamefont {Sheng},
  \citenamefont {He}, \citenamefont {Xu}, \citenamefont {Guo}, \citenamefont
  {Wang}, \citenamefont {Xiong}, \citenamefont {Liu}, \citenamefont {Wang},\
  and\ \citenamefont {Zhan}}]{Sheng2018Highfidelity}%
  \BibitemOpen
  \bibfield  {author} {\bibinfo {author} {\bibfnamefont {C.}~\bibnamefont
  {Sheng}}, \bibinfo {author} {\bibfnamefont {X.}~\bibnamefont {He}}, \bibinfo
  {author} {\bibfnamefont {P.}~\bibnamefont {Xu}}, \bibinfo {author}
  {\bibfnamefont {R.}~\bibnamefont {Guo}}, \bibinfo {author} {\bibfnamefont
  {K.}~\bibnamefont {Wang}}, \bibinfo {author} {\bibfnamefont {Z.}~\bibnamefont
  {Xiong}}, \bibinfo {author} {\bibfnamefont {M.}~\bibnamefont {Liu}}, \bibinfo
  {author} {\bibfnamefont {J.}~\bibnamefont {Wang}}, \ and\ \bibinfo {author}
  {\bibfnamefont {M.}~\bibnamefont {Zhan}},\ }\bibfield  {title} {\emph
  {\bibinfo {title} {High-fidelity single-qubit gates on neutral atoms in a
  two-dimensional magic-intensity optical dipole trap array},\ }}\href
  {\doibase 10.1103/PhysRevLett.121.240501} {\bibfield  {journal} {\bibinfo
  {journal} {Phys. Rev. Lett.}\ }\textbf {\bibinfo {volume} {121}},\ \bibinfo
  {pages} {240501} (\bibinfo {year} {2018})}\BibitemShut {NoStop}%
\bibitem [{\citenamefont {Weiss}\ and\ \citenamefont
  {Saffman}(2017)}]{Weiss2017Quantum}%
  \BibitemOpen
  \bibfield  {author} {\bibinfo {author} {\bibfnamefont {D.~S.}\ \bibnamefont
  {Weiss}}\ and\ \bibinfo {author} {\bibfnamefont {M.}~\bibnamefont
  {Saffman}},\ }\bibfield  {title} {\emph {\bibinfo {title} {Quantum computing
  with neutral atoms},\ }}\href {\doibase 10.1063/PT.3.3626} {\bibfield
  {journal} {\bibinfo  {journal} {Phys. Today}\ }\textbf {\bibinfo {volume}
  {70(7)}},\ \bibinfo {pages} {44} (\bibinfo {year} {2017})}\BibitemShut
  {NoStop}%
\bibitem [{\citenamefont {Levine}\ \emph {et~al.}(2018)\citenamefont {Levine},
  \citenamefont {Keesling}, \citenamefont {Omran}, \citenamefont {Bernien},
  \citenamefont {Schwartz}, \citenamefont {Zibrov}, \citenamefont {Endres},
  \citenamefont {Greiner}, \citenamefont {Vuleti\ifmmode~\acute{c}\else
  \'{c}\fi{}},\ and\ \citenamefont {Lukin}}]{Levine2018Highfidelity}%
  \BibitemOpen
  \bibfield  {author} {\bibinfo {author} {\bibfnamefont {H.}~\bibnamefont
  {Levine}}, \bibinfo {author} {\bibfnamefont {A.}~\bibnamefont {Keesling}},
  \bibinfo {author} {\bibfnamefont {A.}~\bibnamefont {Omran}}, \bibinfo
  {author} {\bibfnamefont {H.}~\bibnamefont {Bernien}}, \bibinfo {author}
  {\bibfnamefont {S.}~\bibnamefont {Schwartz}}, \bibinfo {author}
  {\bibfnamefont {A.~S.}\ \bibnamefont {Zibrov}}, \bibinfo {author}
  {\bibfnamefont {M.}~\bibnamefont {Endres}}, \bibinfo {author} {\bibfnamefont
  {M.}~\bibnamefont {Greiner}}, \bibinfo {author} {\bibfnamefont
  {V.}~\bibnamefont {Vuleti\ifmmode~\acute{c}\else \'{c}\fi{}}}, \ and\
  \bibinfo {author} {\bibfnamefont {M.~D.}\ \bibnamefont {Lukin}},\ }\bibfield
  {title} {\emph {\bibinfo {title} {High-fidelity control and entanglement of
  rydberg-atom qubits},\ }}\href {\doibase 10.1103/PhysRevLett.121.123603}
  {\bibfield  {journal} {\bibinfo  {journal} {Phys. Rev. Lett.}\ }\textbf
  {\bibinfo {volume} {121}},\ \bibinfo {pages} {123603} (\bibinfo {year}
  {2018})}\BibitemShut {NoStop}%
\bibitem [{\citenamefont {Madjarov}\ \emph {et~al.}(2020)\citenamefont
  {Madjarov}, \citenamefont {Covey}, \citenamefont {Shaw}, \citenamefont
  {Choi}, \citenamefont {Kale}, \citenamefont {Cooper}, \citenamefont
  {Pichler}, \citenamefont {Schkolnik}, \citenamefont {Williams},\ and\
  \citenamefont {Endres}}]{Madjarove2020Highfidelity}%
  \BibitemOpen
  \bibfield  {author} {\bibinfo {author} {\bibfnamefont {I.~S.}\ \bibnamefont
  {Madjarov}}, \bibinfo {author} {\bibfnamefont {J.~P.}\ \bibnamefont {Covey}},
  \bibinfo {author} {\bibfnamefont {A.~L.}\ \bibnamefont {Shaw}}, \bibinfo
  {author} {\bibfnamefont {J.}~\bibnamefont {Choi}}, \bibinfo {author}
  {\bibfnamefont {A.}~\bibnamefont {Kale}}, \bibinfo {author} {\bibfnamefont
  {A.}~\bibnamefont {Cooper}}, \bibinfo {author} {\bibfnamefont
  {H.}~\bibnamefont {Pichler}}, \bibinfo {author} {\bibfnamefont
  {V.}~\bibnamefont {Schkolnik}}, \bibinfo {author} {\bibfnamefont {J.~R.}\
  \bibnamefont {Williams}}, \ and\ \bibinfo {author} {\bibfnamefont
  {M.}~\bibnamefont {Endres}},\ }\bibfield  {title} {\emph {\bibinfo {title}
  {High-fidelity entanglement and detection of alkaline-earth {R}ydberg
  atoms},\ }}\href {\doibase 10.1038/s41567-020-0903-z} {\bibfield  {journal}
  {\bibinfo  {journal} {Nat. Phys.}\ }\textbf {\bibinfo {volume} {16}},\
  \bibinfo {pages} {857} (\bibinfo {year} {2020})}\BibitemShut {NoStop}%
\bibitem [{\citenamefont {Omran}\ \emph {et~al.}(2019)\citenamefont {Omran},
  \citenamefont {Levine}, \citenamefont {Keesling}, \citenamefont {Semeghini},
  \citenamefont {Wang}, \citenamefont {Ebadi}, \citenamefont {Bernien},
  \citenamefont {Zibrov}, \citenamefont {Pichler}, \citenamefont {Choi},
  \citenamefont {Cui}, \citenamefont {Rossignolo}, \citenamefont {Rembold},
  \citenamefont {Montangero}, \citenamefont {Calarco}, \citenamefont {Endres},
  \citenamefont {Greiner}, \citenamefont {Vuletić},\ and\ \citenamefont
  {Lukin}}]{Omran2019Generation}%
  \BibitemOpen
  \bibfield  {author} {\bibinfo {author} {\bibfnamefont {A.}~\bibnamefont
  {Omran}}, \bibinfo {author} {\bibfnamefont {H.}~\bibnamefont {Levine}},
  \bibinfo {author} {\bibfnamefont {A.}~\bibnamefont {Keesling}}, \bibinfo
  {author} {\bibfnamefont {G.}~\bibnamefont {Semeghini}}, \bibinfo {author}
  {\bibfnamefont {T.~T.}\ \bibnamefont {Wang}}, \bibinfo {author}
  {\bibfnamefont {S.}~\bibnamefont {Ebadi}}, \bibinfo {author} {\bibfnamefont
  {H.}~\bibnamefont {Bernien}}, \bibinfo {author} {\bibfnamefont {A.~S.}\
  \bibnamefont {Zibrov}}, \bibinfo {author} {\bibfnamefont {H.}~\bibnamefont
  {Pichler}}, \bibinfo {author} {\bibfnamefont {S.}~\bibnamefont {Choi}},
  \bibinfo {author} {\bibfnamefont {J.}~\bibnamefont {Cui}}, \bibinfo {author}
  {\bibfnamefont {M.}~\bibnamefont {Rossignolo}}, \bibinfo {author}
  {\bibfnamefont {P.}~\bibnamefont {Rembold}}, \bibinfo {author} {\bibfnamefont
  {S.}~\bibnamefont {Montangero}}, \bibinfo {author} {\bibfnamefont
  {T.}~\bibnamefont {Calarco}}, \bibinfo {author} {\bibfnamefont
  {M.}~\bibnamefont {Endres}}, \bibinfo {author} {\bibfnamefont
  {M.}~\bibnamefont {Greiner}}, \bibinfo {author} {\bibfnamefont
  {V.}~\bibnamefont {Vuletić}}, \ and\ \bibinfo {author} {\bibfnamefont
  {M.~D.}\ \bibnamefont {Lukin}},\ }\bibfield  {title} {\emph {\bibinfo {title}
  {Generation and manipulation of {S}chr\"odinger cat states in rydberg atom
  arrays},\ }}\href {\doibase 10.1126/science.aax9743} {\bibfield  {journal}
  {\bibinfo  {journal} {Science}\ }\textbf {\bibinfo {volume} {365}},\ \bibinfo
  {pages} {570} (\bibinfo {year} {2019})}\BibitemShut {NoStop}%
\bibitem [{\citenamefont {Barredo}\ \emph {et~al.}(2016)\citenamefont
  {Barredo}, \citenamefont {de~Léséleuc}, \citenamefont {Lienhard},
  \citenamefont {Lahaye},\ and\ \citenamefont
  {Browaeys}}]{Barredo2016Anatombyatom}%
  \BibitemOpen
  \bibfield  {author} {\bibinfo {author} {\bibfnamefont {D.}~\bibnamefont
  {Barredo}}, \bibinfo {author} {\bibfnamefont {S.}~\bibnamefont
  {de~Léséleuc}}, \bibinfo {author} {\bibfnamefont {V.}~\bibnamefont
  {Lienhard}}, \bibinfo {author} {\bibfnamefont {T.}~\bibnamefont {Lahaye}}, \
  and\ \bibinfo {author} {\bibfnamefont {A.}~\bibnamefont {Browaeys}},\
  }\bibfield  {title} {\emph {\bibinfo {title} {An atom-by-atom assembler of
  defect-free arbitrary two-dimensional atomic arrays},\ }}\href {\doibase
  10.1126/science.aah3778} {\bibfield  {journal} {\bibinfo  {journal}
  {Science}\ }\textbf {\bibinfo {volume} {354}},\ \bibinfo {pages} {1021}
  (\bibinfo {year} {2016})}\BibitemShut {NoStop}%
\bibitem [{\citenamefont {Ebadi}\ \emph {et~al.}(2021)\citenamefont {Ebadi},
  \citenamefont {Wang}, \citenamefont {Levine}, \citenamefont {Keesling},
  \citenamefont {Semeghini}, \citenamefont {Omran}, \citenamefont {Bluvstein},
  \citenamefont {Samajdar}, \citenamefont {Pichler}, \citenamefont {Ho},
  \citenamefont {Choi}, \citenamefont {Sachdev}, \citenamefont {Greiner},
  \citenamefont {Vuletić},\ and\ \citenamefont {Lukin}}]{Ebadi2021Quantum}%
  \BibitemOpen
  \bibfield  {author} {\bibinfo {author} {\bibfnamefont {S.}~\bibnamefont
  {Ebadi}}, \bibinfo {author} {\bibfnamefont {T.~T.}\ \bibnamefont {Wang}},
  \bibinfo {author} {\bibfnamefont {H.}~\bibnamefont {Levine}}, \bibinfo
  {author} {\bibfnamefont {A.}~\bibnamefont {Keesling}}, \bibinfo {author}
  {\bibfnamefont {G.}~\bibnamefont {Semeghini}}, \bibinfo {author}
  {\bibfnamefont {A.}~\bibnamefont {Omran}}, \bibinfo {author} {\bibfnamefont
  {D.}~\bibnamefont {Bluvstein}}, \bibinfo {author} {\bibfnamefont
  {R.}~\bibnamefont {Samajdar}}, \bibinfo {author} {\bibfnamefont
  {H.}~\bibnamefont {Pichler}}, \bibinfo {author} {\bibfnamefont {W.~W.}\
  \bibnamefont {Ho}}, \bibinfo {author} {\bibfnamefont {S.}~\bibnamefont
  {Choi}}, \bibinfo {author} {\bibfnamefont {S.}~\bibnamefont {Sachdev}},
  \bibinfo {author} {\bibfnamefont {M.}~\bibnamefont {Greiner}}, \bibinfo
  {author} {\bibfnamefont {V.}~\bibnamefont {Vuletić}}, \ and\ \bibinfo
  {author} {\bibfnamefont {M.~D.}\ \bibnamefont {Lukin}},\ }\bibfield  {title}
  {\emph {\bibinfo {title} {Quantum phases of matter on a 256-atom programmable
  quantum simulator},\ }}\href {\doibase 10.1038/s41586-021-03582-4} {\bibfield
   {journal} {\bibinfo  {journal} {Nature}\ }\textbf {\bibinfo {volume}
  {595}},\ \bibinfo {pages} {227} (\bibinfo {year} {2021})}\BibitemShut
  {NoStop}%
\bibitem [{\citenamefont {Barredo}\ \emph {et~al.}(2018)\citenamefont
  {Barredo}, \citenamefont {Lienhard}, \citenamefont {de~Léséleuc},
  \citenamefont {Lahaye},\ and\ \citenamefont {Browaeys}}]{Barredo2018Barredo}%
  \BibitemOpen
  \bibfield  {author} {\bibinfo {author} {\bibfnamefont {D.}~\bibnamefont
  {Barredo}}, \bibinfo {author} {\bibfnamefont {V.}~\bibnamefont {Lienhard}},
  \bibinfo {author} {\bibfnamefont {S.}~\bibnamefont {de~Léséleuc}}, \bibinfo
  {author} {\bibfnamefont {T.}~\bibnamefont {Lahaye}}, \ and\ \bibinfo {author}
  {\bibfnamefont {A.}~\bibnamefont {Browaeys}},\ }\bibfield  {title} {\emph
  {\bibinfo {title} {Synthetic three-dimensional atomic structures assembled
  atom by atom},\ }}\href {\doibase 10.1038/s41586-018-0450-2} {\bibfield
  {journal} {\bibinfo  {journal} {Nature}\ }\textbf {\bibinfo {volume} {561}},\
  \bibinfo {pages} {79} (\bibinfo {year} {2018})}\BibitemShut {NoStop}%
\bibitem [{\citenamefont {Barredo}\ \emph {et~al.}(2020)\citenamefont
  {Barredo}, \citenamefont {Lienhard}, \citenamefont {Scholl}, \citenamefont
  {de~L\'es\'eleuc}, \citenamefont {Boulier}, \citenamefont {Browaeys},\ and\
  \citenamefont {Lahaye}}]{Barredo2020Threedimensional}%
  \BibitemOpen
  \bibfield  {author} {\bibinfo {author} {\bibfnamefont {D.}~\bibnamefont
  {Barredo}}, \bibinfo {author} {\bibfnamefont {V.}~\bibnamefont {Lienhard}},
  \bibinfo {author} {\bibfnamefont {P.}~\bibnamefont {Scholl}}, \bibinfo
  {author} {\bibfnamefont {S.}~\bibnamefont {de~L\'es\'eleuc}}, \bibinfo
  {author} {\bibfnamefont {T.}~\bibnamefont {Boulier}}, \bibinfo {author}
  {\bibfnamefont {A.}~\bibnamefont {Browaeys}}, \ and\ \bibinfo {author}
  {\bibfnamefont {T.}~\bibnamefont {Lahaye}},\ }\bibfield  {title} {\emph
  {\bibinfo {title} {Three-dimensional trapping of individual {R}ydberg atoms
  in ponderomotive bottle beam traps},\ }}\href {\doibase
  10.1103/PhysRevLett.124.023201} {\bibfield  {journal} {\bibinfo  {journal}
  {Phys. Rev. Lett.}\ }\textbf {\bibinfo {volume} {124}},\ \bibinfo {pages}
  {023201} (\bibinfo {year} {2020})}\BibitemShut {NoStop}%
\bibitem [{\citenamefont {Browaeys}\ \emph {et~al.}(2016)\citenamefont
  {Browaeys}, \citenamefont {Barredo},\ and\ \citenamefont
  {Lahaye}}]{Browaeys2016Experimental}%
  \BibitemOpen
  \bibfield  {author} {\bibinfo {author} {\bibfnamefont {A.}~\bibnamefont
  {Browaeys}}, \bibinfo {author} {\bibfnamefont {D.}~\bibnamefont {Barredo}}, \
  and\ \bibinfo {author} {\bibfnamefont {T.}~\bibnamefont {Lahaye}},\
  }\bibfield  {title} {\emph {\bibinfo {title} {Experimental investigations of
  dipole–dipole interactions between a few {R}ydberg atoms},\ }}\href
  {\doibase 10.1088/0953-4075/49/15/152001} {\bibfield  {journal} {\bibinfo
  {journal} {J. Phys. B: At., Mol. Opt. Phys.}\ }\textbf {\bibinfo {volume}
  {49}},\ \bibinfo {pages} {152001} (\bibinfo {year} {2016})}\BibitemShut
  {NoStop}%
\bibitem [{\citenamefont {Walker}\ and\ \citenamefont
  {Saffman}(2008)}]{Walker2008Consequences}%
  \BibitemOpen
  \bibfield  {author} {\bibinfo {author} {\bibfnamefont {T.~G.}\ \bibnamefont
  {Walker}}\ and\ \bibinfo {author} {\bibfnamefont {M.}~\bibnamefont
  {Saffman}},\ }\bibfield  {title} {\emph {\bibinfo {title} {Consequences of
  {Z}eeman degeneracy for the van der {W}aals blockade between {R}ydberg
  atoms},\ }}\href {\doibase 10.1103/PhysRevA.77.032723} {\bibfield  {journal}
  {\bibinfo  {journal} {Phys. Rev. A}\ }\textbf {\bibinfo {volume} {77}},\
  \bibinfo {pages} {032723} (\bibinfo {year} {2008})}\BibitemShut {NoStop}%
\bibitem [{\citenamefont {Shao}\ \emph
  {et~al.}(2017{\natexlab{a}})\citenamefont {Shao}, \citenamefont {Li},
  \citenamefont {Ji}, \citenamefont {Wu},\ and\ \citenamefont
  {Yi}}]{Shao2017Ground}%
  \BibitemOpen
  \bibfield  {author} {\bibinfo {author} {\bibfnamefont {X.~Q.}\ \bibnamefont
  {Shao}}, \bibinfo {author} {\bibfnamefont {D.~X.}\ \bibnamefont {Li}},
  \bibinfo {author} {\bibfnamefont {Y.~Q.}\ \bibnamefont {Ji}}, \bibinfo
  {author} {\bibfnamefont {J.~H.}\ \bibnamefont {Wu}}, \ and\ \bibinfo {author}
  {\bibfnamefont {X.~X.}\ \bibnamefont {Yi}},\ }\bibfield  {title} {\emph
  {\bibinfo {title} {Ground-state blockade of {R}ydberg atoms and application
  in entanglement generation},\ }}\href {\doibase 10.1103/PhysRevA.96.012328}
  {\bibfield  {journal} {\bibinfo  {journal} {Phys. Rev. A}\ }\textbf {\bibinfo
  {volume} {96}},\ \bibinfo {pages} {012328} (\bibinfo {year}
  {2017}{\natexlab{a}})}\BibitemShut {NoStop}%
\bibitem [{\citenamefont {Shao}\ \emph
  {et~al.}(2017{\natexlab{b}})\citenamefont {Shao}, \citenamefont {Wu},\ and\
  \citenamefont {Yi}}]{Shao2017Dissipation}%
  \BibitemOpen
  \bibfield  {author} {\bibinfo {author} {\bibfnamefont {X.~Q.}\ \bibnamefont
  {Shao}}, \bibinfo {author} {\bibfnamefont {J.~H.}\ \bibnamefont {Wu}}, \ and\
  \bibinfo {author} {\bibfnamefont {X.~X.}\ \bibnamefont {Yi}},\ }\bibfield
  {title} {\emph {\bibinfo {title} {Dissipation-based entanglement via quantum
  zeno dynamics and rydberg antiblockade},\ }}\href {\doibase
  10.1103/PhysRevA.95.062339} {\bibfield  {journal} {\bibinfo  {journal} {Phys.
  Rev. A}\ }\textbf {\bibinfo {volume} {95}},\ \bibinfo {pages} {062339}
  (\bibinfo {year} {2017}{\natexlab{b}})}\BibitemShut {NoStop}%
\bibitem [{\citenamefont {Zhao}\ \emph
  {et~al.}(2017{\natexlab{a}})\citenamefont {Zhao}, \citenamefont {Cui},
  \citenamefont {Xu}, \citenamefont {Sj\"oqvist},\ and\ \citenamefont
  {Tong}}]{Zhao2017Rydberg}%
  \BibitemOpen
  \bibfield  {author} {\bibinfo {author} {\bibfnamefont {P.~Z.}\ \bibnamefont
  {Zhao}}, \bibinfo {author} {\bibfnamefont {X.-D.}\ \bibnamefont {Cui}},
  \bibinfo {author} {\bibfnamefont {G.~F.}\ \bibnamefont {Xu}}, \bibinfo
  {author} {\bibfnamefont {E.}~\bibnamefont {Sj\"oqvist}}, \ and\ \bibinfo
  {author} {\bibfnamefont {D.~M.}\ \bibnamefont {Tong}},\ }\bibfield  {title}
  {\emph {\bibinfo {title} {Rydberg-atom-based scheme of nonadiabatic geometric
  quantum computation},\ }}\href {\doibase 10.1103/PhysRevA.96.052316}
  {\bibfield  {journal} {\bibinfo  {journal} {Phys. Rev. A}\ }\textbf {\bibinfo
  {volume} {96}},\ \bibinfo {pages} {052316} (\bibinfo {year}
  {2017}{\natexlab{a}})}\BibitemShut {NoStop}%
\bibitem [{\citenamefont {Li}\ \emph {et~al.}(2024)\citenamefont {Li},
  \citenamefont {Mu}, \citenamefont {You},\ and\ \citenamefont
  {Shao}}]{Li2024Simulation}%
  \BibitemOpen
  \bibfield  {author} {\bibinfo {author} {\bibfnamefont {S.~X.}\ \bibnamefont
  {Li}}, \bibinfo {author} {\bibfnamefont {W.~L.}\ \bibnamefont {Mu}}, \bibinfo
  {author} {\bibfnamefont {J.~B.}\ \bibnamefont {You}}, \ and\ \bibinfo
  {author} {\bibfnamefont {X.~Q.}\ \bibnamefont {Shao}},\ }\bibfield  {title}
  {\emph {\bibinfo {title} {Simulation of a feedback-based algorithm for
  quantum optimization for a realistic neutral-atom system with an optimized
  small-angle controlled-phase gate},\ }}\href {\doibase
  10.1103/PhysRevA.109.062603} {\bibfield  {journal} {\bibinfo  {journal}
  {Phys. Rev. A}\ }\textbf {\bibinfo {volume} {109}},\ \bibinfo {pages}
  {062603} (\bibinfo {year} {2024})}\BibitemShut {NoStop}%
\bibitem [{\citenamefont {Shao}\ \emph {et~al.}(2024)\citenamefont {Shao},
  \citenamefont {Su}, \citenamefont {Li}, \citenamefont {Nath}, \citenamefont
  {Wu},\ and\ \citenamefont {Li}}]{Shao2024Rydberg}%
  \BibitemOpen
  \bibfield  {author} {\bibinfo {author} {\bibfnamefont {X.-Q.}\ \bibnamefont
  {Shao}}, \bibinfo {author} {\bibfnamefont {S.-L.}\ \bibnamefont {Su}},
  \bibinfo {author} {\bibfnamefont {L.}~\bibnamefont {Li}}, \bibinfo {author}
  {\bibfnamefont {R.}~\bibnamefont {Nath}}, \bibinfo {author} {\bibfnamefont
  {J.-H.}\ \bibnamefont {Wu}}, \ and\ \bibinfo {author} {\bibfnamefont
  {W.}~\bibnamefont {Li}},\ }\bibfield  {title} {\emph {\bibinfo {title}
  {Rydberg superatoms: An artificial quantum system for quantum information
  processing and quantum optics},\ }}\href {\doibase 10.1063/5.0211071}
  {\bibfield  {journal} {\bibinfo  {journal} {Appl. Phys. Rev.}\ }\textbf
  {\bibinfo {volume} {11}},\ \bibinfo {pages} {031320} (\bibinfo {year}
  {2024})}\BibitemShut {NoStop}%
\bibitem [{\citenamefont {Li}\ \emph {et~al.}(2019)\citenamefont {Li},
  \citenamefont {Zheng},\ and\ \citenamefont {Shao}}]{Dong2019Adiabatic}%
  \BibitemOpen
  \bibfield  {author} {\bibinfo {author} {\bibfnamefont {D.-X.}\ \bibnamefont
  {Li}}, \bibinfo {author} {\bibfnamefont {T.-Y.}\ \bibnamefont {Zheng}}, \
  and\ \bibinfo {author} {\bibfnamefont {X.-Q.}\ \bibnamefont {Shao}},\
  }\bibfield  {title} {\emph {\bibinfo {title} {Adiabatic preparation of
  multipartite {GHZ} states via rydberg ground-state blockade},\ }}\href
  {\doibase 10.1364/OE.27.020874} {\bibfield  {journal} {\bibinfo  {journal}
  {Opt. Express}\ }\textbf {\bibinfo {volume} {27}},\ \bibinfo {pages} {20874}
  (\bibinfo {year} {2019})}\BibitemShut {NoStop}%
\bibitem [{\citenamefont {Jaksch}\ \emph {et~al.}(2000)\citenamefont {Jaksch},
  \citenamefont {Cirac}, \citenamefont {Zoller}, \citenamefont {Rolston},
  \citenamefont {C\^ot\'e},\ and\ \citenamefont {Lukin}}]{Jaksch2000Fast}%
  \BibitemOpen
  \bibfield  {author} {\bibinfo {author} {\bibfnamefont {D.}~\bibnamefont
  {Jaksch}}, \bibinfo {author} {\bibfnamefont {J.~I.}\ \bibnamefont {Cirac}},
  \bibinfo {author} {\bibfnamefont {P.}~\bibnamefont {Zoller}}, \bibinfo
  {author} {\bibfnamefont {S.~L.}\ \bibnamefont {Rolston}}, \bibinfo {author}
  {\bibfnamefont {R.}~\bibnamefont {C\^ot\'e}}, \ and\ \bibinfo {author}
  {\bibfnamefont {M.~D.}\ \bibnamefont {Lukin}},\ }\bibfield  {title} {\emph
  {\bibinfo {title} {Fast quantum gates for neutral atoms},\ }}\href {\doibase
  10.1103/PhysRevLett.85.2208} {\bibfield  {journal} {\bibinfo  {journal}
  {Phys. Rev. Lett.}\ }\textbf {\bibinfo {volume} {85}},\ \bibinfo {pages}
  {2208} (\bibinfo {year} {2000})}\BibitemShut {NoStop}%
\bibitem [{\citenamefont {Levine}\ \emph {et~al.}(2019)\citenamefont {Levine},
  \citenamefont {Keesling}, \citenamefont {Semeghini}, \citenamefont {Omran},
  \citenamefont {Wang}, \citenamefont {Ebadi}, \citenamefont {Bernien},
  \citenamefont {Greiner}, \citenamefont {Vuleti\ifmmode~\acute{c}\else
  \'{c}\fi{}}, \citenamefont {Pichler},\ and\ \citenamefont
  {Lukin}}]{Levine2019parallel}%
  \BibitemOpen
  \bibfield  {author} {\bibinfo {author} {\bibfnamefont {H.}~\bibnamefont
  {Levine}}, \bibinfo {author} {\bibfnamefont {A.}~\bibnamefont {Keesling}},
  \bibinfo {author} {\bibfnamefont {G.}~\bibnamefont {Semeghini}}, \bibinfo
  {author} {\bibfnamefont {A.}~\bibnamefont {Omran}}, \bibinfo {author}
  {\bibfnamefont {T.~T.}\ \bibnamefont {Wang}}, \bibinfo {author}
  {\bibfnamefont {S.}~\bibnamefont {Ebadi}}, \bibinfo {author} {\bibfnamefont
  {H.}~\bibnamefont {Bernien}}, \bibinfo {author} {\bibfnamefont
  {M.}~\bibnamefont {Greiner}}, \bibinfo {author} {\bibfnamefont
  {V.}~\bibnamefont {Vuleti\ifmmode~\acute{c}\else \'{c}\fi{}}}, \bibinfo
  {author} {\bibfnamefont {H.}~\bibnamefont {Pichler}}, \ and\ \bibinfo
  {author} {\bibfnamefont {M.~D.}\ \bibnamefont {Lukin}},\ }\bibfield  {title}
  {\emph {\bibinfo {title} {Parallel implementation of high-fidelity multiqubit
  gates with neutral atoms},\ }}\href {\doibase 10.1103/PhysRevLett.123.170503}
  {\bibfield  {journal} {\bibinfo  {journal} {Phys. Rev. Lett.}\ }\textbf
  {\bibinfo {volume} {123}},\ \bibinfo {pages} {170503} (\bibinfo {year}
  {2019})}\BibitemShut {NoStop}%
\bibitem [{\citenamefont {Graham}\ \emph {et~al.}(2019)\citenamefont {Graham},
  \citenamefont {Kwon}, \citenamefont {Grinkemeyer}, \citenamefont {Marra},
  \citenamefont {Jiang}, \citenamefont {Lichtman}, \citenamefont {Sun},
  \citenamefont {Ebert},\ and\ \citenamefont {Saffman}}]{Graham2019Rydberg}%
  \BibitemOpen
  \bibfield  {author} {\bibinfo {author} {\bibfnamefont {T.~M.}\ \bibnamefont
  {Graham}}, \bibinfo {author} {\bibfnamefont {M.}~\bibnamefont {Kwon}},
  \bibinfo {author} {\bibfnamefont {B.}~\bibnamefont {Grinkemeyer}}, \bibinfo
  {author} {\bibfnamefont {Z.}~\bibnamefont {Marra}}, \bibinfo {author}
  {\bibfnamefont {X.}~\bibnamefont {Jiang}}, \bibinfo {author} {\bibfnamefont
  {M.~T.}\ \bibnamefont {Lichtman}}, \bibinfo {author} {\bibfnamefont
  {Y.}~\bibnamefont {Sun}}, \bibinfo {author} {\bibfnamefont {M.}~\bibnamefont
  {Ebert}}, \ and\ \bibinfo {author} {\bibfnamefont {M.}~\bibnamefont
  {Saffman}},\ }\bibfield  {title} {\emph {\bibinfo {title} {Rydberg-mediated
  entanglement in a two-dimensional neutral atom qubit array},\ }}\href
  {\doibase 10.1103/PhysRevLett.123.230501} {\bibfield  {journal} {\bibinfo
  {journal} {Phys. Rev. Lett.}\ }\textbf {\bibinfo {volume} {123}},\ \bibinfo
  {pages} {230501} (\bibinfo {year} {2019})}\BibitemShut {NoStop}%
\bibitem [{\citenamefont {M\o{}ller}\ \emph {et~al.}(2008)\citenamefont
  {M\o{}ller}, \citenamefont {Madsen},\ and\ \citenamefont
  {M\o{}lmer}}]{Moller2008Quantumgates}%
  \BibitemOpen
  \bibfield  {author} {\bibinfo {author} {\bibfnamefont {D.}~\bibnamefont
  {M\o{}ller}}, \bibinfo {author} {\bibfnamefont {L.~B.}\ \bibnamefont
  {Madsen}}, \ and\ \bibinfo {author} {\bibfnamefont {K.}~\bibnamefont
  {M\o{}lmer}},\ }\bibfield  {title} {\emph {\bibinfo {title} {Quantum gates
  and multiparticle entanglement by rydberg excitation blockade and adiabatic
  passage},\ }}\href {\doibase 10.1103/PhysRevLett.100.170504} {\bibfield
  {journal} {\bibinfo  {journal} {Phys. Rev. Lett.}\ }\textbf {\bibinfo
  {volume} {100}},\ \bibinfo {pages} {170504} (\bibinfo {year}
  {2008})}\BibitemShut {NoStop}%
\bibitem [{\citenamefont {Su}\ \emph {et~al.}(2020)\citenamefont {Su},
  \citenamefont {Guo}, \citenamefont {Wu}, \citenamefont {Jin}, \citenamefont
  {Shao},\ and\ \citenamefont {Zhang}}]{Su2020Rydberg}%
  \BibitemOpen
  \bibfield  {author} {\bibinfo {author} {\bibfnamefont {S.-L.}\ \bibnamefont
  {Su}}, \bibinfo {author} {\bibfnamefont {F.-Q.}\ \bibnamefont {Guo}},
  \bibinfo {author} {\bibfnamefont {J.-L.}\ \bibnamefont {Wu}}, \bibinfo
  {author} {\bibfnamefont {Z.}~\bibnamefont {Jin}}, \bibinfo {author}
  {\bibfnamefont {X.~Q.}\ \bibnamefont {Shao}}, \ and\ \bibinfo {author}
  {\bibfnamefont {S.}~\bibnamefont {Zhang}},\ }\bibfield  {title} {\emph
  {\bibinfo {title} {Rydberg antiblockade regimes: Dynamics and applications},\
  }}\href {\doibase 10.1209/0295-5075/131/53001} {\bibfield  {journal}
  {\bibinfo  {journal} {Europhys. Lett.}\ }\textbf {\bibinfo {volume} {131}},\
  \bibinfo {pages} {53001} (\bibinfo {year} {2020})}\BibitemShut {NoStop}%
\bibitem [{\citenamefont {Su}\ and\ \citenamefont {Li}(2021)}]{Su2021Dipole}%
  \BibitemOpen
  \bibfield  {author} {\bibinfo {author} {\bibfnamefont {S.-L.}\ \bibnamefont
  {Su}}\ and\ \bibinfo {author} {\bibfnamefont {W.}~\bibnamefont {Li}},\
  }\bibfield  {title} {\emph {\bibinfo {title}
  {Dipole-dipole-interaction--driven antiblockade of two {R}ydberg atoms},\
  }}\href {\doibase 10.1103/PhysRevA.104.033716} {\bibfield  {journal}
  {\bibinfo  {journal} {Phys. Rev. A}\ }\textbf {\bibinfo {volume} {104}},\
  \bibinfo {pages} {033716} (\bibinfo {year} {2021})}\BibitemShut {NoStop}%
\bibitem [{\citenamefont {Zhao}\ \emph
  {et~al.}(2017{\natexlab{b}})\citenamefont {Zhao}, \citenamefont {Liu},
  \citenamefont {Ji}, \citenamefont {Tang},\ and\ \citenamefont
  {Shao}}]{Zhao2017Robust}%
  \BibitemOpen
  \bibfield  {author} {\bibinfo {author} {\bibfnamefont {Y.~J.}\ \bibnamefont
  {Zhao}}, \bibinfo {author} {\bibfnamefont {B.}~\bibnamefont {Liu}}, \bibinfo
  {author} {\bibfnamefont {Y.~Q.}\ \bibnamefont {Ji}}, \bibinfo {author}
  {\bibfnamefont {S.~Q.}\ \bibnamefont {Tang}}, \ and\ \bibinfo {author}
  {\bibfnamefont {X.~Q.}\ \bibnamefont {Shao}},\ }\bibfield  {title} {\emph
  {\bibinfo {title} {Robust generation of entangled state via ground-state
  antiblockade of {R}ydberg atoms},\ }}\href {\doibase
  10.1038/s41598-017-16533-9} {\bibfield  {journal} {\bibinfo  {journal} {Sci.
  Rep.}\ }\textbf {\bibinfo {volume} {7}},\ \bibinfo {pages} {16489} (\bibinfo
  {year} {2017}{\natexlab{b}})}\BibitemShut {NoStop}%
\bibitem [{\citenamefont {Brion}\ \emph {et~al.}(2007)\citenamefont {Brion},
  \citenamefont {Mouritzen},\ and\ \citenamefont
  {M\o{}lmer}}]{Brion2007Conditional}%
  \BibitemOpen
  \bibfield  {author} {\bibinfo {author} {\bibfnamefont {E.}~\bibnamefont
  {Brion}}, \bibinfo {author} {\bibfnamefont {A.~S.}\ \bibnamefont
  {Mouritzen}}, \ and\ \bibinfo {author} {\bibfnamefont {K.}~\bibnamefont
  {M\o{}lmer}},\ }\bibfield  {title} {\emph {\bibinfo {title} {Conditional
  dynamics induced by new configurations for rydberg dipole-dipole
  interactions},\ }}\href {\doibase 10.1103/PhysRevA.76.022334} {\bibfield
  {journal} {\bibinfo  {journal} {Phys. Rev. A}\ }\textbf {\bibinfo {volume}
  {76}},\ \bibinfo {pages} {022334} (\bibinfo {year} {2007})}\BibitemShut
  {NoStop}%
\bibitem [{\citenamefont {Saffman}\ and\ \citenamefont
  {M\o{}lmer}(2009)}]{Saffman2009Efficient}%
  \BibitemOpen
  \bibfield  {author} {\bibinfo {author} {\bibfnamefont {M.}~\bibnamefont
  {Saffman}}\ and\ \bibinfo {author} {\bibfnamefont {K.}~\bibnamefont
  {M\o{}lmer}},\ }\bibfield  {title} {\emph {\bibinfo {title} {Efficient
  multiparticle entanglement via asymmetric rydberg blockade},\ }}\href
  {\doibase 10.1103/PhysRevLett.102.240502} {\bibfield  {journal} {\bibinfo
  {journal} {Phys. Rev. Lett.}\ }\textbf {\bibinfo {volume} {102}},\ \bibinfo
  {pages} {240502} (\bibinfo {year} {2009})}\BibitemShut {NoStop}%
\bibitem [{\citenamefont {Wu}\ \emph {et~al.}(2010)\citenamefont {Wu},
  \citenamefont {Yang},\ and\ \citenamefont {Zheng}}]{Implementation2010Wu}%
  \BibitemOpen
  \bibfield  {author} {\bibinfo {author} {\bibfnamefont {H.-Z.}\ \bibnamefont
  {Wu}}, \bibinfo {author} {\bibfnamefont {Z.-B.}\ \bibnamefont {Yang}}, \ and\
  \bibinfo {author} {\bibfnamefont {S.-B.}\ \bibnamefont {Zheng}},\ }\bibfield
  {title} {\emph {\bibinfo {title} {Implementation of a multiqubit quantum
  phase gate in a neutral atomic ensemble via the asymmetric rydberg
  blockade},\ }}\href {\doibase 10.1103/PhysRevA.82.034307} {\bibfield
  {journal} {\bibinfo  {journal} {Phys. Rev. A}\ }\textbf {\bibinfo {volume}
  {82}},\ \bibinfo {pages} {034307} (\bibinfo {year} {2010})}\BibitemShut
  {NoStop}%
\bibitem [{\citenamefont {Rao}\ and\ \citenamefont
  {M\o{}lmer}(2014)}]{Rao2014Deterministice}%
  \BibitemOpen
  \bibfield  {author} {\bibinfo {author} {\bibfnamefont {D.~D.~B.}\
  \bibnamefont {Rao}}\ and\ \bibinfo {author} {\bibfnamefont {K.}~\bibnamefont
  {M\o{}lmer}},\ }\bibfield  {title} {\emph {\bibinfo {title} {Deterministic
  entanglement of rydberg ensembles by engineered dissipation},\ }}\href
  {\doibase 10.1103/PhysRevA.90.062319} {\bibfield  {journal} {\bibinfo
  {journal} {Phys. Rev. A}\ }\textbf {\bibinfo {volume} {90}},\ \bibinfo
  {pages} {062319} (\bibinfo {year} {2014})}\BibitemShut {NoStop}%
\bibitem [{\citenamefont {Young}\ \emph {et~al.}(2021)\citenamefont {Young},
  \citenamefont {Bienias}, \citenamefont {Belyansky}, \citenamefont {Kaufman},\
  and\ \citenamefont {Gorshkov}}]{Young2021Asymmetric}%
  \BibitemOpen
  \bibfield  {author} {\bibinfo {author} {\bibfnamefont {J.~T.}\ \bibnamefont
  {Young}}, \bibinfo {author} {\bibfnamefont {P.}~\bibnamefont {Bienias}},
  \bibinfo {author} {\bibfnamefont {R.}~\bibnamefont {Belyansky}}, \bibinfo
  {author} {\bibfnamefont {A.~M.}\ \bibnamefont {Kaufman}}, \ and\ \bibinfo
  {author} {\bibfnamefont {A.~V.}\ \bibnamefont {Gorshkov}},\ }\bibfield
  {title} {\emph {\bibinfo {title} {Asymmetric blockade and multiqubit gates
  via dipole-dipole interactions},\ }}\href {\doibase
  10.1103/PhysRevLett.127.120501} {\bibfield  {journal} {\bibinfo  {journal}
  {Phys. Rev. Lett.}\ }\textbf {\bibinfo {volume} {127}},\ \bibinfo {pages}
  {120501} (\bibinfo {year} {2021})}\BibitemShut {NoStop}%
\bibitem [{\citenamefont {Robicheaux}\ \emph {et~al.}(2021)\citenamefont
  {Robicheaux}, \citenamefont {Graham},\ and\ \citenamefont
  {Saffman}}]{Robicheaux2021Photon}%
  \BibitemOpen
  \bibfield  {author} {\bibinfo {author} {\bibfnamefont {F.}~\bibnamefont
  {Robicheaux}}, \bibinfo {author} {\bibfnamefont {T.~M.}\ \bibnamefont
  {Graham}}, \ and\ \bibinfo {author} {\bibfnamefont {M.}~\bibnamefont
  {Saffman}},\ }\bibfield  {title} {\emph {\bibinfo {title} {Photon-recoil and
  laser-focusing limits to rydberg gate fidelity},\ }}\href {\doibase
  10.1103/PhysRevA.103.022424} {\bibfield  {journal} {\bibinfo  {journal}
  {Phys. Rev. A}\ }\textbf {\bibinfo {volume} {103}},\ \bibinfo {pages}
  {022424} (\bibinfo {year} {2021})}\BibitemShut {NoStop}%
\bibitem [{\citenamefont {Pohl}\ and\ \citenamefont
  {Berman}(2009)}]{Pohl2009Breaking}%
  \BibitemOpen
  \bibfield  {author} {\bibinfo {author} {\bibfnamefont {T.}~\bibnamefont
  {Pohl}}\ and\ \bibinfo {author} {\bibfnamefont {P.~R.}\ \bibnamefont
  {Berman}},\ }\bibfield  {title} {\emph {\bibinfo {title} {Breaking the dipole
  blockade: Nearly resonant dipole interactions in few-atom systems},\ }}\href
  {\doibase 10.1103/PhysRevLett.102.013004} {\bibfield  {journal} {\bibinfo
  {journal} {Phys. Rev. Lett.}\ }\textbf {\bibinfo {volume} {102}},\ \bibinfo
  {pages} {013004} (\bibinfo {year} {2009})}\BibitemShut {NoStop}%
\bibitem [{\citenamefont {Urban}\ \emph {et~al.}(2009)\citenamefont {Urban},
  \citenamefont {Johnson}, \citenamefont {Henage}, \citenamefont {Isenhower},
  \citenamefont {Yavuz}, \citenamefont {Walker},\ and\ \citenamefont
  {Saffman}}]{Urban2009Observation}%
  \BibitemOpen
  \bibfield  {author} {\bibinfo {author} {\bibfnamefont {E.}~\bibnamefont
  {Urban}}, \bibinfo {author} {\bibfnamefont {T.~A.}\ \bibnamefont {Johnson}},
  \bibinfo {author} {\bibfnamefont {T.}~\bibnamefont {Henage}}, \bibinfo
  {author} {\bibfnamefont {L.}~\bibnamefont {Isenhower}}, \bibinfo {author}
  {\bibfnamefont {D.~D.}\ \bibnamefont {Yavuz}}, \bibinfo {author}
  {\bibfnamefont {T.~G.}\ \bibnamefont {Walker}}, \ and\ \bibinfo {author}
  {\bibfnamefont {M.}~\bibnamefont {Saffman}},\ }\bibfield  {title} {\emph
  {\bibinfo {title} {Observation of rydberg blockade between two atoms},\
  }}\href {\doibase 10.1038/nphys1178} {\bibfield  {journal} {\bibinfo
  {journal} {Nat. Phys.}\ }\textbf {\bibinfo {volume} {5}},\ \bibinfo {pages}
  {110} (\bibinfo {year} {2009})}\BibitemShut {NoStop}%
\bibitem [{\citenamefont {James}\ and\ \citenamefont
  {Jerke}(2007)}]{James2007Effective}%
  \BibitemOpen
  \bibfield  {author} {\bibinfo {author} {\bibfnamefont {D.~F.}\ \bibnamefont
  {James}}\ and\ \bibinfo {author} {\bibfnamefont {J.}~\bibnamefont {Jerke}},\
  }\bibfield  {title} {\emph {\bibinfo {title} {Effective {H}amiltonian theory
  and its applications in quantum information},\ }}\href {\doibase
  10.1139/p07-060} {\bibfield  {journal} {\bibinfo  {journal} {Can. J. Phys.}\
  }\textbf {\bibinfo {volume} {85}},\ \bibinfo {pages} {625} (\bibinfo {year}
  {2007})}\BibitemShut {NoStop}%
\bibitem [{\citenamefont {Jin}\ and\ \citenamefont
  {Jing}(2025{\natexlab{a}})}]{Jin2025Universal}%
  \BibitemOpen
  \bibfield  {author} {\bibinfo {author} {\bibfnamefont {Z.-y.}\ \bibnamefont
  {Jin}}\ and\ \bibinfo {author} {\bibfnamefont {J.}~\bibnamefont {Jing}},\
  }\bibfield  {title} {\emph {\bibinfo {title} {Universal perspective on
  nonadiabatic quantum control},\ }}\href {\doibase
  10.1103/PhysRevA.111.012406} {\bibfield  {journal} {\bibinfo  {journal}
  {Phys. Rev. A}\ }\textbf {\bibinfo {volume} {111}},\ \bibinfo {pages}
  {012406} (\bibinfo {year} {2025}{\natexlab{a}})}\BibitemShut {NoStop}%
\bibitem [{\citenamefont {Jin}\ and\ \citenamefont
  {Jing}(2025{\natexlab{b}})}]{Jin2025Entangling}%
  \BibitemOpen
  \bibfield  {author} {\bibinfo {author} {\bibfnamefont {Z.-y.}\ \bibnamefont
  {Jin}}\ and\ \bibinfo {author} {\bibfnamefont {J.}~\bibnamefont {Jing}},\
  }\bibfield  {title} {\emph {\bibinfo {title} {Entangling distant systems via
  universal nonadiabatic passage},\ }}\href {\doibase
  10.1103/PhysRevA.111.022628} {\bibfield  {journal} {\bibinfo  {journal}
  {Phys. Rev. A}\ }\textbf {\bibinfo {volume} {111}},\ \bibinfo {pages}
  {022628} (\bibinfo {year} {2025}{\natexlab{b}})}\BibitemShut {NoStop}%
\bibitem [{\citenamefont {Jin}\ and\ \citenamefont
  {Jing}(2025{\natexlab{c}})}]{Jin2025ErrCorr}%
  \BibitemOpen
  \bibfield  {author} {\bibinfo {author} {\bibfnamefont {Z.-y.}\ \bibnamefont
  {Jin}}\ and\ \bibinfo {author} {\bibfnamefont {J.}~\bibnamefont {Jing}},\
  }\bibfield  {title} {\emph {\bibinfo {title} {Universal quantum control with
  error correction},\ }}\href {\doibase 10.48550/arXiv.2502.19786} {\bibfield
  {journal} {\bibinfo  {journal} {arXiv:}\ }\textbf {\bibinfo {volume}
  {2502}},\ \bibinfo {pages} {19786} (\bibinfo {year}
  {2025}{\natexlab{c}})}\BibitemShut {NoStop}%
\bibitem [{\citenamefont {Blanes}\ \emph {et~al.}(2009)\citenamefont {Blanes},
  \citenamefont {Casas}, \citenamefont {Oteo},\ and\ \citenamefont
  {Ros}}]{Blanes2009Magnus}%
  \BibitemOpen
  \bibfield  {author} {\bibinfo {author} {\bibfnamefont {S.}~\bibnamefont
  {Blanes}}, \bibinfo {author} {\bibfnamefont {F.}~\bibnamefont {Casas}},
  \bibinfo {author} {\bibfnamefont {J.}~\bibnamefont {Oteo}}, \ and\ \bibinfo
  {author} {\bibfnamefont {J.}~\bibnamefont {Ros}},\ }\bibfield  {title} {\emph
  {\bibinfo {title} {The magnus expansion and some of its applications},\
  }}\href {\doibase https://doi.org/10.1016/j.physrep.2008.11.001} {\bibfield
  {journal} {\bibinfo  {journal} {Phys. Rep.}\ }\textbf {\bibinfo {volume}
  {470}},\ \bibinfo {pages} {151} (\bibinfo {year} {2009})}\BibitemShut
  {NoStop}%
\bibitem [{\citenamefont {M\"uller}\ \emph {et~al.}(2011)\citenamefont
  {M\"uller}, \citenamefont {Reich}, \citenamefont {Murphy}, \citenamefont
  {Yuan}, \citenamefont {Vala}, \citenamefont {Whaley}, \citenamefont
  {Calarco},\ and\ \citenamefont {Koch}}]{Muller2011Optimizing}%
  \BibitemOpen
  \bibfield  {author} {\bibinfo {author} {\bibfnamefont {M.~M.}\ \bibnamefont
  {M\"uller}}, \bibinfo {author} {\bibfnamefont {D.~M.}\ \bibnamefont {Reich}},
  \bibinfo {author} {\bibfnamefont {M.}~\bibnamefont {Murphy}}, \bibinfo
  {author} {\bibfnamefont {H.}~\bibnamefont {Yuan}}, \bibinfo {author}
  {\bibfnamefont {J.}~\bibnamefont {Vala}}, \bibinfo {author} {\bibfnamefont
  {K.~B.}\ \bibnamefont {Whaley}}, \bibinfo {author} {\bibfnamefont
  {T.}~\bibnamefont {Calarco}}, \ and\ \bibinfo {author} {\bibfnamefont
  {C.~P.}\ \bibnamefont {Koch}},\ }\bibfield  {title} {\emph {\bibinfo {title}
  {Optimizing entangling quantum gates for physical systems},\ }}\href
  {\doibase 10.1103/PhysRevA.84.042315} {\bibfield  {journal} {\bibinfo
  {journal} {Phys. Rev. A}\ }\textbf {\bibinfo {volume} {84}},\ \bibinfo
  {pages} {042315} (\bibinfo {year} {2011})}\BibitemShut {NoStop}%
\bibitem [{\citenamefont {Goerz}\ \emph {et~al.}(2014)\citenamefont {Goerz},
  \citenamefont {Halperin}, \citenamefont {Aytac}, \citenamefont {Koch},\ and\
  \citenamefont {Whaley}}]{Goerz2014Robustness}%
  \BibitemOpen
  \bibfield  {author} {\bibinfo {author} {\bibfnamefont {M.~H.}\ \bibnamefont
  {Goerz}}, \bibinfo {author} {\bibfnamefont {E.~J.}\ \bibnamefont {Halperin}},
  \bibinfo {author} {\bibfnamefont {J.~M.}\ \bibnamefont {Aytac}}, \bibinfo
  {author} {\bibfnamefont {C.~P.}\ \bibnamefont {Koch}}, \ and\ \bibinfo
  {author} {\bibfnamefont {K.~B.}\ \bibnamefont {Whaley}},\ }\bibfield  {title}
  {\emph {\bibinfo {title} {Robustness of high-fidelity rydberg gates with
  single-site addressability},\ }}\href {\doibase 10.1103/PhysRevA.90.032329}
  {\bibfield  {journal} {\bibinfo  {journal} {Phys. Rev. A}\ }\textbf {\bibinfo
  {volume} {90}},\ \bibinfo {pages} {032329} (\bibinfo {year}
  {2014})}\BibitemShut {NoStop}%
\bibitem [{\citenamefont {M\"uller}\ \emph {et~al.}(2014)\citenamefont
  {M\"uller}, \citenamefont {Murphy}, \citenamefont {Montangero}, \citenamefont
  {Calarco}, \citenamefont {Grangier},\ and\ \citenamefont
  {Browaeys}}]{Muller2014Implementation}%
  \BibitemOpen
  \bibfield  {author} {\bibinfo {author} {\bibfnamefont {M.~M.}\ \bibnamefont
  {M\"uller}}, \bibinfo {author} {\bibfnamefont {M.}~\bibnamefont {Murphy}},
  \bibinfo {author} {\bibfnamefont {S.}~\bibnamefont {Montangero}}, \bibinfo
  {author} {\bibfnamefont {T.}~\bibnamefont {Calarco}}, \bibinfo {author}
  {\bibfnamefont {P.}~\bibnamefont {Grangier}}, \ and\ \bibinfo {author}
  {\bibfnamefont {A.}~\bibnamefont {Browaeys}},\ }\bibfield  {title} {\emph
  {\bibinfo {title} {Implementation of an experimentally feasible
  controlled-phase gate on two blockaded {R}ydberg atoms},\ }}\href {\doibase
  10.1103/PhysRevA.89.032334} {\bibfield  {journal} {\bibinfo  {journal} {Phys.
  Rev. A}\ }\textbf {\bibinfo {volume} {89}},\ \bibinfo {pages} {032334}
  (\bibinfo {year} {2014})}\BibitemShut {NoStop}%
\bibitem [{\citenamefont {Carmichael}(1999)}]{Carmichael1999statistical}%
  \BibitemOpen
  \bibfield  {author} {\bibinfo {author} {\bibfnamefont {H.}~\bibnamefont
  {Carmichael}},\ }\href@noop {} {\emph {\bibinfo {title} {Statistical Methods
  in Quantum Optics}}}\ (\bibinfo  {publisher} {Springer, Berlin},\ \bibinfo
  {year} {1999})\BibitemShut {NoStop}%
\bibitem [{\citenamefont {Scully}\ and\ \citenamefont
  {Zubairy}(1997)}]{Scully1997quantum}%
  \BibitemOpen
  \bibfield  {author} {\bibinfo {author} {\bibfnamefont {M.~O.}\ \bibnamefont
  {Scully}}\ and\ \bibinfo {author} {\bibfnamefont {M.~S.}\ \bibnamefont
  {Zubairy}},\ }\href@noop {} {\emph {\bibinfo {title} {Quantum Optics}}}\
  (\bibinfo  {publisher} {Cambridge University, Cambridge},\ \bibinfo {year}
  {1997})\BibitemShut {NoStop}%
\bibitem [{\citenamefont {Sun}\ \emph {et~al.}(2021)\citenamefont {Sun},
  \citenamefont {Yan}, \citenamefont {Su},\ and\ \citenamefont
  {Jia}}]{Sun2021Onestep}%
  \BibitemOpen
  \bibfield  {author} {\bibinfo {author} {\bibfnamefont {L.-N.}\ \bibnamefont
  {Sun}}, \bibinfo {author} {\bibfnamefont {L.-L.}\ \bibnamefont {Yan}},
  \bibinfo {author} {\bibfnamefont {S.-L.}\ \bibnamefont {Su}}, \ and\ \bibinfo
  {author} {\bibfnamefont {Y.}~\bibnamefont {Jia}},\ }\bibfield  {title} {\emph
  {\bibinfo {title} {One-step implementation of time-optimal-control
  three-qubit nonadiabatic holonomic controlled gates in {R}ydberg atoms},\
  }}\href {\doibase 10.1103/PhysRevApplied.16.064040} {\bibfield  {journal}
  {\bibinfo  {journal} {Phys. Rev. Appl.}\ }\textbf {\bibinfo {volume} {16}},\
  \bibinfo {pages} {064040} (\bibinfo {year} {2021})}\BibitemShut {NoStop}%
\bibitem [{\citenamefont {Beterov}\ \emph {et~al.}(2009)\citenamefont
  {Beterov}, \citenamefont {Ryabtsev}, \citenamefont {Tretyakov},\ and\
  \citenamefont {Entin}}]{Beterov2009Quasiclassical}%
  \BibitemOpen
  \bibfield  {author} {\bibinfo {author} {\bibfnamefont {I.~I.}\ \bibnamefont
  {Beterov}}, \bibinfo {author} {\bibfnamefont {I.~I.}\ \bibnamefont
  {Ryabtsev}}, \bibinfo {author} {\bibfnamefont {D.~B.}\ \bibnamefont
  {Tretyakov}}, \ and\ \bibinfo {author} {\bibfnamefont {V.~M.}\ \bibnamefont
  {Entin}},\ }\bibfield  {title} {\emph {\bibinfo {title} {Quasiclassical
  calculations of blackbody-radiation-induced depopulation rates and effective
  lifetimes of rydberg {$nS$}, {$nP$}, and {$nD$} alkali-metal atoms with
  $n\ensuremath{\le}80$},\ }}\href {\doibase 10.1103/PhysRevA.79.052504}
  {\bibfield  {journal} {\bibinfo  {journal} {Phys. Rev. A}\ }\textbf {\bibinfo
  {volume} {79}},\ \bibinfo {pages} {052504} (\bibinfo {year}
  {2009})}\BibitemShut {NoStop}%
\bibitem [{\citenamefont {Isenhower}\ \emph {et~al.}(2010)\citenamefont
  {Isenhower}, \citenamefont {Urban}, \citenamefont {Zhang}, \citenamefont
  {Gill}, \citenamefont {Henage}, \citenamefont {Johnson}, \citenamefont
  {Walker},\ and\ \citenamefont {Saffman}}]{Isenhower2010Demonstration}%
  \BibitemOpen
  \bibfield  {author} {\bibinfo {author} {\bibfnamefont {L.}~\bibnamefont
  {Isenhower}}, \bibinfo {author} {\bibfnamefont {E.}~\bibnamefont {Urban}},
  \bibinfo {author} {\bibfnamefont {X.~L.}\ \bibnamefont {Zhang}}, \bibinfo
  {author} {\bibfnamefont {A.~T.}\ \bibnamefont {Gill}}, \bibinfo {author}
  {\bibfnamefont {T.}~\bibnamefont {Henage}}, \bibinfo {author} {\bibfnamefont
  {T.~A.}\ \bibnamefont {Johnson}}, \bibinfo {author} {\bibfnamefont {T.~G.}\
  \bibnamefont {Walker}}, \ and\ \bibinfo {author} {\bibfnamefont
  {M.}~\bibnamefont {Saffman}},\ }\bibfield  {title} {\emph {\bibinfo {title}
  {Demonstration of a neutral atom controlled-not quantum gate},\ }}\href
  {\doibase 10.1103/PhysRevLett.104.010503} {\bibfield  {journal} {\bibinfo
  {journal} {Phys. Rev. Lett.}\ }\textbf {\bibinfo {volume} {104}},\ \bibinfo
  {pages} {010503} (\bibinfo {year} {2010})}\BibitemShut {NoStop}%
\bibitem [{\citenamefont {Adams}\ \emph {et~al.}(2020)\citenamefont {Adams},
  \citenamefont {Pritchard},\ and\ \citenamefont {Shaffer}}]{Adams2020Rydberg}%
  \BibitemOpen
  \bibfield  {author} {\bibinfo {author} {\bibfnamefont {C.~S.}\ \bibnamefont
  {Adams}}, \bibinfo {author} {\bibfnamefont {J.~D.}\ \bibnamefont
  {Pritchard}}, \ and\ \bibinfo {author} {\bibfnamefont {J.~P.}\ \bibnamefont
  {Shaffer}},\ }\bibfield  {title} {\emph {\bibinfo {title} {Rydberg atom
  quantum technologies},\ }}\href {\doibase 10.1088/1361-6455/ab52ef}
  {\bibfield  {journal} {\bibinfo  {journal} {J. Phys. B: At. Mol. Opt. Phys.}\
  }\textbf {\bibinfo {volume} {53}},\ \bibinfo {pages} {012002} (\bibinfo
  {year} {2020})}\BibitemShut {NoStop}%
\bibitem [{\citenamefont {Cundiff}\ and\ \citenamefont
  {Ye}(2003)}]{Cundiff2003Colloquium}%
  \BibitemOpen
  \bibfield  {author} {\bibinfo {author} {\bibfnamefont {S.~T.}\ \bibnamefont
  {Cundiff}}\ and\ \bibinfo {author} {\bibfnamefont {J.}~\bibnamefont {Ye}},\
  }\bibfield  {title} {\emph {\bibinfo {title} {Colloquium: Femtosecond optical
  frequency combs},\ }}\href {\doibase 10.1103/RevModPhys.75.325} {\bibfield
  {journal} {\bibinfo  {journal} {Rev. Mod. Phys.}\ }\textbf {\bibinfo {volume}
  {75}},\ \bibinfo {pages} {325} (\bibinfo {year} {2003})}\BibitemShut
  {NoStop}%
\bibitem [{\citenamefont {Vaillant}\ \emph {et~al.}(2012)\citenamefont
  {Vaillant}, \citenamefont {Jones},\ and\ \citenamefont
  {Potvliege}}]{Vaillant2012Longrange}%
  \BibitemOpen
  \bibfield  {author} {\bibinfo {author} {\bibfnamefont {C.~L.}\ \bibnamefont
  {Vaillant}}, \bibinfo {author} {\bibfnamefont {M.~P.~A.}\ \bibnamefont
  {Jones}}, \ and\ \bibinfo {author} {\bibfnamefont {R.~M.}\ \bibnamefont
  {Potvliege}},\ }\bibfield  {title} {\emph {\bibinfo {title} {Long-range
  rydberg–rydberg interactions in calcium, strontium and ytterbium},\ }}\href
  {\doibase 10.1088/0953-4075/45/13/135004} {\bibfield  {journal} {\bibinfo
  {journal} {J. Phys. B: At. Mol. Opt. Phys.}\ }\textbf {\bibinfo {volume}
  {45}},\ \bibinfo {pages} {135004} (\bibinfo {year} {2012})}\BibitemShut
  {NoStop}%
\bibitem [{\citenamefont {Zeng}\ \emph {et~al.}(2017)\citenamefont {Zeng},
  \citenamefont {Xu}, \citenamefont {He}, \citenamefont {Liu}, \citenamefont
  {Liu}, \citenamefont {Wang}, \citenamefont {Papoular}, \citenamefont
  {Shlyapnikov},\ and\ \citenamefont {Zhan}}]{Zheng2017Entangling}%
  \BibitemOpen
  \bibfield  {author} {\bibinfo {author} {\bibfnamefont {Y.}~\bibnamefont
  {Zeng}}, \bibinfo {author} {\bibfnamefont {P.}~\bibnamefont {Xu}}, \bibinfo
  {author} {\bibfnamefont {X.}~\bibnamefont {He}}, \bibinfo {author}
  {\bibfnamefont {Y.}~\bibnamefont {Liu}}, \bibinfo {author} {\bibfnamefont
  {M.}~\bibnamefont {Liu}}, \bibinfo {author} {\bibfnamefont {J.}~\bibnamefont
  {Wang}}, \bibinfo {author} {\bibfnamefont {D.~J.}\ \bibnamefont {Papoular}},
  \bibinfo {author} {\bibfnamefont {G.~V.}\ \bibnamefont {Shlyapnikov}}, \ and\
  \bibinfo {author} {\bibfnamefont {M.}~\bibnamefont {Zhan}},\ }\bibfield
  {title} {\emph {\bibinfo {title} {Entangling two individual atoms of
  different isotopes via rydberg blockade},\ }}\href {\doibase
  10.1103/PhysRevLett.119.160502} {\bibfield  {journal} {\bibinfo  {journal}
  {Phys. Rev. Lett.}\ }\textbf {\bibinfo {volume} {119}},\ \bibinfo {pages}
  {160502} (\bibinfo {year} {2017})}\BibitemShut {NoStop}%
\bibitem [{\citenamefont {Lorenz}\ \emph {et~al.}(2021)\citenamefont {Lorenz},
  \citenamefont {Festa}, \citenamefont {Steinert},\ and\ \citenamefont
  {Gross}}]{Nikolaus2021Raman}%
  \BibitemOpen
  \bibfield  {author} {\bibinfo {author} {\bibfnamefont {N.}~\bibnamefont
  {Lorenz}}, \bibinfo {author} {\bibfnamefont {L.}~\bibnamefont {Festa}},
  \bibinfo {author} {\bibfnamefont {L.-M.}\ \bibnamefont {Steinert}}, \ and\
  \bibinfo {author} {\bibfnamefont {C.}~\bibnamefont {Gross}},\ }\bibfield
  {title} {\emph {\bibinfo {title} {Raman sideband cooling in optical tweezer
  arrays for rydberg dressing},\ }}\href {\doibase
  10.21468/SciPostPhys.10.3.052} {\bibfield  {journal} {\bibinfo  {journal}
  {Sci Post Phys.}\ }\textbf {\bibinfo {volume} {10}},\ \bibinfo {pages} {052}
  (\bibinfo {year} {2021})}\BibitemShut {NoStop}%
\bibitem [{\citenamefont {Evered}\ \emph {et~al.}(2023)\citenamefont {Evered},
  \citenamefont {Bluvstein}, \citenamefont {Kalinowski}, \citenamefont {Ebadi},
  \citenamefont {Manovitz}, \citenamefont {Zhou}, \citenamefont {Li},
  \citenamefont {Geim}, \citenamefont {Wang}, \citenamefont {Maskara},
  \citenamefont {Levine}, \citenamefont {Semeghini}, \citenamefont {Greiner},
  \citenamefont {Vuletić},\ and\ \citenamefont
  {Lukin}}]{Evered2023Highfidelity}%
  \BibitemOpen
  \bibfield  {author} {\bibinfo {author} {\bibfnamefont {S.~J.}\ \bibnamefont
  {Evered}}, \bibinfo {author} {\bibfnamefont {D.}~\bibnamefont {Bluvstein}},
  \bibinfo {author} {\bibfnamefont {M.}~\bibnamefont {Kalinowski}}, \bibinfo
  {author} {\bibfnamefont {S.}~\bibnamefont {Ebadi}}, \bibinfo {author}
  {\bibfnamefont {T.}~\bibnamefont {Manovitz}}, \bibinfo {author}
  {\bibfnamefont {H.}~\bibnamefont {Zhou}}, \bibinfo {author} {\bibfnamefont
  {S.~H.}\ \bibnamefont {Li}}, \bibinfo {author} {\bibfnamefont {A.~A.}\
  \bibnamefont {Geim}}, \bibinfo {author} {\bibfnamefont {T.~T.}\ \bibnamefont
  {Wang}}, \bibinfo {author} {\bibfnamefont {N.}~\bibnamefont {Maskara}},
  \bibinfo {author} {\bibfnamefont {H.}~\bibnamefont {Levine}}, \bibinfo
  {author} {\bibfnamefont {G.}~\bibnamefont {Semeghini}}, \bibinfo {author}
  {\bibfnamefont {M.}~\bibnamefont {Greiner}}, \bibinfo {author} {\bibfnamefont
  {V.}~\bibnamefont {Vuletić}}, \ and\ \bibinfo {author} {\bibfnamefont
  {M.~D.}\ \bibnamefont {Lukin}},\ }\bibfield  {title} {\emph {\bibinfo {title}
  {High-fidelity parallel entangling gates on a neutral-atom quantum
  computer},\ }}\href {\doibase 10.1038/s41586-023-06481-y} {\bibfield
  {journal} {\bibinfo  {journal} {Nature}\ }\textbf {\bibinfo {volume} {622}},\
  \bibinfo {pages} {268} (\bibinfo {year} {2023})}\BibitemShut {NoStop}%
\bibitem [{\citenamefont {Bohlouli-Zanjani}\ \emph {et~al.}(2006)\citenamefont
  {Bohlouli-Zanjani}, \citenamefont {Afrousheh},\ and\ \citenamefont
  {Martin}}]{Bohlouli2006Optical}%
  \BibitemOpen
  \bibfield  {author} {\bibinfo {author} {\bibfnamefont {P.}~\bibnamefont
  {Bohlouli-Zanjani}}, \bibinfo {author} {\bibfnamefont {K.}~\bibnamefont
  {Afrousheh}}, \ and\ \bibinfo {author} {\bibfnamefont {J.~D.~D.}\
  \bibnamefont {Martin}},\ }\bibfield  {title} {\emph {\bibinfo {title}
  {Optical transfer cavity stabilization using current-modulated
  injection-locked diode lasers},\ }}\href {\doibase 10.1063/1.2337094}
  {\bibfield  {journal} {\bibinfo  {journal} {Rev. Sci. Instrum.}\ }\textbf
  {\bibinfo {volume} {77}},\ \bibinfo {pages} {093105} (\bibinfo {year}
  {2006})}\BibitemShut {NoStop}%
\bibitem [{\citenamefont {Johnson}\ \emph {et~al.}(2008)\citenamefont
  {Johnson}, \citenamefont {Urban}, \citenamefont {Henage}, \citenamefont
  {Isenhower}, \citenamefont {Yavuz}, \citenamefont {Walker},\ and\
  \citenamefont {Saffman}}]{Johnson2008Rabi}%
  \BibitemOpen
  \bibfield  {author} {\bibinfo {author} {\bibfnamefont {T.~A.}\ \bibnamefont
  {Johnson}}, \bibinfo {author} {\bibfnamefont {E.}~\bibnamefont {Urban}},
  \bibinfo {author} {\bibfnamefont {T.}~\bibnamefont {Henage}}, \bibinfo
  {author} {\bibfnamefont {L.}~\bibnamefont {Isenhower}}, \bibinfo {author}
  {\bibfnamefont {D.~D.}\ \bibnamefont {Yavuz}}, \bibinfo {author}
  {\bibfnamefont {T.~G.}\ \bibnamefont {Walker}}, \ and\ \bibinfo {author}
  {\bibfnamefont {M.}~\bibnamefont {Saffman}},\ }\bibfield  {title} {\emph
  {\bibinfo {title} {Rabi oscillations between ground and rydberg states with
  dipole-dipole atomic interactions},\ }}\href {\doibase
  10.1103/PhysRevLett.100.113003} {\bibfield  {journal} {\bibinfo  {journal}
  {Phys. Rev. Lett.}\ }\textbf {\bibinfo {volume} {100}},\ \bibinfo {pages}
  {113003} (\bibinfo {year} {2008})}\BibitemShut {NoStop}%
\bibitem [{\citenamefont {Abel}\ \emph {et~al.}(2009)\citenamefont {Abel},
  \citenamefont {Mohapatra}, \citenamefont {Bason}, \citenamefont {Pritchard},
  \citenamefont {Weatherill}, \citenamefont {Raitzsch},\ and\ \citenamefont
  {Adams}}]{Abel2009Laser}%
  \BibitemOpen
  \bibfield  {author} {\bibinfo {author} {\bibfnamefont {R.~P.}\ \bibnamefont
  {Abel}}, \bibinfo {author} {\bibfnamefont {A.~K.}\ \bibnamefont {Mohapatra}},
  \bibinfo {author} {\bibfnamefont {M.~G.}\ \bibnamefont {Bason}}, \bibinfo
  {author} {\bibfnamefont {J.~D.}\ \bibnamefont {Pritchard}}, \bibinfo {author}
  {\bibfnamefont {K.~J.}\ \bibnamefont {Weatherill}}, \bibinfo {author}
  {\bibfnamefont {U.}~\bibnamefont {Raitzsch}}, \ and\ \bibinfo {author}
  {\bibfnamefont {C.~S.}\ \bibnamefont {Adams}},\ }\bibfield  {title} {\emph
  {\bibinfo {title} {Laser frequency stabilization to excited state transitions
  using electromagnetically induced transparency in a cascade system},\ }}\href
  {\doibase 10.1063/1.3086305} {\bibfield  {journal} {\bibinfo  {journal}
  {Appl. Phys. Lett.}\ }\textbf {\bibinfo {volume} {94}},\ \bibinfo {pages}
  {071107} (\bibinfo {year} {2009})}\BibitemShut {NoStop}%
\bibitem [{\citenamefont {Weichman}(2024)}]{Weichman2024Doppler}%
  \BibitemOpen
  \bibfield  {author} {\bibinfo {author} {\bibfnamefont {P.~B.}\ \bibnamefont
  {Weichman}},\ }\bibfield  {title} {\emph {\bibinfo {title} {Doppler
  sensitivity and resonant tuning of rydberg atom-based antennas},\ }}\href
  {\doibase 10.1088/1361-6455/ad6385} {\bibfield  {journal} {\bibinfo
  {journal} {J. Phys. B: At. Mol. Opt. Phys.}\ }\textbf {\bibinfo {volume}
  {57}},\ \bibinfo {pages} {165501} (\bibinfo {year} {2024})}\BibitemShut
  {NoStop}%
\bibitem [{\citenamefont {de~Oliveira}\ \emph {et~al.}(2025)\citenamefont
  {de~Oliveira}, \citenamefont {Diamond-Hitchcock}, \citenamefont {Walker},
  \citenamefont {Wells-Pestell}, \citenamefont {Pelegr\'{\i}}, \citenamefont
  {Picken}, \citenamefont {Malcolm}, \citenamefont {Daley}, \citenamefont
  {Bass},\ and\ \citenamefont {Pritchard}}]{Oliveira2025Demonstration}%
  \BibitemOpen
  \bibfield  {author} {\bibinfo {author} {\bibfnamefont {A.~G.}\ \bibnamefont
  {de~Oliveira}}, \bibinfo {author} {\bibfnamefont {E.}~\bibnamefont
  {Diamond-Hitchcock}}, \bibinfo {author} {\bibfnamefont {D.~M.}\ \bibnamefont
  {Walker}}, \bibinfo {author} {\bibfnamefont {M.~T.}\ \bibnamefont
  {Wells-Pestell}}, \bibinfo {author} {\bibfnamefont {G.}~\bibnamefont
  {Pelegr\'{\i}}}, \bibinfo {author} {\bibfnamefont {C.~J.}\ \bibnamefont
  {Picken}}, \bibinfo {author} {\bibfnamefont {G.~P.~A.}\ \bibnamefont
  {Malcolm}}, \bibinfo {author} {\bibfnamefont {A.~J.}\ \bibnamefont {Daley}},
  \bibinfo {author} {\bibfnamefont {J.}~\bibnamefont {Bass}}, \ and\ \bibinfo
  {author} {\bibfnamefont {J.~D.}\ \bibnamefont {Pritchard}},\ }\bibfield
  {title} {\emph {\bibinfo {title} {Demonstration of weighted-graph
  optimization on a rydberg-atom array using local light shifts},\ }}\href
  {\doibase 10.1103/PRXQuantum.6.010301} {\bibfield  {journal} {\bibinfo
  {journal} {PRX Quantum}\ }\textbf {\bibinfo {volume} {6}},\ \bibinfo {pages}
  {010301} (\bibinfo {year} {2025})}\BibitemShut {NoStop}%
\bibitem [{\citenamefont {Zeiher}\ \emph {et~al.}(2016)\citenamefont {Zeiher},
  \citenamefont {van Bijnen}, \citenamefont {Schauß}, \citenamefont {Hild},
  \citenamefont {Choi}, \citenamefont {Pohl}, \citenamefont {Bloch},\ and\
  \citenamefont {Gross}}]{Zeiher2016Manybody}%
  \BibitemOpen
  \bibfield  {author} {\bibinfo {author} {\bibfnamefont {J.}~\bibnamefont
  {Zeiher}}, \bibinfo {author} {\bibfnamefont {R.}~\bibnamefont {van Bijnen}},
  \bibinfo {author} {\bibfnamefont {P.}~\bibnamefont {Schauß}}, \bibinfo
  {author} {\bibfnamefont {S.}~\bibnamefont {Hild}}, \bibinfo {author}
  {\bibfnamefont {J.-y.}\ \bibnamefont {Choi}}, \bibinfo {author}
  {\bibfnamefont {T.}~\bibnamefont {Pohl}}, \bibinfo {author} {\bibfnamefont
  {I.}~\bibnamefont {Bloch}}, \ and\ \bibinfo {author} {\bibfnamefont
  {C.}~\bibnamefont {Gross}},\ }\bibfield  {title} {\emph {\bibinfo {title}
  {Many-body interferometry of a rydberg-dressed spin lattice},\ }}\href
  {\doibase 10.1038/nphys3835} {\bibfield  {journal} {\bibinfo  {journal} {Nat.
  Phys.}\ }\textbf {\bibinfo {volume} {12}},\ \bibinfo {pages} {1095} (\bibinfo
  {year} {2016})}\BibitemShut {NoStop}%
\bibitem [{\citenamefont {Vitanov}\ \emph {et~al.}(2017)\citenamefont
  {Vitanov}, \citenamefont {Rangelov}, \citenamefont {Shore},\ and\
  \citenamefont {Bergmann}}]{Vitanove2017Stimulated}%
  \BibitemOpen
  \bibfield  {author} {\bibinfo {author} {\bibfnamefont {N.~V.}\ \bibnamefont
  {Vitanov}}, \bibinfo {author} {\bibfnamefont {A.~A.}\ \bibnamefont
  {Rangelov}}, \bibinfo {author} {\bibfnamefont {B.~W.}\ \bibnamefont {Shore}},
  \ and\ \bibinfo {author} {\bibfnamefont {K.}~\bibnamefont {Bergmann}},\
  }\bibfield  {title} {\emph {\bibinfo {title} {Stimulated raman adiabatic
  passage in physics, chemistry, and beyond},\ }}\href {\doibase
  10.1103/RevModPhys.89.015006} {\bibfield  {journal} {\bibinfo  {journal}
  {Rev. Mod. Phys.}\ }\textbf {\bibinfo {volume} {89}},\ \bibinfo {pages}
  {015006} (\bibinfo {year} {2017})}\BibitemShut {NoStop}%
\bibitem [{\citenamefont {Jing}\ \emph {et~al.}(2013)\citenamefont {Jing},
  \citenamefont {Wu}, \citenamefont {You},\ and\ \citenamefont
  {Yu}}]{Jing2013Nonperturbative}%
  \BibitemOpen
  \bibfield  {author} {\bibinfo {author} {\bibfnamefont {J.}~\bibnamefont
  {Jing}}, \bibinfo {author} {\bibfnamefont {L.-A.}\ \bibnamefont {Wu}},
  \bibinfo {author} {\bibfnamefont {J.~Q.}\ \bibnamefont {You}}, \ and\
  \bibinfo {author} {\bibfnamefont {T.}~\bibnamefont {Yu}},\ }\bibfield
  {title} {\emph {\bibinfo {title} {Nonperturbative quantum dynamical
  decoupling},\ }}\href {\doibase 10.1103/PhysRevA.88.022333} {\bibfield
  {journal} {\bibinfo  {journal} {Phys. Rev. A}\ }\textbf {\bibinfo {volume}
  {88}},\ \bibinfo {pages} {022333} (\bibinfo {year} {2013})}\BibitemShut
  {NoStop}%
\bibitem [{\citenamefont {Jing}\ \emph {et~al.}(2015)\citenamefont {Jing},
  \citenamefont {Wu}, \citenamefont {Byrd}, \citenamefont {You}, \citenamefont
  {Yu},\ and\ \citenamefont {Wang}}]{Jing2015Nonperturbative}%
  \BibitemOpen
  \bibfield  {author} {\bibinfo {author} {\bibfnamefont {J.}~\bibnamefont
  {Jing}}, \bibinfo {author} {\bibfnamefont {L.-A.}\ \bibnamefont {Wu}},
  \bibinfo {author} {\bibfnamefont {M.}~\bibnamefont {Byrd}}, \bibinfo {author}
  {\bibfnamefont {J.~Q.}\ \bibnamefont {You}}, \bibinfo {author} {\bibfnamefont
  {T.}~\bibnamefont {Yu}}, \ and\ \bibinfo {author} {\bibfnamefont {Z.-M.}\
  \bibnamefont {Wang}},\ }\bibfield  {title} {\emph {\bibinfo {title}
  {Nonperturbative leakage elimination operators and control of a three-level
  system},\ }}\href {\doibase 10.1103/PhysRevLett.114.190502} {\bibfield
  {journal} {\bibinfo  {journal} {Phys. Rev. Lett.}\ }\textbf {\bibinfo
  {volume} {114}},\ \bibinfo {pages} {190502} (\bibinfo {year}
  {2015})}\BibitemShut {NoStop}%
\end{thebibliography}%

\end{document}